\documentclass[reprint,superscriptaddress,preprintnumbers,nofootinbib,showpacs, amsmath,amssymb,aps,prl,floatfix,]{revtex4-2}

\usepackage[colorlinks = true,
            linkcolor = blue,
            urlcolor  = blue,
            citecolor = blue,
            anchorcolor = blue]{hyperref}

\usepackage{placeins}

\usepackage{graphicx}
\usepackage{dcolumn}
\usepackage{bm}
\usepackage{hyperref}
\usepackage{booktabs}
\AtBeginDocument{
\heavyrulewidth=.08em
\lightrulewidth=.05em
\cmidrulewidth=.03em
\belowrulesep=.65ex
\belowbottomsep=0pt
\aboverulesep=.4ex
\abovetopsep=0pt
\cmidrulesep=\doublerulesep
\cmidrulekern=.5em
\defaultaddspace=.5em
}

\usepackage{textcomp}
\usepackage{siunitx}
\DeclareSIUnit{\litre}{\ell}

\begin{document}

\title{First Axion-Like Particle Results from a Broadband Search for Wave-Like Dark Matter in the 44\,to\,52\,$\mu$eV Range with a Coaxial Dish Antenna}

\author{Gabe Hoshino}\email[Correspondence to: gyhoshino@uchicago.edu]{}
\affiliation{Department of Physics, University of Chicago, Chicago, IL 60637, USA}
\affiliation{Enrico Fermi Institute, University of Chicago, Chicago, IL 60637, USA}
\affiliation{Kavli Institute for Cosmological Physics, University of Chicago, Chicago, IL 60637, USA}

\author{Stefan Knirck}
\affiliation{Fermi National Accelerator Laboratory, Batavia, IL 60510, USA}
\affiliation{Laboratory for Particle Physics and Cosmology, Department of Physics, Harvard University, Cambridge, MA 02138, USA}

\author{Mohamed H. Awida}
\affiliation{Fermi National Accelerator Laboratory, Batavia, IL 60510, USA}

\author{Gustavo I. Cancelo}
\affiliation{Fermi National Accelerator Laboratory, Batavia, IL 60510, USA}

\author{Simon Corrodi}
\affiliation{Argonne National Laboratory, Lemont, IL 60439, USA}

\author{Martin Di Federico}
\affiliation{Fermi National Accelerator Laboratory, Batavia, IL 60510, USA}
\affiliation{Universidad Nacional del Sur, IIIE-CONICET, Argentina}

\author{Benjamin~Knepper}
\affiliation{Fermi National Accelerator Laboratory, Batavia, IL 60510, USA}
\affiliation{Enrico Fermi Institute, University of Chicago, Chicago, IL 60637, USA}

\author{Alex Lapuente} 
\affiliation{Department of Physics, University of Chicago, Chicago, IL 60637, USA}

\author{Mira Littmann} 
\affiliation{Department of Physics, University of Chicago, Chicago, IL 60637, USA}

\author{David W. Miller} 
\affiliation{Department of Physics, University of Chicago, Chicago, IL 60637, USA}
\affiliation{Enrico Fermi Institute, University of Chicago, Chicago, IL 60637, USA}
\affiliation{Kavli Institute for Cosmological Physics, University of Chicago, Chicago, IL 60637, USA}

\author{Donald V. Mitchell} 
\affiliation{Fermi National Accelerator Laboratory, Batavia, IL 60510, USA}

\author{Derrick Rodriguez} 
\affiliation{Department of Physics, University of Chicago, Chicago, IL 60637, USA}

\author{Mark K. Ruschman} 
\affiliation{Fermi National Accelerator Laboratory, Batavia, IL 60510, USA}

\author{Chiara P. Salemi}
  \affiliation{SLAC National Accelerator Laboratory/Kavli Institute for Particle Astrophysics and Cosmology, Menlo Park, Stanford University, Stanford, CA 94025, USA}

\author{Matthew A. Sawtell} 
\affiliation{Fermi National Accelerator Laboratory, Batavia, IL 60510, USA}

\author{Leandro Stefanazzi} 
\affiliation{Fermi National Accelerator Laboratory, Batavia, IL 60510, USA}

\author{Andrew Sonnenschein} 
\affiliation{Fermi National Accelerator Laboratory, Batavia, IL 60510, USA}
\affiliation{Enrico Fermi Institute, University of Chicago, Chicago, IL 60637, USA}

\author{Gary W. Teafoe} 
\affiliation{Fermi National Accelerator Laboratory, Batavia, IL 60510, USA}

\author{Peter Winter}
\affiliation{Argonne National Laboratory, Lemont, IL 60439, USA}

\collaboration{GigaBREAD}

\date{\today}

\begin{abstract}
  We present the results from the first axion-like particle search conducted using a dish antenna. The experiment was conducted at room temperature and sensitive to axion-like particles in the $44-52\,\mu\mathrm{eV}$ range ($10.7 - 12.5\,\mathrm{GHz}$). The novel dish antenna geometry was proposed by the BREAD collaboration and previously used to conduct a dark photon search in the same mass range. To allow for axion-like particle sensitivity, the BREAD dish antenna was placed in a $3.9\,\mathrm{T}$ solenoid magnet at Argonne National Laboratory. In the presence of a magnetic field, axion-like dark matter converts to photons at the conductive surface of the reflector. The signal is focused onto a custom coaxial horn antenna and read out with a low-noise radio-frequency receiver. No evidence of axion-like dark matter was observed in this mass range and we place the most stringent laboratory constraints on the axion-photon coupling strength, $g_{a\gamma\gamma}$, in this mass range at 90\% confidence.
\end{abstract}


\maketitle

\begin{figure*}
	\centering
	\includegraphics[width=\linewidth]{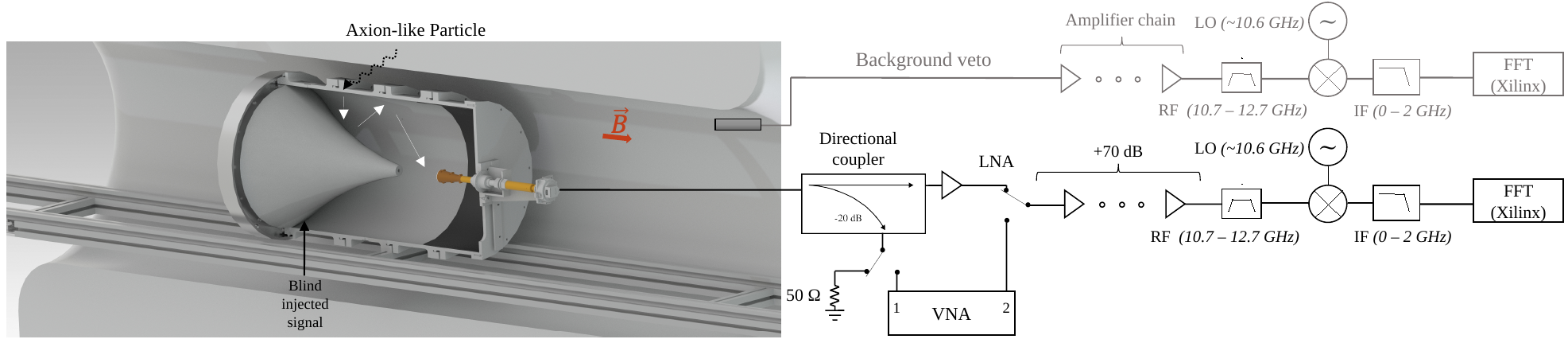}
	\caption{\label{fig:setup} Experimental Setup. ALPs stimulate the emission of photons perpendicular to the conductive walls of the reflector in the presence of a magnetic field. The signal is focused and read out with a horn antenna before being sent through a low-noise receiver chain and being mixed down to and digitized with a DAQ board. A similar receiver chain is used in parallel but which is not connected to the reflector in order to monitor backgrounds. Additionally, a vector network analyzer is connected to the main data acquisition receiver through a radio-frequency (RF) switch and a directional coupler in order to calibrate the antenna position based on reflection measurements.}
\end{figure*}
\emph{Introduction --}
Cosmological and astrophysical observations suggest that the matter content of the universe is dominated by cold dark matter~\cite{Rubin:1970zza,Tyson:1998vp,Tegmark:2003ud,Clowe:2006eq,Akrami:2018vks,Bertone:2004pz}. Roughly 85\% of the matter content of the universe is dark matter, yet the Standard Model of Particle Physics has no compelling candidates to provide a particle model which agrees with the observed properties of dark matter at larger scales. Pseudoscalar dark matter encompasses a wide class of dark matter candidates which may be added in extensions to the Standard Model of Particle Physics \cite{Arvanitaki:2009fg, Jaeckel:2010ni, Arias:2012az, Essig:2013lka, Baker:2013zta, Battaglieri:2017aum}. The QCD axion is an especially compelling pseudoscalar dark matter candidate because, in addition to providing a particle model of dark matter, it can also solve a fine-tuning problem in the Standard Model called the Strong CP problem \cite{Peccei:1977hh, Wilczek:1977pj, Weinberg:1977ma}.

Quantum effects give rise to a relatively generic effective coupling between pseudoscalar dark matter and photons,

\begin{equation}
\label{eq:photon_coupling}
\begin{aligned}
  \mathcal{L} \supset -\frac{1}{4}g_{a\gamma\gamma}aF_{\mu\nu}\tilde{F}^{\mu\nu},
\end{aligned}
\end{equation}
where $g_{a\gamma\gamma}$ is the effective coupling strength, $a$ is the pseudoscalar axion or axion-like field, $F_{\mu\nu}$ is the electromagnetic field strength tensor, and $\tilde{F}^{\mu\nu} = \varepsilon^{\mu\nu\rho\sigma}F_{\rho\sigma}$ with Levi-Civita tensor $\varepsilon^{\mu\nu\rho\sigma}$. Many experiments~\cite{adams2023axiondarkmatter} seek to detect a signal generated by this coupling or place constraints on the coupling strength, $g_{a\gamma\gamma}$~\cite{PhysRevLett.51.1415}. Thus far, the most established laboratory method for measuring $g_{a\gamma\gamma}$ is to use a resonant cavity to enhance the coupling of the dark matter to photons~\cite{PhysRevLett.118.061302, PhysRevD.97.092001, backes2021quantum, haystaccollaboration2023newresultshaystacsphase, haystaccollaboration2024darkmatteraxionsearch, TASEH, PhysRevLett.59.839, PhysRevD.40.3153, PhysRevD.42.1297, HAGMANN1996209}. This method has been very successful for constraining QCD axions in the $\sim 1 \, \mu\mathrm{eV}$ regime and resonant cavity experiments like ADMX~\cite{PhysRevLett.104.041301, PhysRevLett.120.151301, PhysRevLett.124.101303, ADMX:2021nhd, bartram2024axiondarkmatterexperiment} and CAPP\cite{PhysRevLett.124.101802, PhysRevLett.125.221302, PhysRevLett.126.191802, PhysRevD.106.092007, PhysRevLett.128.241805, Kim_2023, Yi_2023, Yang_2023, kim2024experimentalsearchinvisibledark, ahn2024extensivesearchaxiondark} have reached sensitivities high enough to constrain the Kim-Shifman-Vainshtein-Zakharov~\cite{PhysRevLett.43.103, SHIFMAN1980493} and Dine-Fischler-Srednicki-Zhitnitsky~\cite{DINE1981199} benchmark QCD axion models. Even with these successes, QCD axions and pseudoscalar dark matter remains relatively unprobed for masses well above $\sim 1 \, \mu\mathrm{eV}$. Resonant cavities are narrow-band and have volumes, and thus signal strengths, which decrease with shorter wavelength, making them impractical to measure the orders of magnitude of unprobed pseudoscalar dark matter masses above the regime where resonant cavities are currently operating.

To probe these higher masses, dish antennas have been proposed which can detect a dark matter signal through non-resonant dark matter conversion to photons. Dish antenna experiments use a large reflector area with magnetic field parallel to the conductive surface of the reflector, allowing axion-like particles to convert into a photon signal at the reflector surface. The geometry of the reflector surface is designed to focus the signal from the large area of the reflector onto a detector. This geometric focusing can occur regardless of the signal wavelength, making it both broadband and suitable to probe dark matter signals at wavelengths even approaching the optical regime \cite{Horns:2012jf, BREAD:2021tpx}. The converted photons from an axion-like particle produce a signal power of
\begin{equation}
\label{eq:power}
\begin{aligned}
	P\,=&\  4.4 \times10^{-23} \, \mathrm{W} \left(\frac{\eta}{0.5}\right)\left(\frac{g_{a\gamma\gamma}}{10^{-11} \, \mathrm{GeV}^{-1}}\frac{\mathrm{meV}}{m_{a}}\right)^{2}\\ &\times \left(\frac{B_{\mathrm{ext}}}{10 \, \mathrm{T}}\right)^{2}\left(\frac{\rho_{\mathrm{DM}}}{0.45 \, \mathrm{GeV}/\mathrm{cm}^{3}}\right)\left(\frac{A_{\mathrm{dish}}}{10 \, \mathrm{m^{2}}}\right)
\end{aligned}
\end{equation}
where $\eta$ is the efficiency of the detector, $m_{a}$ is the mass of the axion-like particle, $B_{\mathrm{ext}}$ is the strength of an external magnetic field in which the reflector is placed, $\rho_{\mathrm{DM}}$ is the local dark matter halo density, and $A_{\mathrm{dish}}$ is the surface area of the reflector where the photon signal can be generated. The signal frequency $f$ given by the mass of the axion-like particle $m_{\mathrm{a}}$ and a signal lineshape in frequency space caused by the velocity distribution of a non-relativistic dark matter halo $f = m_{\mathrm{a}} c^2 / h + \mathcal{O}(10^{-6})$~\cite{Turner:1990qx,Knirck:2018knd}.

There are a number of existing experiments which have used dish antennas to constrain dark photon dark matter~\cite{Suzuki:2015sza,Brun:2019kak,Tomita:2020usq,Ramanathan:2022egk,DOSUE-RR:2022ise,Bajjali:2023uis,Adachi:2023wuo, knirck2024first}. Dark photon dark matter generates a photon signal through kinetic mixing, a mechanism which does not need an external magnetic field like the photon coupling of axion-like particles. A challenge which any dish antenna must face in order to search for axion-like particles is combining the reflector geometry with a strong magnetic field such that the field is parallel to a large fraction of the reflector's surface, providing a larger DM-photon conversion area and thus a stronger signal. Many of the existing reflector geometries are difficult to combine with a strong magnetic field, but the proposed Broadband Reflector Experiment for Axion Detection (BREAD) offers a novel geometry with a cylindrically symmetric DM-photon conversion surface which can be readily combined with high-field solenoid magnets~\cite{BREAD:2021tpx}. Recently, GigaBREAD, a room temperature pilot experiment, was conducted to search for dark photon dark matter in the $44-52 \, \mu\mathrm{eV}$ range ($10.7-12.5 \, \mathrm{GHz}$)~\cite{knirck2024first}. This work discusses axion-like particle results in the same mass range using a slightly modified version of the GigaBREAD apparatus and a $3.9 \ \mathrm{T}$ solenoid magnet at Argonne National Laboratory. This work represents a milestone for the BREAD reflector concept and the axion detection field as a whole as it is the first axion-like particle search conducted with a dish antenna, opening up this technique for searching the unprobed parameter space of high-mass axions.

\emph{Experimental Setup --}
  Figure~\ref{fig:setup} is a diagram of the apparatus used in this work. An aluminum reflector is placed in the 3.9 Tesla magnetic field of a superconducting solenoid magnet. The reflector barrel is surrounded by foam radio-frequency (RF) absorber and foil shielding (not pictured in Figure~\ref{fig:setup}) to attenuate environmental backgrounds. ALPs interact with the high magnetic field to cause the emission of photons perpendicular to the outer walls of the reflector. This photon signal is focused onto the aperture of a custom coaxial horn antenna which was optimized to couple efficiently to the reflector over the signal band for this work. The horn antenna used in this work was previously used in the first science run of GigaBREAD to look for the dark photon \cite{knirck2024first,Barros:2013coax,Bykov:2008coax}. The antenna is mounted to a piezo motor-driven rod which can move the antenna into and out of the focal spot (up to $\sim 1\,\mathrm{cm}$ above the focus and $\sim 1\,\mathrm{cm}$ below the focus). The signal received by the antenna is then read out with a low-noise amplifier chain and data acquisition board~\cite{xilinx} capable of real-time Fourier transforms and averaging. A pin antenna placed outside of the reflector and a second receiver chain similar to the one just described are used to monitor backgrounds from the environment. Finally, a pin antenna is inserted through a small hole in the cylindrical reflector wall in order to inject test signals.

  Measurements of the reflectivity of power emitted by the antenna can be performed using a network analyzer connected through an RF switch and a directional coupler as shown in Figure~\ref{fig:setup}. These reflectivity measurements can be used as an \emph{in situ} calibration of the antenna position within the barrel. Simulations of reflectivity have been performed using a full-wave azimuthally-symmetric model in COMSOL\textsuperscript{\textregistered}~\cite{COMSOL}. Similar simulations were performed of the signal which would be produced by the modified Maxwell's equations due to the presence of a space-filling effective current induced by a wave-like dark matter field. As discussed in our dark photon result using the GigaBREAD apparatus~\cite{knirck2024first}, the simulations show that the dark matter signal is highest where the reflectivity is highest because radiation emitted from the antenna aperture on-focus will refocus onto the antenna aperture whereas waves emitted off-focus will not receive the same geometric enhancement. By maximizing the frequency-averaged reflectivity, the antenna can be placed on focus.

  The $y$-factor method~\cite{y_factor} was used to characterize the gain and added noise of the receiver chain. The signal is amplified by about $70 \, \mathrm{dB}$ before passing through a $10.7-12.7 \, \mathrm{GHz}$ band-pass filter and down-converted using a broadband mixer to a frequency band that can be read out using a DAQ board. The down-converted intermediate frequency (IF) band is given by $f_{\mathrm{IF}} = |f_{\mathrm{RF}} - f_{\mathrm{LO}}| < 2 \, \mathrm{GHz}$ where $f_{\mathrm{LO}} \approx 10.6 \, \mathrm{GHz}$ is the local oscillator frequency. The IF frequency signal is read out using a Xilinx RFSoC FPGA~\cite{xilinx} board with firmware based on an open source platform called QICK which was developed at Fermilab~\cite{Stefanazzi:2021otz}. The board is able to digitize a $2 \, \mathrm{GHz}$ band with an analog-to-digital converter (ADC) sampling at $4 \, \mathrm{GHz}$. The firmware is able to perform real-time fast Fourier transforms (FFTs), creating power spectra with a resolution that roughly matches the expected linewidth of a dark matter signal, $\Delta f = 7.8 \, \mathrm{kHz}$, providing optimal signal-to-noise performance over a large bandwidth. The average of $10,000$ spectra is performed in firmware before the data is transferred and further processed using a python script on the board's Pynq~\cite{pynq} subsystem.

  There are many sources of radio-frequency interference (RFI) backgrounds which can enter our receiver and potentially appear as a signal in the final averaged spectra. Unlike the previous GigaBREAD dark photon search~\cite{knirck2024first}, the data here could not be taken in an anechoic chamber as we were constrained by the location of the solenoid magnet. Additionally, the solenoid magnet was located in close proximity to the Argonne Wakefield Accelerator which was potentially a significant source of backgrounds.

  Backgrounds which appear in the IF band of the receiver are reduced using an LO frequency hopping scheme~\cite{knirck2024first} in which the LO frequency is randomly shifted during data taking to distribute any backgrounds over many frequency bins, rendering them insignificant. Backgrounds which couple into the receiver before down-conversion and appear in the signal band of the receiver will not be affected by the LO frequency hopping because they are in the same frequency band as potential dark matter signals. These higher frequency backgrounds are attenuated by the reflector barrel itself which is well-isolated from the outside environment. Additional attenuation was provided by RF absorber foam and foil which was placed around the reflector barrel. In order to further mitigate backgrounds which are detectable even after the attenuation from the reflector barrel and shielding, a second antenna and receiver chain are used to veto frequency bins which contain significant backgrounds. This second antenna is a pin antenna placed outside of the reflector barrel but behind some of the RF shielding and is expected to couple to backgrounds but not a dark matter signal as it is not inside of the reflector geometry where the signal is enhanced by focusing. The background monitoring receiver chain is nearly identical to the one used to collect signal data with a low-noise amplifier chain, a broadband mixer, and an RFSoC FPGA board. As part of the data analysis procedure, roughly 1 day of data is averaged at a time from both the signal receiver and background monitoring receiver. For bins which have large backgrounds ($> 5 \sigma$ local significance) in the background monitoring receiver, corresponding bins are masked in the signal receiver data. This masking is performed for the data on a day-by-day basis in order to allow for sensitivity in bins where the backgrounds are transient. The background monitoring receiver was able to mitigate 10 backgrounds which would have appeared as significant signals in our final data.

  \emph{Data Taking Run --} Data was taken in a $3.9 \, \mathrm{T}$ magnetic field at Argonne National Laboratory. Roughly 3 days of data were taken with a system noise temperature of $\sim 400 \, \mathrm{K}$ and 1 month of data were taken with a higher system noise temperature of $\sim 600 \, \mathrm{K}$ due to suboptimal operation of the HEMT low-noise amplifier. Data was taken in 5 sub-runs. Before each sub-run the antenna position was calibrated using reflectivity measurements. The antenna was swept into and out of the focus (from $\sim 1 \, \mathrm{cm}$ above the focus to $\sim 1 \, \mathrm{cm}$ below the focus) for the duration of the experiment with one full cycle taking roughly 4 hours. A blind test signal was injected using a pin antenna inserted into the side of the reflector and a signal generator. The power and frequency of the test signal were unknown to the person performing the initial analysis.

\begin{figure*}
    \centering
    \includegraphics[width=\linewidth]{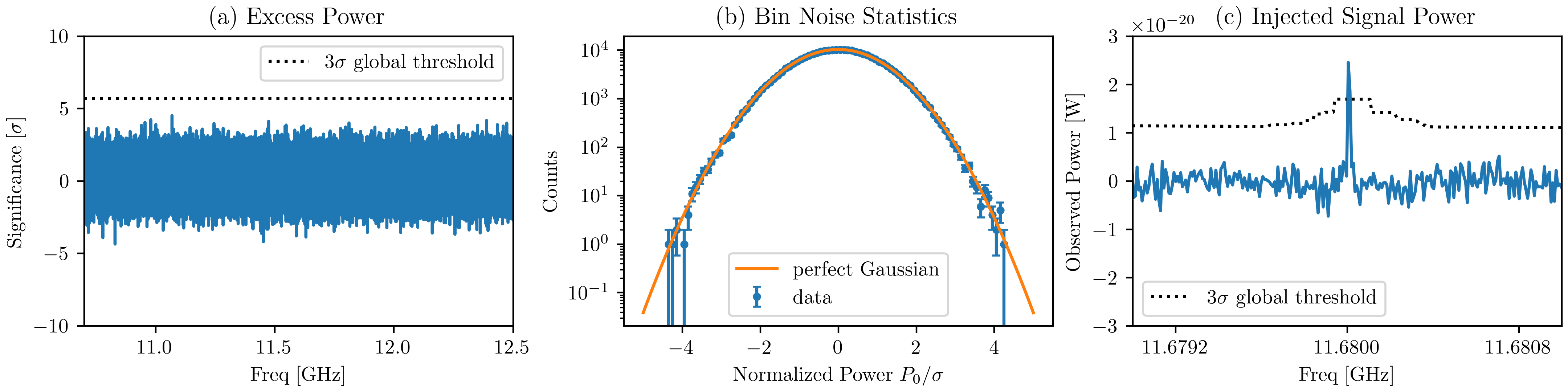}
    \caption{(a) Observed excess power scaled by the standard deviation in each bin in the signal band after accounting for the frequency response of the antenna and reflector and the gain of the receiver chain. The 3$\sigma$ global significance threshold is also shown. None of the frequency bins were observed to have an excess power which passes this threshold. (b) Distribution of the excess powers observed per bin scaled by the corresponding standard deviations which is well-modeled by a Gaussian. (c) A blind test signal which was injected at $11.68 \, \mathrm{GHz}$ and was observed at the expected power of $\sim 10^{-20} \, \mathrm{W}$.}
    \label{fig:data}
\end{figure*}
\emph{Data Analysis --}
The data were analyzed using methods which are standard in cavity haloscope experiments~\cite{ADMX:2020hay}. The baseline of each spectrum was calculated using a fourth-order Savitsky-Golay filter~\cite{doi:10.1021/ac60214a047} and subtracted so that each spectrum is an excess power above the receiver baseline power in each bin, $P_t$. The standard deviation in each bin, $\sigma_t$, is given by $\sigma_{t} = P_{t}/\sqrt{N_{t}}$ where $N_{t}$ is the number of averages, which can vary from bin to bin due to background masking. The antenna position changes the efficiency with which a signal would be received $\eta(z_t)$. $\eta(z_t)$ is determined by comparing reflectivity measurements with COMSOL\textsuperscript{\textregistered}~\cite{COMSOL} simulations of the reflectivity and expected signal at each antenna position. The baseline can also drift slightly over time, changing the expected noise level. To account for these variations over time, the spectra taken over different antenna positions, $z_t$, and at different times, $t$, can be averaged using the optimally weighted average~\cite{ADMX:2001dbg, Brubaker:2017rna},
\begin{equation}
    P_0 = \frac{ \sum_t  \left( {\eta(z_{t})}/{\sigma_{t}} \right)^2 P_{t} / \eta(z_{t}) }{\sum_t  \left( {\eta(z_{t})}/{\sigma_t} \right)^2 },
\end{equation}
The above weighted average was first performed in $\sim 20 \, \mathrm{hour}$ intervals for both the main receiver and compared to an averaged spectrum over the same 20 hour interval for the background monitoring receiver chain. Any bins with backgrounds in the background monitoring data which appear with a significance of more than $5\sigma$ were masked in the science data for that 20 hour interval. After all the 20 hour intervals were averaged together, the injected test signal was identified at $11.68 \, \mathrm{GHz}$ with the expected power of $\sim 10^{-20} \, \mathrm{W}$ and is shown in Fig.~\ref{fig:data}\,(c). The frequency bin containing this signal was removed from the remaining results presented here. To account for signal spillover between frequency bins, signals were searched for by using a cross-correlation of the observed excess powers with the expected lineshape from the Maxwell-Boltzmann velocity distribution of a dark matter halo~\cite{Turner:1990qx}.

If signals are significantly reflected at the first amplifier in our receiver chain, a signal may be attenuated due to interference with itself. This is accounted for using measurements of the reflectivity of the reflector and antenna and reflectivity measurements of the first amplifier and all the components between the first amplifier and the horn antenna. These measurements were combined into an rf circuit simulation using the scikit-rf~\cite{skrf} Python package to estimate the degree of interference a signal would have with itself. Across all frequencies and antenna positions, this effect attenuates our signal by a factor of no more than $67\%$.

\begin{figure*}
    \centering
    \includegraphics[width=\linewidth]{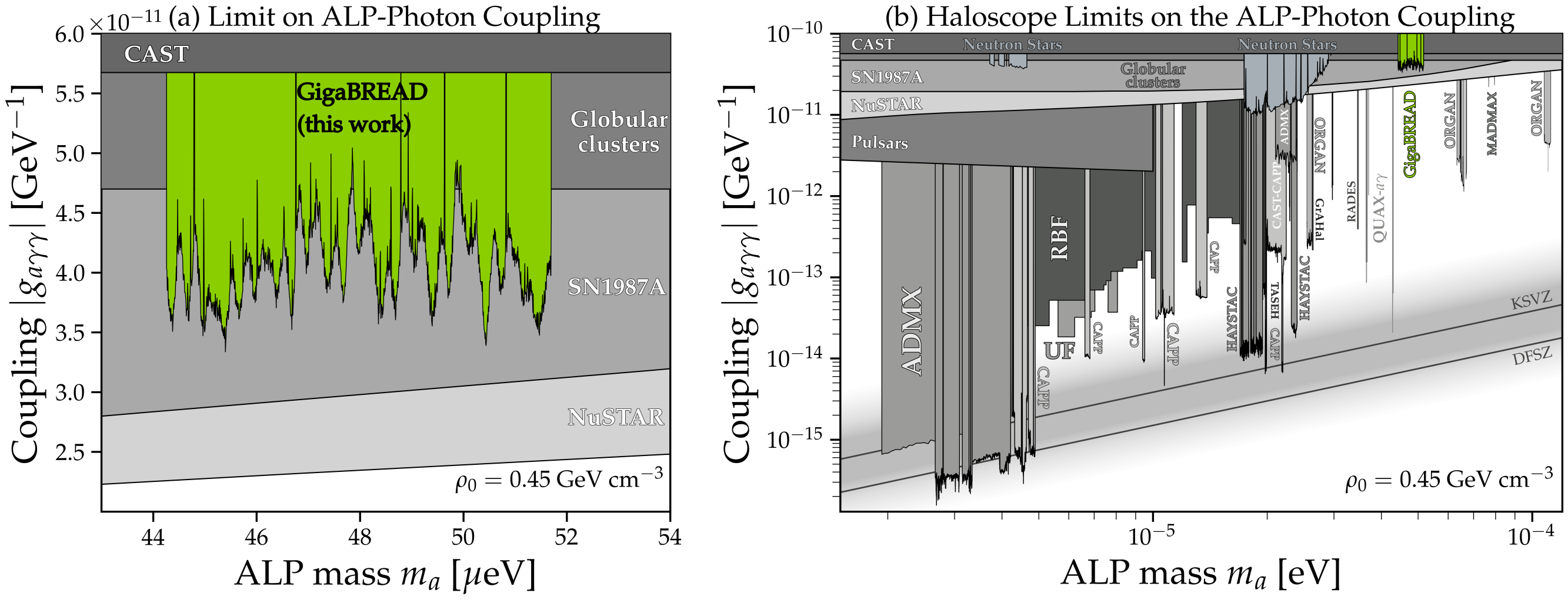}
    \caption{Axion-like particle photon coupling parameter space~\cite{AxionLimits} excluded in this work (GigaBREAD), compared to CAST~\cite{CAST:2007jps, Anastassopoulos:2017ftl, CAST:2024eil}, astrophysical constraints~\cite{Hoof:2022xbe, Manzari:2024jns, Ayala:2014pea, Dolan:2022kul, ruz2024nustaraxionhelioscope, PhysRevLett.131.111004, PhysRevLett.125.171301, Darling_2020, PhysRevD.105.L021305, PhysRevLett.129.251102, Battye_2023}, and other haloscopes~\cite{PhysRevLett.104.041301, PhysRevLett.120.151301, PhysRevLett.124.101303, ADMX:2021nhd, bartram2024axiondarkmatterexperiment, PhysRevLett.121.261302, Bartram_2023, PhysRevLett.124.101802, PhysRevLett.125.221302, PhysRevLett.126.191802, PhysRevD.106.092007, PhysRevLett.128.241805, Kim_2023, Yi_2023, Yang_2023, kim2024experimentalsearchinvisibledark, ahn2024extensivesearchaxiondark, Adair_2022, PhysRevD.42.1297, HAGMANN1996209, PhysRevLett.59.839, PhysRevD.40.3153, PhysRevLett.118.061302, PhysRevD.97.092001, backes2021quantum, haystaccollaboration2023newresultshaystacsphase, haystaccollaboration2024darkmatteraxionsearch, TASEH, grenet2021grenobleaxionhaloscopeplatform, Alesini:2019ajt, Alesini:2020vny, Alesini:2022lnp, QUAX:2023gop, QUAX:2024fut, alvarez2021first, ahyoune2024radesaxionsearchresults, McAllister:2017lkb, Quiskamp:2022pks, quiskamp2023exclusionalpcogenesisdark, quiskamp2024nearquantumlimitedaxiondark, garcia2024searchaxiondarkmatter}.}
    \label{fig:limit}
\end{figure*}

As can be seen in Fig.~\ref{fig:data}\,(a) no signal surpasses a $3\sigma$ global threshold, so a 90\% confidence level upper limit on the dark matter power was calculated using the observed positive excess powers in each bin. From the upper limit on DM signal power, a limit on $g_{a\gamma\gamma}$ was calculated using Eq.~\ref{eq:power} and the result is shown in Fig.~\ref{fig:limit}.

\begin{table}[]
\caption{\label{tab:systematics} Frequency-averaged systematic uncertainties on the ALP signal power.}
\vspace{0.2cm}
\begin{tabular}{@{}lr@{}}
\toprule
Effect                                  & Uncertainty on $P_{0}$ \\ \midrule
Non-zero DM velocity & $< \SI{1}{\percent}$\\
Detection Efficiency Simulation  & \SI{17}{\percent}\\[2pt]
System Noise Temperature & \SI{9}{\percent}  \\
Gain Variations & \SI{8}{\percent} \\[2pt] 
Baseline Removal & \SI{3}{\percent} \\
\midrule
Total & \SI{21}{\percent}\\
\bottomrule
\end{tabular}
\end{table}

Table~\ref{tab:systematics} shows the estimated systematic uncertainties. Due to the non-zero velocity of dark matter, the momentum of the dark matter can be transferred to the photon signal and cause the focal spot to become spread out~\cite{Jaeckel:2013sqa,Jaeckel:2015kea,Jaeckel:2017sjb}. In this relatively light mass regime the de~Broglie wavelength of the axion-like particles is significantly larger than our reflector $\lambda_{\mathrm{DB}} \sim 30 \ \mathrm{m}$. Thus, in this work, velocity effects are negligible because the dark matter wave is approximately coherent and excites the lowest mode of our reflector.

As in the GigaBREAD dark photon experiment~\cite{knirck2024first}, the largest systematic uncertainty is in the simulated efficiency of the coupling between the coaxial antenna and signals generated at the surface of the reflector. The uncertainty is estimated by comparing in-lab reflectivity measurements with simulation. Ultimately these discrepancies are consistent with the impact of known geometric imperfections in the fabricated reflector which could not be practically accounted for in simulations. Another systematic uncertainty comes from fluctuations in the amplifier gain and noise level over time. The Savitsky-Golay filter used to estimate the amplifier baseline power can introduce systematic uncertainty by attenuating signals~\cite{Brubaker:2017rna,Diehl:2023fuk}. This attenuation for the filter parameters used in this work is estimated to be less than $3\%$.
All systematic uncertainties were added in quadrature to obtain a total systematic uncertainty at each frequency and the most conservative $P_{0}$ within this uncertainty was taken to calculate the limit.

\emph{Conclusion --} Ultimately no power excess above the $3\sigma$ discovery threshold was observed. The 90\% confidence exclusion limit we are able to place on axion-like dark matter in the $44-52 \ \mu\mathrm{eV}$ range is shown in Fig.~\ref{fig:limit}. This is the most sensitive laboratory measurement in this mass range. Additionally, this work is the first axion-like particle result from a dish antenna experiment, demonstrating the relative advantage of the BREAD reflector geometry which is optimal for use with solenoid magnets. As a pilot experiment, GigaBREAD paves the way for future dish antenna axion experiments. An upgraded version of GigaBREAD using quantum-limited amplifiers and a $9.4 \ \mathrm{T}$ field could be performed with relatively small modifications in the ADMX-EFR magnet at Fermilab~\cite{ADMX-EFR}. BREAD can also be implemented for higher frequency experiments using terahertz and infrared photon sensors capable of single photon sensitivity \cite{BREAD:2021tpx}.

\emph{Acknowledgments --}
We thank Argonne National Laboratory for use of the solenoid magnet facility which enabled the search for axion-like particles, under contract DE-AC02-06CH11357 (Argonne National Laboratory). We also thank the University of Chicago Joint Task Force Initiative (JTFI) for its generous support for the resources needed to ramp the magnet to its full field strength, thereby maximizing the sensitivity of the experiment.  G.H. would also like to thank JTFI for its support. 
C. P. Salemi is supported by the Kavli Institute for Particle Astrophysics and Cosmology Porat Fellowship.
We thank Abigail Vieregg and her group for the use of the RF absorber materials.
We thank Dan Zhang and the ADMX collaboration for helpful discussions about analysis. 
We thank all BREAD collaborators for inspiring and helpful discussions.
This work is funded by the by the Department of Energy through the resources of the Fermi National Accelerator Laboratory (Fermilab), a U.S. Department of Energy, Office of Science, HEP User Facility. Fermi National Accelerator Laboratory is managed by Fermi Forward Discovery Group, LLC (FFDG), acting under Contract No. 89243024CSC000002.

\bibliography{refs}

%apsrev4-2.bst 2019-01-14 (MD) hand-edited version of apsrev4-1.bst
%Control: key (0)
%Control: author (8) initials jnrlst
%Control: editor formatted (1) identically to author
%Control: production of article title (0) allowed
%Control: page (0) single
%Control: year (1) truncated
%Control: production of eprint (0) enabled
\begin{thebibliography}{105}%
\makeatletter
\providecommand \@ifxundefined [1]{%
 \@ifx{#1\undefined}
}%
\providecommand \@ifnum [1]{%
 \ifnum #1\expandafter \@firstoftwo
 \else \expandafter \@secondoftwo
 \fi
}%
\providecommand \@ifx [1]{%
 \ifx #1\expandafter \@firstoftwo
 \else \expandafter \@secondoftwo
 \fi
}%
\providecommand \natexlab [1]{#1}%
\providecommand \enquote  [1]{``#1''}%
\providecommand \bibnamefont  [1]{#1}%
\providecommand \bibfnamefont [1]{#1}%
\providecommand \citenamefont [1]{#1}%
\providecommand \href@noop [0]{\@secondoftwo}%
\providecommand \href [0]{\begingroup \@sanitize@url \@href}%
\providecommand \@href[1]{\@@startlink{#1}\@@href}%
\providecommand \@@href[1]{\endgroup#1\@@endlink}%
\providecommand \@sanitize@url [0]{\catcode `\\12\catcode `\$12\catcode `\&12\catcode `\#12\catcode `\^12\catcode `\_12\catcode `\%12\relax}%
\providecommand \@@startlink[1]{}%
\providecommand \@@endlink[0]{}%
\providecommand \url  [0]{\begingroup\@sanitize@url \@url }%
\providecommand \@url [1]{\endgroup\@href {#1}{\urlprefix }}%
\providecommand \urlprefix  [0]{URL }%
\providecommand \Eprint [0]{\href }%
\providecommand \doibase [0]{https://doi.org/}%
\providecommand \selectlanguage [0]{\@gobble}%
\providecommand \bibinfo  [0]{\@secondoftwo}%
\providecommand \bibfield  [0]{\@secondoftwo}%
\providecommand \translation [1]{[#1]}%
\providecommand \BibitemOpen [0]{}%
\providecommand \bibitemStop [0]{}%
\providecommand \bibitemNoStop [0]{.\EOS\space}%
\providecommand \EOS [0]{\spacefactor3000\relax}%
\providecommand \BibitemShut  [1]{\csname bibitem#1\endcsname}%
\let\auto@bib@innerbib\@empty
%</preamble>
\bibitem [{\citenamefont {Rubin}\ and\ \citenamefont {Ford}(1970)}]{Rubin:1970zza}%
  \BibitemOpen
  \bibfield  {author} {\bibinfo {author} {\bibfnamefont {V.~C.}\ \bibnamefont {Rubin}}\ and\ \bibinfo {author} {\bibfnamefont {W.~K.}\ \bibnamefont {Ford}, \bibfnamefont {Jr.}},\ }\bibfield  {title} {\bibinfo {title} {{Rotation of the Andromeda Nebula from a Spectroscopic Survey of Emission Regions}},\ }\href {https://doi.org/10.1086/150317} {\bibfield  {journal} {\bibinfo  {journal} {Astrophys. J.}\ }\textbf {\bibinfo {volume} {159}},\ \bibinfo {pages} {379} (\bibinfo {year} {1970})}\BibitemShut {NoStop}%
%%CITATION = ASJOA,159,379;%%
\bibitem [{\citenamefont {Tyson}\ \emph {et~al.}(1998)\citenamefont {Tyson}, \citenamefont {Kochanski},\ and\ \citenamefont {Dell'Antonio}}]{Tyson:1998vp}%
  \BibitemOpen
  \bibfield  {author} {\bibinfo {author} {\bibfnamefont {J.~A.}\ \bibnamefont {Tyson}}, \bibinfo {author} {\bibfnamefont {G.~P.}\ \bibnamefont {Kochanski}},\ and\ \bibinfo {author} {\bibfnamefont {I.~P.}\ \bibnamefont {Dell'Antonio}},\ }\bibfield  {title} {\bibinfo {title} {{Detailed mass map of CL0024+1654 from strong lensing}},\ }\href {https://doi.org/10.1086/311314} {\bibfield  {journal} {\bibinfo  {journal} {Astrophys. J.}\ }\textbf {\bibinfo {volume} {498}},\ \bibinfo {pages} {L107} (\bibinfo {year} {1998})},\ \Eprint {https://arxiv.org/abs/astro-ph/9801193} {arXiv:astro-ph/9801193 [astro-ph]} \BibitemShut {NoStop}%
%%CITATION = ASTRO-PH/9801193;%%
\bibitem [{\citenamefont {Tegmark}\ \emph {et~al.}(2004)\citenamefont {Tegmark} \emph {et~al.}}]{Tegmark:2003ud}%
  \BibitemOpen
  \bibfield  {author} {\bibinfo {author} {\bibfnamefont {M.}~\bibnamefont {Tegmark}} \emph {et~al.} (\bibinfo {collaboration} {SDSS}),\ }\bibfield  {title} {\bibinfo {title} {{Cosmological parameters from SDSS and WMAP}},\ }\href {https://doi.org/10.1103/PhysRevD.69.103501} {\bibfield  {journal} {\bibinfo  {journal} {Phys. Rev. D}\ }\textbf {\bibinfo {volume} {69}},\ \bibinfo {pages} {103501} (\bibinfo {year} {2004})},\ \Eprint {https://arxiv.org/abs/astro-ph/0310723} {arXiv:astro-ph/0310723} \BibitemShut {NoStop}%
\bibitem [{\citenamefont {Clowe}\ \emph {et~al.}(2006)\citenamefont {Clowe}, \citenamefont {Bradac}, \citenamefont {Gonzalez}, \citenamefont {Markevitch}, \citenamefont {Randall} \emph {et~al.}}]{Clowe:2006eq}%
  \BibitemOpen
  \bibfield  {author} {\bibinfo {author} {\bibfnamefont {D.}~\bibnamefont {Clowe}}, \bibinfo {author} {\bibfnamefont {M.}~\bibnamefont {Bradac}}, \bibinfo {author} {\bibfnamefont {A.~H.}\ \bibnamefont {Gonzalez}}, \bibinfo {author} {\bibfnamefont {M.}~\bibnamefont {Markevitch}}, \bibinfo {author} {\bibfnamefont {S.~W.}\ \bibnamefont {Randall}}, \emph {et~al.},\ }\bibfield  {title} {\bibinfo {title} {{A direct empirical proof of the existence of dark matter}},\ }\href {https://doi.org/10.1086/508162} {\bibfield  {journal} {\bibinfo  {journal} {Astrophys. J.}\ }\textbf {\bibinfo {volume} {648}},\ \bibinfo {pages} {L109} (\bibinfo {year} {2006})},\ \Eprint {https://arxiv.org/abs/astro-ph/0608407} {arXiv:astro-ph/0608407 [astro-ph]} \BibitemShut {NoStop}%
%%CITATION = ASTRO-PH/0608407;%%
\bibitem [{\citenamefont {Aghanim}\ \emph {et~al.}(2020)\citenamefont {Aghanim} \emph {et~al.}}]{Akrami:2018vks}%
  \BibitemOpen
  \bibfield  {author} {\bibinfo {author} {\bibfnamefont {N.}~\bibnamefont {Aghanim}} \emph {et~al.} (\bibinfo {collaboration} {Planck}),\ }\bibfield  {title} {\bibinfo {title} {{Planck 2018 results. I. Overview and the cosmological legacy of Planck}},\ }\href {https://doi.org/10.1051/0004-6361/201833880} {\bibfield  {journal} {\bibinfo  {journal} {Astron. Astrophys.}\ }\textbf {\bibinfo {volume} {641}},\ \bibinfo {pages} {A1} (\bibinfo {year} {2020})},\ \Eprint {https://arxiv.org/abs/1807.06205} {arXiv:1807.06205 [astro-ph.CO]} \BibitemShut {NoStop}%
\bibitem [{\citenamefont {Bertone}\ \emph {et~al.}(2005)\citenamefont {Bertone}, \citenamefont {Hooper},\ and\ \citenamefont {Silk}}]{Bertone:2004pz}%
  \BibitemOpen
  \bibfield  {author} {\bibinfo {author} {\bibfnamefont {G.}~\bibnamefont {Bertone}}, \bibinfo {author} {\bibfnamefont {D.}~\bibnamefont {Hooper}},\ and\ \bibinfo {author} {\bibfnamefont {J.}~\bibnamefont {Silk}},\ }\bibfield  {title} {\bibinfo {title} {{Particle dark matter: Evidence, candidates and constraints}},\ }\href {https://doi.org/10.1016/j.physrep.2004.08.031} {\bibfield  {journal} {\bibinfo  {journal} {Phys. Rept.}\ }\textbf {\bibinfo {volume} {405}},\ \bibinfo {pages} {279} (\bibinfo {year} {2005})},\ \Eprint {https://arxiv.org/abs/hep-ph/0404175} {arXiv:hep-ph/0404175 [hep-ph]} \BibitemShut {NoStop}%
%%CITATION = HEP-PH/0404175;%%
\bibitem [{\citenamefont {Arvanitaki}\ \emph {et~al.}(2010)\citenamefont {Arvanitaki}, \citenamefont {Dimopoulos}, \citenamefont {Dubovsky}, \citenamefont {Kaloper},\ and\ \citenamefont {March-Russell}}]{Arvanitaki:2009fg}%
  \BibitemOpen
  \bibfield  {author} {\bibinfo {author} {\bibfnamefont {A.}~\bibnamefont {Arvanitaki}}, \bibinfo {author} {\bibfnamefont {S.}~\bibnamefont {Dimopoulos}}, \bibinfo {author} {\bibfnamefont {S.}~\bibnamefont {Dubovsky}}, \bibinfo {author} {\bibfnamefont {N.}~\bibnamefont {Kaloper}},\ and\ \bibinfo {author} {\bibfnamefont {J.}~\bibnamefont {March-Russell}},\ }\bibfield  {title} {\bibinfo {title} {{String Axiverse}},\ }\href {https://doi.org/10.1103/PhysRevD.81.123530} {\bibfield  {journal} {\bibinfo  {journal} {Phys. Rev. D}\ }\textbf {\bibinfo {volume} {81}},\ \bibinfo {pages} {123530} (\bibinfo {year} {2010})},\ \Eprint {https://arxiv.org/abs/0905.4720} {arXiv:0905.4720 [hep-th]} \BibitemShut {NoStop}%
\bibitem [{\citenamefont {Jaeckel}\ and\ \citenamefont {Ringwald}(2010)}]{Jaeckel:2010ni}%
  \BibitemOpen
  \bibfield  {author} {\bibinfo {author} {\bibfnamefont {J.}~\bibnamefont {Jaeckel}}\ and\ \bibinfo {author} {\bibfnamefont {A.}~\bibnamefont {Ringwald}},\ }\bibfield  {title} {\bibinfo {title} {{The Low-Energy Frontier of Particle Physics}},\ }\href {https://doi.org/10.1146/annurev.nucl.012809.104433} {\bibfield  {journal} {\bibinfo  {journal} {Ann. Rev. Nucl. Part. Sci.}\ }\textbf {\bibinfo {volume} {60}},\ \bibinfo {pages} {405} (\bibinfo {year} {2010})},\ \Eprint {https://arxiv.org/abs/1002.0329} {arXiv:1002.0329 [hep-ph]} \BibitemShut {NoStop}%
\bibitem [{\citenamefont {Arias}\ \emph {et~al.}(2012)\citenamefont {Arias}, \citenamefont {Cadamuro}, \citenamefont {Goodsell}, \citenamefont {Jaeckel}, \citenamefont {Redondo},\ and\ \citenamefont {Ringwald}}]{Arias:2012az}%
  \BibitemOpen
  \bibfield  {author} {\bibinfo {author} {\bibfnamefont {P.}~\bibnamefont {Arias}}, \bibinfo {author} {\bibfnamefont {D.}~\bibnamefont {Cadamuro}}, \bibinfo {author} {\bibfnamefont {M.}~\bibnamefont {Goodsell}}, \bibinfo {author} {\bibfnamefont {J.}~\bibnamefont {Jaeckel}}, \bibinfo {author} {\bibfnamefont {J.}~\bibnamefont {Redondo}},\ and\ \bibinfo {author} {\bibfnamefont {A.}~\bibnamefont {Ringwald}},\ }\bibfield  {title} {\bibinfo {title} {{WISPy Cold Dark Matter}},\ }\href {https://doi.org/10.1088/1475-7516/2012/06/013} {\bibfield  {journal} {\bibinfo  {journal} {JCAP}\ }\textbf {\bibinfo {volume} {06}},\ \bibinfo {pages} {013}},\ \Eprint {https://arxiv.org/abs/1201.5902} {arXiv:1201.5902 [hep-ph]} \BibitemShut {NoStop}%
%%CITATION = ARXIV:1201.5902;%%
\bibitem [{\citenamefont {Essig}\ \emph {et~al.}(2013)\citenamefont {Essig} \emph {et~al.}}]{Essig:2013lka}%
  \BibitemOpen
  \bibfield  {author} {\bibinfo {author} {\bibfnamefont {R.}~\bibnamefont {Essig}} \emph {et~al.},\ }\bibfield  {title} {\bibinfo {title} {{Working Group Report: New Light Weakly Coupled Particles}},\ }in\ \href {https://inspirehep.net/record/1263039/files/arXiv:1311.0029.pdf} {\emph {\bibinfo {booktitle} {{Community Summer Study 2013: Snowmass on the Mississippi}}}}\ (\bibinfo {year} {2013})\ \Eprint {https://arxiv.org/abs/1311.0029} {arXiv:1311.0029 [hep-ph]} \BibitemShut {NoStop}%
%%CITATION = ARXIV:1311.0029;%%
\bibitem [{\citenamefont {Baker}\ \emph {et~al.}(2013)\citenamefont {Baker} \emph {et~al.}}]{Baker:2013zta}%
  \BibitemOpen
  \bibfield  {author} {\bibinfo {author} {\bibfnamefont {K.}~\bibnamefont {Baker}} \emph {et~al.},\ }\bibfield  {title} {\bibinfo {title} {{The quest for axions and other new light particles}},\ }\href {https://doi.org/10.1002/andp.201300727} {\bibfield  {journal} {\bibinfo  {journal} {Ann. Phys. (Berlin)}\ }\textbf {\bibinfo {volume} {525}},\ \bibinfo {pages} {A93} (\bibinfo {year} {2013})},\ \Eprint {https://arxiv.org/abs/1306.2841} {arXiv:1306.2841 [hep-ph]} \BibitemShut {NoStop}%
%%CITATION = ARXIV:1306.2841;%%
\bibitem [{\citenamefont {Battaglieri}\ \emph {et~al.}(2017)\citenamefont {Battaglieri} \emph {et~al.}}]{Battaglieri:2017aum}%
  \BibitemOpen
  \bibfield  {author} {\bibinfo {author} {\bibfnamefont {M.}~\bibnamefont {Battaglieri}} \emph {et~al.},\ }\bibfield  {title} {\bibinfo {title} {{New Ideas in Dark Matter 2017: Community Report}},\ }in\ \href {http://lss.fnal.gov/archive/2017/conf/fermilab-conf-17-282-ae-ppd-t.pdf} {\emph {\bibinfo {booktitle} {{U.S. Cosmic Visions}}}}\ (\bibinfo {year} {2017})\ \Eprint {https://arxiv.org/abs/1707.04591} {arXiv:1707.04591 [hep-ph]} \BibitemShut {NoStop}%
%%CITATION = ARXIV:1707.04591;%%
\bibitem [{\citenamefont {Peccei}\ and\ \citenamefont {Quinn}(1977)}]{Peccei:1977hh}%
  \BibitemOpen
  \bibfield  {author} {\bibinfo {author} {\bibfnamefont {R.~D.}\ \bibnamefont {Peccei}}\ and\ \bibinfo {author} {\bibfnamefont {H.~R.}\ \bibnamefont {Quinn}},\ }\bibfield  {title} {\bibinfo {title} {{CP Conservation in the Presence of Instantons}},\ }\href {https://doi.org/10.1103/PhysRevLett.38.1440} {\bibfield  {journal} {\bibinfo  {journal} {Phys.\ Rev.\ Lett.}\ }\textbf {\bibinfo {volume} {38}},\ \bibinfo {pages} {1440} (\bibinfo {year} {1977})}\BibitemShut {NoStop}%
\bibitem [{\citenamefont {Wilczek}(1978)}]{Wilczek:1977pj}%
  \BibitemOpen
  \bibfield  {author} {\bibinfo {author} {\bibfnamefont {F.}~\bibnamefont {Wilczek}},\ }\bibfield  {title} {\bibinfo {title} {{Problem of Strong $P$ and $T$ Invariance in the Presence of Instantons}},\ }\href {https://doi.org/10.1103/PhysRevLett.40.279} {\bibfield  {journal} {\bibinfo  {journal} {Phys.\ Rev.\ Lett.}\ }\textbf {\bibinfo {volume} {40}},\ \bibinfo {pages} {279} (\bibinfo {year} {1978})}\BibitemShut {NoStop}%
\bibitem [{\citenamefont {Weinberg}(1978)}]{Weinberg:1977ma}%
  \BibitemOpen
  \bibfield  {author} {\bibinfo {author} {\bibfnamefont {S.}~\bibnamefont {Weinberg}},\ }\bibfield  {title} {\bibinfo {title} {{A New Light Boson?}},\ }\href {https://doi.org/10.1103/PhysRevLett.40.223} {\bibfield  {journal} {\bibinfo  {journal} {Phys.\ Rev.\ Lett.}\ }\textbf {\bibinfo {volume} {40}},\ \bibinfo {pages} {223} (\bibinfo {year} {1978})}\BibitemShut {NoStop}%
\bibitem [{\citenamefont {Adams}\ \emph {et~al.}(2023)\citenamefont {Adams}, \citenamefont {Aggarwal}, \citenamefont {Agrawal}, \citenamefont {Balafendiev}, \citenamefont {Bartram}, \citenamefont {Baryakhtar}, \citenamefont {Bekker}, \citenamefont {Belov}, \citenamefont {Berggren}, \citenamefont {Berlin} \emph {et~al.}}]{adams2023axiondarkmatter}%
  \BibitemOpen
  \bibfield  {author} {\bibinfo {author} {\bibfnamefont {C.}~\bibnamefont {Adams}}, \bibinfo {author} {\bibfnamefont {N.}~\bibnamefont {Aggarwal}}, \bibinfo {author} {\bibfnamefont {A.}~\bibnamefont {Agrawal}}, \bibinfo {author} {\bibfnamefont {R.}~\bibnamefont {Balafendiev}}, \bibinfo {author} {\bibfnamefont {C.}~\bibnamefont {Bartram}}, \bibinfo {author} {\bibfnamefont {M.}~\bibnamefont {Baryakhtar}}, \bibinfo {author} {\bibfnamefont {H.}~\bibnamefont {Bekker}}, \bibinfo {author} {\bibfnamefont {P.}~\bibnamefont {Belov}}, \bibinfo {author} {\bibfnamefont {K.}~\bibnamefont {Berggren}}, \bibinfo {author} {\bibfnamefont {A.}~\bibnamefont {Berlin}}, \emph {et~al.},\ }\href {https://arxiv.org/abs/2203.14923} {\bibinfo {title} {Axion dark matter}} (\bibinfo {year} {2023}),\ \Eprint {https://arxiv.org/abs/2203.14923} {arXiv:2203.14923 [hep-ex]} \BibitemShut {NoStop}%
\bibitem [{\citenamefont {Sikivie}(1983)}]{PhysRevLett.51.1415}%
  \BibitemOpen
  \bibfield  {author} {\bibinfo {author} {\bibfnamefont {P.}~\bibnamefont {Sikivie}},\ }\bibfield  {title} {\bibinfo {title} {Experimental tests of the "invisible" axion},\ }\href {https://doi.org/10.1103/PhysRevLett.51.1415} {\bibfield  {journal} {\bibinfo  {journal} {Phys. Rev. Lett.}\ }\textbf {\bibinfo {volume} {51}},\ \bibinfo {pages} {1415} (\bibinfo {year} {1983})}\BibitemShut {NoStop}%
\bibitem [{\citenamefont {Brubaker}\ \emph {et~al.}(2017{\natexlab{a}})\citenamefont {Brubaker}, \citenamefont {Zhong}, \citenamefont {Gurevich}, \citenamefont {Cahn}, \citenamefont {Lamoreaux}, \citenamefont {Simanovskaia}, \citenamefont {Root}, \citenamefont {Lewis}, \citenamefont {Al~Kenany}, \citenamefont {Backes}, \citenamefont {Urdinaran}, \citenamefont {Rapidis}, \citenamefont {Shokair}, \citenamefont {van Bibber}, \citenamefont {Palken}, \citenamefont {Malnou}, \citenamefont {Kindel}, \citenamefont {Anil}, \citenamefont {Lehnert},\ and\ \citenamefont {Carosi}}]{PhysRevLett.118.061302}%
  \BibitemOpen
  \bibfield  {author} {\bibinfo {author} {\bibfnamefont {B.~M.}\ \bibnamefont {Brubaker}}, \bibinfo {author} {\bibfnamefont {L.}~\bibnamefont {Zhong}}, \bibinfo {author} {\bibfnamefont {Y.~V.}\ \bibnamefont {Gurevich}}, \bibinfo {author} {\bibfnamefont {S.~B.}\ \bibnamefont {Cahn}}, \bibinfo {author} {\bibfnamefont {S.~K.}\ \bibnamefont {Lamoreaux}}, \bibinfo {author} {\bibfnamefont {M.}~\bibnamefont {Simanovskaia}}, \bibinfo {author} {\bibfnamefont {J.~R.}\ \bibnamefont {Root}}, \bibinfo {author} {\bibfnamefont {S.~M.}\ \bibnamefont {Lewis}}, \bibinfo {author} {\bibfnamefont {S.}~\bibnamefont {Al~Kenany}}, \bibinfo {author} {\bibfnamefont {K.~M.}\ \bibnamefont {Backes}}, \bibinfo {author} {\bibfnamefont {I.}~\bibnamefont {Urdinaran}}, \bibinfo {author} {\bibfnamefont {N.~M.}\ \bibnamefont {Rapidis}}, \bibinfo {author} {\bibfnamefont {T.~M.}\ \bibnamefont {Shokair}}, \bibinfo {author} {\bibfnamefont {K.~A.}\ \bibnamefont {van Bibber}}, \bibinfo {author} {\bibfnamefont {D.~A.}\ \bibnamefont {Palken}}, \bibinfo
  {author} {\bibfnamefont {M.}~\bibnamefont {Malnou}}, \bibinfo {author} {\bibfnamefont {W.~F.}\ \bibnamefont {Kindel}}, \bibinfo {author} {\bibfnamefont {M.~A.}\ \bibnamefont {Anil}}, \bibinfo {author} {\bibfnamefont {K.~W.}\ \bibnamefont {Lehnert}},\ and\ \bibinfo {author} {\bibfnamefont {G.}~\bibnamefont {Carosi}},\ }\bibfield  {title} {\bibinfo {title} {First results from a microwave cavity axion search at $24\text{ }\text{ }\ensuremath{\mu}\mathrm{eV}$},\ }\href {https://doi.org/10.1103/PhysRevLett.118.061302} {\bibfield  {journal} {\bibinfo  {journal} {Phys. Rev. Lett.}\ }\textbf {\bibinfo {volume} {118}},\ \bibinfo {pages} {061302} (\bibinfo {year} {2017}{\natexlab{a}})}\BibitemShut {NoStop}%
\bibitem [{\citenamefont {Zhong}\ \emph {et~al.}(2018)\citenamefont {Zhong}, \citenamefont {Al~Kenany}, \citenamefont {Backes}, \citenamefont {Brubaker}, \citenamefont {Cahn}, \citenamefont {Carosi}, \citenamefont {Gurevich}, \citenamefont {Kindel}, \citenamefont {Lamoreaux}, \citenamefont {Lehnert}, \citenamefont {Lewis}, \citenamefont {Malnou}, \citenamefont {Maruyama}, \citenamefont {Palken}, \citenamefont {Rapidis}, \citenamefont {Root}, \citenamefont {Simanovskaia}, \citenamefont {Shokair}, \citenamefont {Speller}, \citenamefont {Urdinaran},\ and\ \citenamefont {van Bibber}}]{PhysRevD.97.092001}%
  \BibitemOpen
  \bibfield  {author} {\bibinfo {author} {\bibfnamefont {L.}~\bibnamefont {Zhong}}, \bibinfo {author} {\bibfnamefont {S.}~\bibnamefont {Al~Kenany}}, \bibinfo {author} {\bibfnamefont {K.~M.}\ \bibnamefont {Backes}}, \bibinfo {author} {\bibfnamefont {B.~M.}\ \bibnamefont {Brubaker}}, \bibinfo {author} {\bibfnamefont {S.~B.}\ \bibnamefont {Cahn}}, \bibinfo {author} {\bibfnamefont {G.}~\bibnamefont {Carosi}}, \bibinfo {author} {\bibfnamefont {Y.~V.}\ \bibnamefont {Gurevich}}, \bibinfo {author} {\bibfnamefont {W.~F.}\ \bibnamefont {Kindel}}, \bibinfo {author} {\bibfnamefont {S.~K.}\ \bibnamefont {Lamoreaux}}, \bibinfo {author} {\bibfnamefont {K.~W.}\ \bibnamefont {Lehnert}}, \bibinfo {author} {\bibfnamefont {S.~M.}\ \bibnamefont {Lewis}}, \bibinfo {author} {\bibfnamefont {M.}~\bibnamefont {Malnou}}, \bibinfo {author} {\bibfnamefont {R.~H.}\ \bibnamefont {Maruyama}}, \bibinfo {author} {\bibfnamefont {D.~A.}\ \bibnamefont {Palken}}, \bibinfo {author} {\bibfnamefont {N.~M.}\ \bibnamefont {Rapidis}}, \bibinfo {author}
  {\bibfnamefont {J.~R.}\ \bibnamefont {Root}}, \bibinfo {author} {\bibfnamefont {M.}~\bibnamefont {Simanovskaia}}, \bibinfo {author} {\bibfnamefont {T.~M.}\ \bibnamefont {Shokair}}, \bibinfo {author} {\bibfnamefont {D.~H.}\ \bibnamefont {Speller}}, \bibinfo {author} {\bibfnamefont {I.}~\bibnamefont {Urdinaran}},\ and\ \bibinfo {author} {\bibfnamefont {K.~A.}\ \bibnamefont {van Bibber}},\ }\bibfield  {title} {\bibinfo {title} {Results from phase 1 of the haystac microwave cavity axion experiment},\ }\href {https://doi.org/10.1103/PhysRevD.97.092001} {\bibfield  {journal} {\bibinfo  {journal} {Phys. Rev. D}\ }\textbf {\bibinfo {volume} {97}},\ \bibinfo {pages} {092001} (\bibinfo {year} {2018})}\BibitemShut {NoStop}%
\bibitem [{\citenamefont {Backes}\ \emph {et~al.}(2021)\citenamefont {Backes}, \citenamefont {Palken}, \citenamefont {Kenany}, \citenamefont {Brubaker}, \citenamefont {Cahn}, \citenamefont {Droster}, \citenamefont {Hilton}, \citenamefont {Ghosh}, \citenamefont {Jackson}, \citenamefont {Lamoreaux} \emph {et~al.}}]{backes2021quantum}%
  \BibitemOpen
  \bibfield  {author} {\bibinfo {author} {\bibfnamefont {K.~M.}\ \bibnamefont {Backes}}, \bibinfo {author} {\bibfnamefont {D.~A.}\ \bibnamefont {Palken}}, \bibinfo {author} {\bibfnamefont {S.~A.}\ \bibnamefont {Kenany}}, \bibinfo {author} {\bibfnamefont {B.~M.}\ \bibnamefont {Brubaker}}, \bibinfo {author} {\bibfnamefont {S.}~\bibnamefont {Cahn}}, \bibinfo {author} {\bibfnamefont {A.}~\bibnamefont {Droster}}, \bibinfo {author} {\bibfnamefont {G.~C.}\ \bibnamefont {Hilton}}, \bibinfo {author} {\bibfnamefont {S.}~\bibnamefont {Ghosh}}, \bibinfo {author} {\bibfnamefont {H.}~\bibnamefont {Jackson}}, \bibinfo {author} {\bibfnamefont {S.~K.}\ \bibnamefont {Lamoreaux}}, \emph {et~al.},\ }\bibfield  {title} {\bibinfo {title} {A quantum enhanced search for dark matter axions},\ }\href@noop {} {\bibfield  {journal} {\bibinfo  {journal} {Nature}\ }\textbf {\bibinfo {volume} {590}},\ \bibinfo {pages} {238} (\bibinfo {year} {2021})}\BibitemShut {NoStop}%
\bibitem [{\citenamefont {Collaboration}\ \emph {et~al.}(2023)\citenamefont {Collaboration}, \citenamefont {Jewell}, \citenamefont {Leder}, \citenamefont {Backes}, \citenamefont {Bai}, \citenamefont {van Bibber}, \citenamefont {Brubaker}, \citenamefont {Cahn}, \citenamefont {Droster}, \citenamefont {Esmat}, \citenamefont {Ghosh}, \citenamefont {Graham}, \citenamefont {Hilton}, \citenamefont {Jackson}, \citenamefont {Laffan}, \citenamefont {Lamoreaux}, \citenamefont {Lehnert}, \citenamefont {Lewis}, \citenamefont {Malnou}, \citenamefont {Maruyama}, \citenamefont {Palken}, \citenamefont {Rapidis}, \citenamefont {Ruddy}, \citenamefont {Simanovskaia}, \citenamefont {Singh}, \citenamefont {Speller}, \citenamefont {Vale}, \citenamefont {Wang},\ and\ \citenamefont {Zhu}}]{haystaccollaboration2023newresultshaystacsphase}%
  \BibitemOpen
  \bibfield  {author} {\bibinfo {author} {\bibfnamefont {H.}~\bibnamefont {Collaboration}}, \bibinfo {author} {\bibfnamefont {M.~J.}\ \bibnamefont {Jewell}}, \bibinfo {author} {\bibfnamefont {A.~F.}\ \bibnamefont {Leder}}, \bibinfo {author} {\bibfnamefont {K.~M.}\ \bibnamefont {Backes}}, \bibinfo {author} {\bibfnamefont {X.}~\bibnamefont {Bai}}, \bibinfo {author} {\bibfnamefont {K.}~\bibnamefont {van Bibber}}, \bibinfo {author} {\bibfnamefont {B.~M.}\ \bibnamefont {Brubaker}}, \bibinfo {author} {\bibfnamefont {S.~B.}\ \bibnamefont {Cahn}}, \bibinfo {author} {\bibfnamefont {A.}~\bibnamefont {Droster}}, \bibinfo {author} {\bibfnamefont {M.~H.}\ \bibnamefont {Esmat}}, \bibinfo {author} {\bibfnamefont {S.}~\bibnamefont {Ghosh}}, \bibinfo {author} {\bibfnamefont {E.}~\bibnamefont {Graham}}, \bibinfo {author} {\bibfnamefont {G.~C.}\ \bibnamefont {Hilton}}, \bibinfo {author} {\bibfnamefont {H.}~\bibnamefont {Jackson}}, \bibinfo {author} {\bibfnamefont {C.}~\bibnamefont {Laffan}}, \bibinfo {author} {\bibfnamefont
  {S.~K.}\ \bibnamefont {Lamoreaux}}, \bibinfo {author} {\bibfnamefont {K.~W.}\ \bibnamefont {Lehnert}}, \bibinfo {author} {\bibfnamefont {S.~M.}\ \bibnamefont {Lewis}}, \bibinfo {author} {\bibfnamefont {M.}~\bibnamefont {Malnou}}, \bibinfo {author} {\bibfnamefont {R.~H.}\ \bibnamefont {Maruyama}}, \bibinfo {author} {\bibfnamefont {D.~A.}\ \bibnamefont {Palken}}, \bibinfo {author} {\bibfnamefont {N.~M.}\ \bibnamefont {Rapidis}}, \bibinfo {author} {\bibfnamefont {E.~P.}\ \bibnamefont {Ruddy}}, \bibinfo {author} {\bibfnamefont {M.}~\bibnamefont {Simanovskaia}}, \bibinfo {author} {\bibfnamefont {S.}~\bibnamefont {Singh}}, \bibinfo {author} {\bibfnamefont {D.~H.}\ \bibnamefont {Speller}}, \bibinfo {author} {\bibfnamefont {L.~R.}\ \bibnamefont {Vale}}, \bibinfo {author} {\bibfnamefont {H.}~\bibnamefont {Wang}},\ and\ \bibinfo {author} {\bibfnamefont {Y.}~\bibnamefont {Zhu}},\ }\href {https://arxiv.org/abs/2301.09721} {\bibinfo {title} {New results from haystac's phase ii operation with a squeezed state receiver}}
  (\bibinfo {year} {2023}),\ \Eprint {https://arxiv.org/abs/2301.09721} {arXiv:2301.09721 [hep-ex]} \BibitemShut {NoStop}%
\bibitem [{\citenamefont {Collaboration}\ \emph {et~al.}(2024)\citenamefont {Collaboration}, \citenamefont {Bai}, \citenamefont {Jewell}, \citenamefont {Echevers}, \citenamefont {van Bibber}, \citenamefont {Droster}, \citenamefont {Esmat}, \citenamefont {Ghosh}, \citenamefont {Graham}, \citenamefont {Jackson}, \citenamefont {Laffan}, \citenamefont {Lamoreaux}, \citenamefont {Leder}, \citenamefont {Lehnert}, \citenamefont {Lewis}, \citenamefont {Maruyama}, \citenamefont {Nath}, \citenamefont {Rapidis}, \citenamefont {Ruddy}, \citenamefont {Silva-Feaver}, \citenamefont {Simanovskaia}, \citenamefont {Singh}, \citenamefont {Speller}, \citenamefont {Zacarias},\ and\ \citenamefont {Zhu}}]{haystaccollaboration2024darkmatteraxionsearch}%
  \BibitemOpen
  \bibfield  {author} {\bibinfo {author} {\bibfnamefont {H.}~\bibnamefont {Collaboration}}, \bibinfo {author} {\bibfnamefont {X.}~\bibnamefont {Bai}}, \bibinfo {author} {\bibfnamefont {M.~J.}\ \bibnamefont {Jewell}}, \bibinfo {author} {\bibfnamefont {J.}~\bibnamefont {Echevers}}, \bibinfo {author} {\bibfnamefont {K.}~\bibnamefont {van Bibber}}, \bibinfo {author} {\bibfnamefont {A.}~\bibnamefont {Droster}}, \bibinfo {author} {\bibfnamefont {M.~H.}\ \bibnamefont {Esmat}}, \bibinfo {author} {\bibfnamefont {S.}~\bibnamefont {Ghosh}}, \bibinfo {author} {\bibfnamefont {E.}~\bibnamefont {Graham}}, \bibinfo {author} {\bibfnamefont {H.}~\bibnamefont {Jackson}}, \bibinfo {author} {\bibfnamefont {C.}~\bibnamefont {Laffan}}, \bibinfo {author} {\bibfnamefont {S.~K.}\ \bibnamefont {Lamoreaux}}, \bibinfo {author} {\bibfnamefont {A.~F.}\ \bibnamefont {Leder}}, \bibinfo {author} {\bibfnamefont {K.~W.}\ \bibnamefont {Lehnert}}, \bibinfo {author} {\bibfnamefont {S.~M.}\ \bibnamefont {Lewis}}, \bibinfo {author} {\bibfnamefont
  {R.~H.}\ \bibnamefont {Maruyama}}, \bibinfo {author} {\bibfnamefont {R.~D.}\ \bibnamefont {Nath}}, \bibinfo {author} {\bibfnamefont {N.~M.}\ \bibnamefont {Rapidis}}, \bibinfo {author} {\bibfnamefont {E.~P.}\ \bibnamefont {Ruddy}}, \bibinfo {author} {\bibfnamefont {M.}~\bibnamefont {Silva-Feaver}}, \bibinfo {author} {\bibfnamefont {M.}~\bibnamefont {Simanovskaia}}, \bibinfo {author} {\bibfnamefont {S.}~\bibnamefont {Singh}}, \bibinfo {author} {\bibfnamefont {D.~H.}\ \bibnamefont {Speller}}, \bibinfo {author} {\bibfnamefont {S.}~\bibnamefont {Zacarias}},\ and\ \bibinfo {author} {\bibfnamefont {Y.}~\bibnamefont {Zhu}},\ }\href {https://arxiv.org/abs/2409.08998} {\bibinfo {title} {Dark matter axion search with haystac phase ii}} (\bibinfo {year} {2024}),\ \Eprint {https://arxiv.org/abs/2409.08998} {arXiv:2409.08998 [hep-ex]} \BibitemShut {NoStop}%
\bibitem [{\citenamefont {Chang}\ \emph {et~al.}(2022)\citenamefont {Chang}, \citenamefont {Chang}, \citenamefont {Chang}, \citenamefont {Chang}, \citenamefont {Chang}, \citenamefont {Chen}, \citenamefont {Chen}, \citenamefont {Chen}, \citenamefont {Chen}, \citenamefont {Chiang}, \citenamefont {Chien}, \citenamefont {Doan}, \citenamefont {Hung}, \citenamefont {Kuo}, \citenamefont {Lai}, \citenamefont {Liu}, \citenamefont {OuYang}, \citenamefont {Wu},\ and\ \citenamefont {Yu}}]{TASEH}%
  \BibitemOpen
  \bibfield  {author} {\bibinfo {author} {\bibfnamefont {H.}~\bibnamefont {Chang}}, \bibinfo {author} {\bibfnamefont {J.-Y.}\ \bibnamefont {Chang}}, \bibinfo {author} {\bibfnamefont {Y.-C.}\ \bibnamefont {Chang}}, \bibinfo {author} {\bibfnamefont {Y.-H.}\ \bibnamefont {Chang}}, \bibinfo {author} {\bibfnamefont {Y.-H.}\ \bibnamefont {Chang}}, \bibinfo {author} {\bibfnamefont {C.-H.}\ \bibnamefont {Chen}}, \bibinfo {author} {\bibfnamefont {C.-F.}\ \bibnamefont {Chen}}, \bibinfo {author} {\bibfnamefont {K.-Y.}\ \bibnamefont {Chen}}, \bibinfo {author} {\bibfnamefont {Y.-F.}\ \bibnamefont {Chen}}, \bibinfo {author} {\bibfnamefont {W.-Y.}\ \bibnamefont {Chiang}}, \bibinfo {author} {\bibfnamefont {W.-C.}\ \bibnamefont {Chien}}, \bibinfo {author} {\bibfnamefont {H.~T.}\ \bibnamefont {Doan}}, \bibinfo {author} {\bibfnamefont {W.-C.}\ \bibnamefont {Hung}}, \bibinfo {author} {\bibfnamefont {W.}~\bibnamefont {Kuo}}, \bibinfo {author} {\bibfnamefont {S.-B.}\ \bibnamefont {Lai}}, \bibinfo {author} {\bibfnamefont {H.-W.}\
  \bibnamefont {Liu}}, \bibinfo {author} {\bibfnamefont {M.-W.}\ \bibnamefont {OuYang}}, \bibinfo {author} {\bibfnamefont {P.-I.}\ \bibnamefont {Wu}},\ and\ \bibinfo {author} {\bibfnamefont {S.-S.}\ \bibnamefont {Yu}} (\bibinfo {collaboration} {TASEH Collaboration}),\ }\bibfield  {title} {\bibinfo {title} {First results from the taiwan axion search experiment with a haloscope at $19.6\text{ }\text{ }\ensuremath{\mu}\mathrm{eV}$},\ }\href {https://doi.org/10.1103/PhysRevLett.129.111802} {\bibfield  {journal} {\bibinfo  {journal} {Phys. Rev. Lett.}\ }\textbf {\bibinfo {volume} {129}},\ \bibinfo {pages} {111802} (\bibinfo {year} {2022})}\BibitemShut {NoStop}%
\bibitem [{\citenamefont {DePanfilis}\ \emph {et~al.}(1987)\citenamefont {DePanfilis}, \citenamefont {Melissinos}, \citenamefont {Moskowitz}, \citenamefont {Rogers}, \citenamefont {Semertzidis}, \citenamefont {Wuensch}, \citenamefont {Halama}, \citenamefont {Prodell}, \citenamefont {Fowler},\ and\ \citenamefont {Nezrick}}]{PhysRevLett.59.839}%
  \BibitemOpen
  \bibfield  {author} {\bibinfo {author} {\bibfnamefont {S.}~\bibnamefont {DePanfilis}}, \bibinfo {author} {\bibfnamefont {A.}~\bibnamefont {Melissinos}}, \bibinfo {author} {\bibfnamefont {B.}~\bibnamefont {Moskowitz}}, \bibinfo {author} {\bibfnamefont {J.}~\bibnamefont {Rogers}}, \bibinfo {author} {\bibfnamefont {Y.~K.}\ \bibnamefont {Semertzidis}}, \bibinfo {author} {\bibfnamefont {W.}~\bibnamefont {Wuensch}}, \bibinfo {author} {\bibfnamefont {H.}~\bibnamefont {Halama}}, \bibinfo {author} {\bibfnamefont {A.}~\bibnamefont {Prodell}}, \bibinfo {author} {\bibfnamefont {W.}~\bibnamefont {Fowler}},\ and\ \bibinfo {author} {\bibfnamefont {F.}~\bibnamefont {Nezrick}},\ }\bibfield  {title} {\bibinfo {title} {Limits on the abundance and coupling of cosmic axions at 4. 5< ma< 5. 0 $\mu$ev},\ }\href {https://doi.org/10.1103/PhysRevLett.59.839} {\bibfield  {journal} {\bibinfo  {journal} {Phys. Rev. Lett.}\ }\textbf {\bibinfo {volume} {59}},\ \bibinfo {pages} {839} (\bibinfo {year} {1987})}\BibitemShut {NoStop}%
\bibitem [{\citenamefont {Wuensch}\ \emph {et~al.}(1989)\citenamefont {Wuensch}, \citenamefont {De~Panfilis-Wuensch}, \citenamefont {Semertzidis}, \citenamefont {Rogers}, \citenamefont {Melissinos}, \citenamefont {Halama}, \citenamefont {Moskowitz}, \citenamefont {Prodell}, \citenamefont {Fowler},\ and\ \citenamefont {Nezrick}}]{PhysRevD.40.3153}%
  \BibitemOpen
  \bibfield  {author} {\bibinfo {author} {\bibfnamefont {W.}~\bibnamefont {Wuensch}}, \bibinfo {author} {\bibfnamefont {S.}~\bibnamefont {De~Panfilis-Wuensch}}, \bibinfo {author} {\bibfnamefont {Y.~K.}\ \bibnamefont {Semertzidis}}, \bibinfo {author} {\bibfnamefont {J.}~\bibnamefont {Rogers}}, \bibinfo {author} {\bibfnamefont {A.}~\bibnamefont {Melissinos}}, \bibinfo {author} {\bibfnamefont {H.}~\bibnamefont {Halama}}, \bibinfo {author} {\bibfnamefont {B.}~\bibnamefont {Moskowitz}}, \bibinfo {author} {\bibfnamefont {A.}~\bibnamefont {Prodell}}, \bibinfo {author} {\bibfnamefont {W.}~\bibnamefont {Fowler}},\ and\ \bibinfo {author} {\bibfnamefont {F.}~\bibnamefont {Nezrick}},\ }\bibfield  {title} {\bibinfo {title} {Results of a laboratory search for cosmic axions and other weakly coupled light particles},\ }\href {https://doi.org/10.1103/PhysRevD.40.3153} {\bibfield  {journal} {\bibinfo  {journal} {Phys. Rev. D}\ }\textbf {\bibinfo {volume} {40}},\ \bibinfo {pages} {3153} (\bibinfo {year} {1989})}\BibitemShut
  {NoStop}%
\bibitem [{\citenamefont {Hagmann}\ \emph {et~al.}(1990)\citenamefont {Hagmann}, \citenamefont {Sikivie}, \citenamefont {Sullivan},\ and\ \citenamefont {Tanner}}]{PhysRevD.42.1297}%
  \BibitemOpen
  \bibfield  {author} {\bibinfo {author} {\bibfnamefont {C.}~\bibnamefont {Hagmann}}, \bibinfo {author} {\bibfnamefont {P.}~\bibnamefont {Sikivie}}, \bibinfo {author} {\bibfnamefont {N.~S.}\ \bibnamefont {Sullivan}},\ and\ \bibinfo {author} {\bibfnamefont {D.~B.}\ \bibnamefont {Tanner}},\ }\bibfield  {title} {\bibinfo {title} {Results from a search for cosmic axions},\ }\href {https://doi.org/10.1103/PhysRevD.42.1297} {\bibfield  {journal} {\bibinfo  {journal} {Phys. Rev. D}\ }\textbf {\bibinfo {volume} {42}},\ \bibinfo {pages} {1297} (\bibinfo {year} {1990})}\BibitemShut {NoStop}%
\bibitem [{\citenamefont {Hagmann}\ \emph {et~al.}(1996)\citenamefont {Hagmann}, \citenamefont {Kinion}, \citenamefont {Stoeffl}, \citenamefont {{van Bibber}}, \citenamefont {Daw}, \citenamefont {McBride}, \citenamefont {Peng}, \citenamefont {Rosenberg}, \citenamefont {Xin}, \citenamefont {Laveigne}, \citenamefont {Sikivie}, \citenamefont {Sullivan}, \citenamefont {Tanner}, \citenamefont {Moltz}, \citenamefont {Nezrick}, \citenamefont {Turner}, \citenamefont {Golubev},\ and\ \citenamefont {Kravchuk}}]{HAGMANN1996209}%
  \BibitemOpen
  \bibfield  {author} {\bibinfo {author} {\bibfnamefont {C.}~\bibnamefont {Hagmann}}, \bibinfo {author} {\bibfnamefont {D.}~\bibnamefont {Kinion}}, \bibinfo {author} {\bibfnamefont {W.}~\bibnamefont {Stoeffl}}, \bibinfo {author} {\bibfnamefont {K.}~\bibnamefont {{van Bibber}}}, \bibinfo {author} {\bibfnamefont {E.}~\bibnamefont {Daw}}, \bibinfo {author} {\bibfnamefont {J.}~\bibnamefont {McBride}}, \bibinfo {author} {\bibfnamefont {H.}~\bibnamefont {Peng}}, \bibinfo {author} {\bibfnamefont {L.}~\bibnamefont {Rosenberg}}, \bibinfo {author} {\bibfnamefont {H.}~\bibnamefont {Xin}}, \bibinfo {author} {\bibfnamefont {J.}~\bibnamefont {Laveigne}}, \bibinfo {author} {\bibfnamefont {P.}~\bibnamefont {Sikivie}}, \bibinfo {author} {\bibfnamefont {N.}~\bibnamefont {Sullivan}}, \bibinfo {author} {\bibfnamefont {D.}~\bibnamefont {Tanner}}, \bibinfo {author} {\bibfnamefont {D.}~\bibnamefont {Moltz}}, \bibinfo {author} {\bibfnamefont {F.}~\bibnamefont {Nezrick}}, \bibinfo {author} {\bibfnamefont {M.}~\bibnamefont {Turner}},
  \bibinfo {author} {\bibfnamefont {N.}~\bibnamefont {Golubev}},\ and\ \bibinfo {author} {\bibfnamefont {L.}~\bibnamefont {Kravchuk}},\ }\bibfield  {title} {\bibinfo {title} {First results from a second generation galactic axion experiment},\ }\href {https://doi.org/https://doi.org/10.1016/S0920-5632(96)00516-6} {\bibfield  {journal} {\bibinfo  {journal} {Nuclear Physics B - Proceedings Supplements}\ }\textbf {\bibinfo {volume} {51}},\ \bibinfo {pages} {209} (\bibinfo {year} {1996})},\ \bibinfo {note} {proceedings of the International Symposium on Sources and Detection of Dark Matter in the Universe}\BibitemShut {NoStop}%
\bibitem [{\citenamefont {Asztalos}\ \emph {et~al.}(2010)\citenamefont {Asztalos}, \citenamefont {Carosi}, \citenamefont {Hagmann}, \citenamefont {Kinion}, \citenamefont {van Bibber}, \citenamefont {Hotz}, \citenamefont {Rosenberg}, \citenamefont {Rybka}, \citenamefont {Hoskins}, \citenamefont {Hwang}, \citenamefont {Sikivie}, \citenamefont {Tanner}, \citenamefont {Bradley},\ and\ \citenamefont {Clarke}}]{PhysRevLett.104.041301}%
  \BibitemOpen
  \bibfield  {author} {\bibinfo {author} {\bibfnamefont {S.~J.}\ \bibnamefont {Asztalos}}, \bibinfo {author} {\bibfnamefont {G.}~\bibnamefont {Carosi}}, \bibinfo {author} {\bibfnamefont {C.}~\bibnamefont {Hagmann}}, \bibinfo {author} {\bibfnamefont {D.}~\bibnamefont {Kinion}}, \bibinfo {author} {\bibfnamefont {K.}~\bibnamefont {van Bibber}}, \bibinfo {author} {\bibfnamefont {M.}~\bibnamefont {Hotz}}, \bibinfo {author} {\bibfnamefont {L.~J.}\ \bibnamefont {Rosenberg}}, \bibinfo {author} {\bibfnamefont {G.}~\bibnamefont {Rybka}}, \bibinfo {author} {\bibfnamefont {J.}~\bibnamefont {Hoskins}}, \bibinfo {author} {\bibfnamefont {J.}~\bibnamefont {Hwang}}, \bibinfo {author} {\bibfnamefont {P.}~\bibnamefont {Sikivie}}, \bibinfo {author} {\bibfnamefont {D.~B.}\ \bibnamefont {Tanner}}, \bibinfo {author} {\bibfnamefont {R.}~\bibnamefont {Bradley}},\ and\ \bibinfo {author} {\bibfnamefont {J.}~\bibnamefont {Clarke}},\ }\bibfield  {title} {\bibinfo {title} {Squid-based microwave cavity search for dark-matter axions},\
  }\href {https://doi.org/10.1103/PhysRevLett.104.041301} {\bibfield  {journal} {\bibinfo  {journal} {Phys. Rev. Lett.}\ }\textbf {\bibinfo {volume} {104}},\ \bibinfo {pages} {041301} (\bibinfo {year} {2010})}\BibitemShut {NoStop}%
\bibitem [{\citenamefont {Du}\ \emph {et~al.}(2018)\citenamefont {Du}, \citenamefont {Force}, \citenamefont {Khatiwada}, \citenamefont {Lentz}, \citenamefont {Ottens}, \citenamefont {Rosenberg}, \citenamefont {Rybka}, \citenamefont {Carosi}, \citenamefont {Woollett}, \citenamefont {Bowring}, \citenamefont {Chou}, \citenamefont {Sonnenschein}, \citenamefont {Wester}, \citenamefont {Boutan}, \citenamefont {Oblath}, \citenamefont {Bradley}, \citenamefont {Daw}, \citenamefont {Dixit}, \citenamefont {Clarke}, \citenamefont {O'Kelley}, \citenamefont {Crisosto}, \citenamefont {Gleason}, \citenamefont {Jois}, \citenamefont {Sikivie}, \citenamefont {Stern}, \citenamefont {Sullivan}, \citenamefont {Tanner},\ and\ \citenamefont {Hilton}}]{PhysRevLett.120.151301}%
  \BibitemOpen
  \bibfield  {author} {\bibinfo {author} {\bibfnamefont {N.}~\bibnamefont {Du}}, \bibinfo {author} {\bibfnamefont {N.}~\bibnamefont {Force}}, \bibinfo {author} {\bibfnamefont {R.}~\bibnamefont {Khatiwada}}, \bibinfo {author} {\bibfnamefont {E.}~\bibnamefont {Lentz}}, \bibinfo {author} {\bibfnamefont {R.}~\bibnamefont {Ottens}}, \bibinfo {author} {\bibfnamefont {L.~J.}\ \bibnamefont {Rosenberg}}, \bibinfo {author} {\bibfnamefont {G.}~\bibnamefont {Rybka}}, \bibinfo {author} {\bibfnamefont {G.}~\bibnamefont {Carosi}}, \bibinfo {author} {\bibfnamefont {N.}~\bibnamefont {Woollett}}, \bibinfo {author} {\bibfnamefont {D.}~\bibnamefont {Bowring}}, \bibinfo {author} {\bibfnamefont {A.~S.}\ \bibnamefont {Chou}}, \bibinfo {author} {\bibfnamefont {A.}~\bibnamefont {Sonnenschein}}, \bibinfo {author} {\bibfnamefont {W.}~\bibnamefont {Wester}}, \bibinfo {author} {\bibfnamefont {C.}~\bibnamefont {Boutan}}, \bibinfo {author} {\bibfnamefont {N.~S.}\ \bibnamefont {Oblath}}, \bibinfo {author} {\bibfnamefont {R.}~\bibnamefont
  {Bradley}}, \bibinfo {author} {\bibfnamefont {E.~J.}\ \bibnamefont {Daw}}, \bibinfo {author} {\bibfnamefont {A.~V.}\ \bibnamefont {Dixit}}, \bibinfo {author} {\bibfnamefont {J.}~\bibnamefont {Clarke}}, \bibinfo {author} {\bibfnamefont {S.~R.}\ \bibnamefont {O'Kelley}}, \bibinfo {author} {\bibfnamefont {N.}~\bibnamefont {Crisosto}}, \bibinfo {author} {\bibfnamefont {J.~R.}\ \bibnamefont {Gleason}}, \bibinfo {author} {\bibfnamefont {S.}~\bibnamefont {Jois}}, \bibinfo {author} {\bibfnamefont {P.}~\bibnamefont {Sikivie}}, \bibinfo {author} {\bibfnamefont {I.}~\bibnamefont {Stern}}, \bibinfo {author} {\bibfnamefont {N.~S.}\ \bibnamefont {Sullivan}}, \bibinfo {author} {\bibfnamefont {D.~B.}\ \bibnamefont {Tanner}},\ and\ \bibinfo {author} {\bibfnamefont {G.~C.}\ \bibnamefont {Hilton}} (\bibinfo {collaboration} {ADMX Collaboration}),\ }\bibfield  {title} {\bibinfo {title} {Search for invisible axion dark matter with the axion dark matter experiment},\ }\href {https://doi.org/10.1103/PhysRevLett.120.151301}
  {\bibfield  {journal} {\bibinfo  {journal} {Phys. Rev. Lett.}\ }\textbf {\bibinfo {volume} {120}},\ \bibinfo {pages} {151301} (\bibinfo {year} {2018})}\BibitemShut {NoStop}%
\bibitem [{\citenamefont {Braine}\ \emph {et~al.}(2020)\citenamefont {Braine}, \citenamefont {Cervantes}, \citenamefont {Crisosto}, \citenamefont {Du}, \citenamefont {Kimes}, \citenamefont {Rosenberg}, \citenamefont {Rybka}, \citenamefont {Yang}, \citenamefont {Bowring}, \citenamefont {Chou}, \citenamefont {Khatiwada}, \citenamefont {Sonnenschein}, \citenamefont {Wester}, \citenamefont {Carosi}, \citenamefont {Woollett}, \citenamefont {Duffy}, \citenamefont {Bradley}, \citenamefont {Boutan}, \citenamefont {Jones}, \citenamefont {LaRoque}, \citenamefont {Oblath}, \citenamefont {Taubman}, \citenamefont {Clarke}, \citenamefont {Dove}, \citenamefont {Eddins}, \citenamefont {O'Kelley}, \citenamefont {Nawaz}, \citenamefont {Siddiqi}, \citenamefont {Stevenson}, \citenamefont {Agrawal}, \citenamefont {Dixit}, \citenamefont {Gleason}, \citenamefont {Jois}, \citenamefont {Sikivie}, \citenamefont {Solomon}, \citenamefont {Sullivan}, \citenamefont {Tanner}, \citenamefont {Lentz}, \citenamefont {Daw}, \citenamefont
  {Buckley}, \citenamefont {Harrington}, \citenamefont {Henriksen},\ and\ \citenamefont {Murch}}]{PhysRevLett.124.101303}%
  \BibitemOpen
  \bibfield  {author} {\bibinfo {author} {\bibfnamefont {T.}~\bibnamefont {Braine}}, \bibinfo {author} {\bibfnamefont {R.}~\bibnamefont {Cervantes}}, \bibinfo {author} {\bibfnamefont {N.}~\bibnamefont {Crisosto}}, \bibinfo {author} {\bibfnamefont {N.}~\bibnamefont {Du}}, \bibinfo {author} {\bibfnamefont {S.}~\bibnamefont {Kimes}}, \bibinfo {author} {\bibfnamefont {L.~J.}\ \bibnamefont {Rosenberg}}, \bibinfo {author} {\bibfnamefont {G.}~\bibnamefont {Rybka}}, \bibinfo {author} {\bibfnamefont {J.}~\bibnamefont {Yang}}, \bibinfo {author} {\bibfnamefont {D.}~\bibnamefont {Bowring}}, \bibinfo {author} {\bibfnamefont {A.~S.}\ \bibnamefont {Chou}}, \bibinfo {author} {\bibfnamefont {R.}~\bibnamefont {Khatiwada}}, \bibinfo {author} {\bibfnamefont {A.}~\bibnamefont {Sonnenschein}}, \bibinfo {author} {\bibfnamefont {W.}~\bibnamefont {Wester}}, \bibinfo {author} {\bibfnamefont {G.}~\bibnamefont {Carosi}}, \bibinfo {author} {\bibfnamefont {N.}~\bibnamefont {Woollett}}, \bibinfo {author} {\bibfnamefont {L.~D.}\
  \bibnamefont {Duffy}}, \bibinfo {author} {\bibfnamefont {R.}~\bibnamefont {Bradley}}, \bibinfo {author} {\bibfnamefont {C.}~\bibnamefont {Boutan}}, \bibinfo {author} {\bibfnamefont {M.}~\bibnamefont {Jones}}, \bibinfo {author} {\bibfnamefont {B.~H.}\ \bibnamefont {LaRoque}}, \bibinfo {author} {\bibfnamefont {N.~S.}\ \bibnamefont {Oblath}}, \bibinfo {author} {\bibfnamefont {M.~S.}\ \bibnamefont {Taubman}}, \bibinfo {author} {\bibfnamefont {J.}~\bibnamefont {Clarke}}, \bibinfo {author} {\bibfnamefont {A.}~\bibnamefont {Dove}}, \bibinfo {author} {\bibfnamefont {A.}~\bibnamefont {Eddins}}, \bibinfo {author} {\bibfnamefont {S.~R.}\ \bibnamefont {O'Kelley}}, \bibinfo {author} {\bibfnamefont {S.}~\bibnamefont {Nawaz}}, \bibinfo {author} {\bibfnamefont {I.}~\bibnamefont {Siddiqi}}, \bibinfo {author} {\bibfnamefont {N.}~\bibnamefont {Stevenson}}, \bibinfo {author} {\bibfnamefont {A.}~\bibnamefont {Agrawal}}, \bibinfo {author} {\bibfnamefont {A.~V.}\ \bibnamefont {Dixit}}, \bibinfo {author} {\bibfnamefont {J.~R.}\
  \bibnamefont {Gleason}}, \bibinfo {author} {\bibfnamefont {S.}~\bibnamefont {Jois}}, \bibinfo {author} {\bibfnamefont {P.}~\bibnamefont {Sikivie}}, \bibinfo {author} {\bibfnamefont {J.~A.}\ \bibnamefont {Solomon}}, \bibinfo {author} {\bibfnamefont {N.~S.}\ \bibnamefont {Sullivan}}, \bibinfo {author} {\bibfnamefont {D.~B.}\ \bibnamefont {Tanner}}, \bibinfo {author} {\bibfnamefont {E.}~\bibnamefont {Lentz}}, \bibinfo {author} {\bibfnamefont {E.~J.}\ \bibnamefont {Daw}}, \bibinfo {author} {\bibfnamefont {J.~H.}\ \bibnamefont {Buckley}}, \bibinfo {author} {\bibfnamefont {P.~M.}\ \bibnamefont {Harrington}}, \bibinfo {author} {\bibfnamefont {E.~A.}\ \bibnamefont {Henriksen}},\ and\ \bibinfo {author} {\bibfnamefont {K.~W.}\ \bibnamefont {Murch}} (\bibinfo {collaboration} {ADMX Collaboration}),\ }\bibfield  {title} {\bibinfo {title} {Extended search for the invisible axion with the axion dark matter experiment},\ }\href {https://doi.org/10.1103/PhysRevLett.124.101303} {\bibfield  {journal} {\bibinfo  {journal}
  {Phys. Rev. Lett.}\ }\textbf {\bibinfo {volume} {124}},\ \bibinfo {pages} {101303} (\bibinfo {year} {2020})}\BibitemShut {NoStop}%
\bibitem [{\citenamefont {Bartram}\ \emph {et~al.}(2021{\natexlab{a}})\citenamefont {Bartram}, \citenamefont {Braine}, \citenamefont {Burns}, \citenamefont {Cervantes}, \citenamefont {Crisosto}, \citenamefont {Du}, \citenamefont {Korandla}, \citenamefont {Leum}, \citenamefont {Mohapatra}, \citenamefont {Nitta}, \citenamefont {Rosenberg}, \citenamefont {Rybka}, \citenamefont {Yang}, \citenamefont {Clarke}, \citenamefont {Siddiqi}, \citenamefont {Agrawal}, \citenamefont {Dixit}, \citenamefont {Awida}, \citenamefont {Chou}, \citenamefont {Hollister}, \citenamefont {Knirck}, \citenamefont {Sonnenschein}, \citenamefont {Wester}, \citenamefont {Gleason}, \citenamefont {Hipp}, \citenamefont {Jois}, \citenamefont {Sikivie}, \citenamefont {Sullivan}, \citenamefont {Tanner}, \citenamefont {Lentz}, \citenamefont {Khatiwada}, \citenamefont {Carosi}, \citenamefont {Robertson}, \citenamefont {Woollett}, \citenamefont {Duffy}, \citenamefont {Boutan}, \citenamefont {Jones}, \citenamefont {LaRoque}, \citenamefont {Oblath},
  \citenamefont {Taubman}, \citenamefont {Daw}, \citenamefont {Perry}, \citenamefont {Buckley}, \citenamefont {Gaikwad}, \citenamefont {Hoffman}, \citenamefont {Murch}, \citenamefont {Goryachev}, \citenamefont {McAllister}, \citenamefont {Quiskamp}, \citenamefont {Thomson},\ and\ \citenamefont {Tobar}}]{ADMX:2021nhd}%
  \BibitemOpen
  \bibfield  {author} {\bibinfo {author} {\bibfnamefont {C.}~\bibnamefont {Bartram}}, \bibinfo {author} {\bibfnamefont {T.}~\bibnamefont {Braine}}, \bibinfo {author} {\bibfnamefont {E.}~\bibnamefont {Burns}}, \bibinfo {author} {\bibfnamefont {R.}~\bibnamefont {Cervantes}}, \bibinfo {author} {\bibfnamefont {N.}~\bibnamefont {Crisosto}}, \bibinfo {author} {\bibfnamefont {N.}~\bibnamefont {Du}}, \bibinfo {author} {\bibfnamefont {H.}~\bibnamefont {Korandla}}, \bibinfo {author} {\bibfnamefont {G.}~\bibnamefont {Leum}}, \bibinfo {author} {\bibfnamefont {P.}~\bibnamefont {Mohapatra}}, \bibinfo {author} {\bibfnamefont {T.}~\bibnamefont {Nitta}}, \bibinfo {author} {\bibfnamefont {L.~J.}\ \bibnamefont {Rosenberg}}, \bibinfo {author} {\bibfnamefont {G.}~\bibnamefont {Rybka}}, \bibinfo {author} {\bibfnamefont {J.}~\bibnamefont {Yang}}, \bibinfo {author} {\bibfnamefont {J.}~\bibnamefont {Clarke}}, \bibinfo {author} {\bibfnamefont {I.}~\bibnamefont {Siddiqi}}, \bibinfo {author} {\bibfnamefont {A.}~\bibnamefont {Agrawal}},
  \bibinfo {author} {\bibfnamefont {A.~V.}\ \bibnamefont {Dixit}}, \bibinfo {author} {\bibfnamefont {M.~H.}\ \bibnamefont {Awida}}, \bibinfo {author} {\bibfnamefont {A.~S.}\ \bibnamefont {Chou}}, \bibinfo {author} {\bibfnamefont {M.}~\bibnamefont {Hollister}}, \bibinfo {author} {\bibfnamefont {S.}~\bibnamefont {Knirck}}, \bibinfo {author} {\bibfnamefont {A.}~\bibnamefont {Sonnenschein}}, \bibinfo {author} {\bibfnamefont {W.}~\bibnamefont {Wester}}, \bibinfo {author} {\bibfnamefont {J.~R.}\ \bibnamefont {Gleason}}, \bibinfo {author} {\bibfnamefont {A.~T.}\ \bibnamefont {Hipp}}, \bibinfo {author} {\bibfnamefont {S.}~\bibnamefont {Jois}}, \bibinfo {author} {\bibfnamefont {P.}~\bibnamefont {Sikivie}}, \bibinfo {author} {\bibfnamefont {N.~S.}\ \bibnamefont {Sullivan}}, \bibinfo {author} {\bibfnamefont {D.~B.}\ \bibnamefont {Tanner}}, \bibinfo {author} {\bibfnamefont {E.}~\bibnamefont {Lentz}}, \bibinfo {author} {\bibfnamefont {R.}~\bibnamefont {Khatiwada}}, \bibinfo {author} {\bibfnamefont {G.}~\bibnamefont
  {Carosi}}, \bibinfo {author} {\bibfnamefont {N.}~\bibnamefont {Robertson}}, \bibinfo {author} {\bibfnamefont {N.}~\bibnamefont {Woollett}}, \bibinfo {author} {\bibfnamefont {L.~D.}\ \bibnamefont {Duffy}}, \bibinfo {author} {\bibfnamefont {C.}~\bibnamefont {Boutan}}, \bibinfo {author} {\bibfnamefont {M.}~\bibnamefont {Jones}}, \bibinfo {author} {\bibfnamefont {B.~H.}\ \bibnamefont {LaRoque}}, \bibinfo {author} {\bibfnamefont {N.~S.}\ \bibnamefont {Oblath}}, \bibinfo {author} {\bibfnamefont {M.~S.}\ \bibnamefont {Taubman}}, \bibinfo {author} {\bibfnamefont {E.~J.}\ \bibnamefont {Daw}}, \bibinfo {author} {\bibfnamefont {M.~G.}\ \bibnamefont {Perry}}, \bibinfo {author} {\bibfnamefont {J.~H.}\ \bibnamefont {Buckley}}, \bibinfo {author} {\bibfnamefont {C.}~\bibnamefont {Gaikwad}}, \bibinfo {author} {\bibfnamefont {J.}~\bibnamefont {Hoffman}}, \bibinfo {author} {\bibfnamefont {K.~W.}\ \bibnamefont {Murch}}, \bibinfo {author} {\bibfnamefont {M.}~\bibnamefont {Goryachev}}, \bibinfo {author} {\bibfnamefont {B.~T.}\
  \bibnamefont {McAllister}}, \bibinfo {author} {\bibfnamefont {A.}~\bibnamefont {Quiskamp}}, \bibinfo {author} {\bibfnamefont {C.}~\bibnamefont {Thomson}},\ and\ \bibinfo {author} {\bibfnamefont {M.~E.}\ \bibnamefont {Tobar}} (\bibinfo {collaboration} {ADMX Collaboration}),\ }\bibfield  {title} {\bibinfo {title} {Search for invisible axion dark matter in the $3.3--4.2\text{ }\text{ }\ensuremath{\mu}\mathrm{eV}$ mass range},\ }\href {https://doi.org/10.1103/PhysRevLett.127.261803} {\bibfield  {journal} {\bibinfo  {journal} {Phys. Rev. Lett.}\ }\textbf {\bibinfo {volume} {127}},\ \bibinfo {pages} {261803} (\bibinfo {year} {2021}{\natexlab{a}})}\BibitemShut {NoStop}%
\bibitem [{\citenamefont {Bartram}\ \emph {et~al.}(2024)\citenamefont {Bartram}, \citenamefont {Boutan}, \citenamefont {Braine}, \citenamefont {Buckley}, \citenamefont {Caligiure}, \citenamefont {Carosi}, \citenamefont {Chou}, \citenamefont {Cisneros}, \citenamefont {Clarke}, \citenamefont {Daw} \emph {et~al.}}]{bartram2024axiondarkmatterexperiment}%
  \BibitemOpen
  \bibfield  {author} {\bibinfo {author} {\bibfnamefont {C.}~\bibnamefont {Bartram}}, \bibinfo {author} {\bibfnamefont {C.}~\bibnamefont {Boutan}}, \bibinfo {author} {\bibfnamefont {T.}~\bibnamefont {Braine}}, \bibinfo {author} {\bibfnamefont {J.}~\bibnamefont {Buckley}}, \bibinfo {author} {\bibfnamefont {T.}~\bibnamefont {Caligiure}}, \bibinfo {author} {\bibfnamefont {G.}~\bibnamefont {Carosi}}, \bibinfo {author} {\bibfnamefont {A.}~\bibnamefont {Chou}}, \bibinfo {author} {\bibfnamefont {C.}~\bibnamefont {Cisneros}}, \bibinfo {author} {\bibfnamefont {J.}~\bibnamefont {Clarke}}, \bibinfo {author} {\bibfnamefont {E.}~\bibnamefont {Daw}}, \emph {et~al.},\ }\bibfield  {title} {\bibinfo {title} {Axion dark matter experiment around 3.3 $\{$$\backslash$mu$\}$ ev with dine-fischler-srednicki-zhitnitsky discovery ability},\ }\href@noop {} {\bibfield  {journal} {\bibinfo  {journal} {arXiv preprint arXiv:2408.15227}\ } (\bibinfo {year} {2024})}\BibitemShut {NoStop}%
\bibitem [{\citenamefont {Lee}\ \emph {et~al.}(2020)\citenamefont {Lee}, \citenamefont {Ahn}, \citenamefont {Choi}, \citenamefont {Ko},\ and\ \citenamefont {Semertzidis}}]{PhysRevLett.124.101802}%
  \BibitemOpen
  \bibfield  {author} {\bibinfo {author} {\bibfnamefont {S.}~\bibnamefont {Lee}}, \bibinfo {author} {\bibfnamefont {S.}~\bibnamefont {Ahn}}, \bibinfo {author} {\bibfnamefont {J.}~\bibnamefont {Choi}}, \bibinfo {author} {\bibfnamefont {B.~R.}\ \bibnamefont {Ko}},\ and\ \bibinfo {author} {\bibfnamefont {Y.~K.}\ \bibnamefont {Semertzidis}},\ }\bibfield  {title} {\bibinfo {title} {Axion dark matter search around 6.7 $\mu$ ev},\ }\href {https://doi.org/10.1103/PhysRevLett.124.101802} {\bibfield  {journal} {\bibinfo  {journal} {Phys. Rev. Lett.}\ }\textbf {\bibinfo {volume} {124}},\ \bibinfo {pages} {101802} (\bibinfo {year} {2020})}\BibitemShut {NoStop}%
\bibitem [{\citenamefont {Jeong}\ \emph {et~al.}(2020)\citenamefont {Jeong}, \citenamefont {Youn}, \citenamefont {Bae}, \citenamefont {Kim}, \citenamefont {Seong}, \citenamefont {Kim},\ and\ \citenamefont {Semertzidis}}]{PhysRevLett.125.221302}%
  \BibitemOpen
  \bibfield  {author} {\bibinfo {author} {\bibfnamefont {J.}~\bibnamefont {Jeong}}, \bibinfo {author} {\bibfnamefont {S.}~\bibnamefont {Youn}}, \bibinfo {author} {\bibfnamefont {S.}~\bibnamefont {Bae}}, \bibinfo {author} {\bibfnamefont {J.}~\bibnamefont {Kim}}, \bibinfo {author} {\bibfnamefont {T.}~\bibnamefont {Seong}}, \bibinfo {author} {\bibfnamefont {J.~E.}\ \bibnamefont {Kim}},\ and\ \bibinfo {author} {\bibfnamefont {Y.~K.}\ \bibnamefont {Semertzidis}},\ }\bibfield  {title} {\bibinfo {title} {Search for invisible axion dark matter with a multiple-cell haloscope},\ }\href {https://doi.org/10.1103/PhysRevLett.125.221302} {\bibfield  {journal} {\bibinfo  {journal} {Phys. Rev. Lett.}\ }\textbf {\bibinfo {volume} {125}},\ \bibinfo {pages} {221302} (\bibinfo {year} {2020})}\BibitemShut {NoStop}%
\bibitem [{\citenamefont {Kwon}\ \emph {et~al.}(2021)\citenamefont {Kwon}, \citenamefont {Lee}, \citenamefont {Chung}, \citenamefont {Ahn}, \citenamefont {Byun}, \citenamefont {Caspers}, \citenamefont {Choi}, \citenamefont {Choi}, \citenamefont {Chong}, \citenamefont {Jeong} \emph {et~al.}}]{PhysRevLett.126.191802}%
  \BibitemOpen
  \bibfield  {author} {\bibinfo {author} {\bibfnamefont {O.}~\bibnamefont {Kwon}}, \bibinfo {author} {\bibfnamefont {D.}~\bibnamefont {Lee}}, \bibinfo {author} {\bibfnamefont {W.}~\bibnamefont {Chung}}, \bibinfo {author} {\bibfnamefont {D.}~\bibnamefont {Ahn}}, \bibinfo {author} {\bibfnamefont {H.}~\bibnamefont {Byun}}, \bibinfo {author} {\bibfnamefont {F.}~\bibnamefont {Caspers}}, \bibinfo {author} {\bibfnamefont {H.}~\bibnamefont {Choi}}, \bibinfo {author} {\bibfnamefont {J.}~\bibnamefont {Choi}}, \bibinfo {author} {\bibfnamefont {Y.}~\bibnamefont {Chong}}, \bibinfo {author} {\bibfnamefont {H.}~\bibnamefont {Jeong}}, \emph {et~al.},\ }\bibfield  {title} {\bibinfo {title} {First results from an axion haloscope at capp around 10.7 $\mu$ ev},\ }\href {https://doi.org/10.1103/PhysRevLett.126.191802} {\bibfield  {journal} {\bibinfo  {journal} {Phys. Rev. Lett.}\ }\textbf {\bibinfo {volume} {126}},\ \bibinfo {pages} {191802} (\bibinfo {year} {2021})}\BibitemShut {NoStop}%
\bibitem [{\citenamefont {Yoon}\ \emph {et~al.}(2022)\citenamefont {Yoon}, \citenamefont {Ahn}, \citenamefont {Yang}, \citenamefont {Lee}, \citenamefont {Kim}, \citenamefont {Park}, \citenamefont {Min},\ and\ \citenamefont {Yoo}}]{PhysRevD.106.092007}%
  \BibitemOpen
  \bibfield  {author} {\bibinfo {author} {\bibfnamefont {H.}~\bibnamefont {Yoon}}, \bibinfo {author} {\bibfnamefont {M.}~\bibnamefont {Ahn}}, \bibinfo {author} {\bibfnamefont {B.}~\bibnamefont {Yang}}, \bibinfo {author} {\bibfnamefont {Y.}~\bibnamefont {Lee}}, \bibinfo {author} {\bibfnamefont {D.}~\bibnamefont {Kim}}, \bibinfo {author} {\bibfnamefont {H.}~\bibnamefont {Park}}, \bibinfo {author} {\bibfnamefont {B.}~\bibnamefont {Min}},\ and\ \bibinfo {author} {\bibfnamefont {J.}~\bibnamefont {Yoo}},\ }\bibfield  {title} {\bibinfo {title} {Axion haloscope using an 18 t high temperature superconducting magnet},\ }\href {https://doi.org/10.1103/PhysRevD.106.092007} {\bibfield  {journal} {\bibinfo  {journal} {Phys. Rev. D}\ }\textbf {\bibinfo {volume} {106}},\ \bibinfo {pages} {092007} (\bibinfo {year} {2022})}\BibitemShut {NoStop}%
\bibitem [{\citenamefont {Lee}\ \emph {et~al.}(2022)\citenamefont {Lee}, \citenamefont {Yang}, \citenamefont {Yoon}, \citenamefont {Ahn}, \citenamefont {Park}, \citenamefont {Min}, \citenamefont {Kim},\ and\ \citenamefont {Yoo}}]{PhysRevLett.128.241805}%
  \BibitemOpen
  \bibfield  {author} {\bibinfo {author} {\bibfnamefont {Y.}~\bibnamefont {Lee}}, \bibinfo {author} {\bibfnamefont {B.}~\bibnamefont {Yang}}, \bibinfo {author} {\bibfnamefont {H.}~\bibnamefont {Yoon}}, \bibinfo {author} {\bibfnamefont {M.}~\bibnamefont {Ahn}}, \bibinfo {author} {\bibfnamefont {H.}~\bibnamefont {Park}}, \bibinfo {author} {\bibfnamefont {B.}~\bibnamefont {Min}}, \bibinfo {author} {\bibfnamefont {D.}~\bibnamefont {Kim}},\ and\ \bibinfo {author} {\bibfnamefont {J.}~\bibnamefont {Yoo}},\ }\bibfield  {title} {\bibinfo {title} {Searching for invisible axion dark matter with an 18 t magnet haloscope},\ }\href {https://doi.org/10.1103/PhysRevLett.128.241805} {\bibfield  {journal} {\bibinfo  {journal} {Phys. Rev. Lett.}\ }\textbf {\bibinfo {volume} {128}},\ \bibinfo {pages} {241805} (\bibinfo {year} {2022})}\BibitemShut {NoStop}%
\bibitem [{\citenamefont {Kim}\ \emph {et~al.}(2023)\citenamefont {Kim}, \citenamefont {Kwon}, \citenamefont {Kutlu}, \citenamefont {Chung}, \citenamefont {Matlashov}, \citenamefont {Uchaikin}, \citenamefont {Van~Loo}, \citenamefont {Nakamura}, \citenamefont {Oh}, \citenamefont {Byun} \emph {et~al.}}]{Kim_2023}%
  \BibitemOpen
  \bibfield  {author} {\bibinfo {author} {\bibfnamefont {J.}~\bibnamefont {Kim}}, \bibinfo {author} {\bibfnamefont {O.}~\bibnamefont {Kwon}}, \bibinfo {author} {\bibfnamefont {{\c{C}}.}~\bibnamefont {Kutlu}}, \bibinfo {author} {\bibfnamefont {W.}~\bibnamefont {Chung}}, \bibinfo {author} {\bibfnamefont {A.}~\bibnamefont {Matlashov}}, \bibinfo {author} {\bibfnamefont {S.}~\bibnamefont {Uchaikin}}, \bibinfo {author} {\bibfnamefont {A.~F.}\ \bibnamefont {Van~Loo}}, \bibinfo {author} {\bibfnamefont {Y.}~\bibnamefont {Nakamura}}, \bibinfo {author} {\bibfnamefont {S.}~\bibnamefont {Oh}}, \bibinfo {author} {\bibfnamefont {H.}~\bibnamefont {Byun}}, \emph {et~al.},\ }\bibfield  {title} {\bibinfo {title} {Near-quantum-noise axion dark matter search at capp around 9.5 $\mu$ ev},\ }\bibfield  {journal} {\bibinfo  {journal} {Physical Review Letters}\ }\textbf {\bibinfo {volume} {130}},\ \href {https://doi.org/10.1103/physrevlett.130.091602} {10.1103/physrevlett.130.091602} (\bibinfo {year} {2023})\BibitemShut {NoStop}%
\bibitem [{\citenamefont {Yi}\ \emph {et~al.}(2023)\citenamefont {Yi}, \citenamefont {Ahn}, \citenamefont {Kutlu}, \citenamefont {Kim}, \citenamefont {Ko}, \citenamefont {Ivanov}, \citenamefont {Byun}, \citenamefont {Van~Loo}, \citenamefont {Park}, \citenamefont {Jeong} \emph {et~al.}}]{Yi_2023}%
  \BibitemOpen
  \bibfield  {author} {\bibinfo {author} {\bibfnamefont {A.~K.}\ \bibnamefont {Yi}}, \bibinfo {author} {\bibfnamefont {S.}~\bibnamefont {Ahn}}, \bibinfo {author} {\bibfnamefont {{\c{C}}.}~\bibnamefont {Kutlu}}, \bibinfo {author} {\bibfnamefont {J.}~\bibnamefont {Kim}}, \bibinfo {author} {\bibfnamefont {B.~R.}\ \bibnamefont {Ko}}, \bibinfo {author} {\bibfnamefont {B.~I.}\ \bibnamefont {Ivanov}}, \bibinfo {author} {\bibfnamefont {H.}~\bibnamefont {Byun}}, \bibinfo {author} {\bibfnamefont {A.~F.}\ \bibnamefont {Van~Loo}}, \bibinfo {author} {\bibfnamefont {S.}~\bibnamefont {Park}}, \bibinfo {author} {\bibfnamefont {J.}~\bibnamefont {Jeong}}, \emph {et~al.},\ }\bibfield  {title} {\bibinfo {title} {Axion dark matter search around 4.55 $\mu$ ev with dine-fischler-srednicki-zhitnitskii sensitivity},\ }\bibfield  {journal} {\bibinfo  {journal} {Physical Review Letters}\ }\textbf {\bibinfo {volume} {130}},\ \href {https://doi.org/10.1103/physrevlett.130.071002} {10.1103/physrevlett.130.071002} (\bibinfo {year}
  {2023})\BibitemShut {NoStop}%
\bibitem [{\citenamefont {Yang}\ \emph {et~al.}(2023)\citenamefont {Yang}, \citenamefont {Yoon}, \citenamefont {Ahn}, \citenamefont {Lee},\ and\ \citenamefont {Yoo}}]{Yang_2023}%
  \BibitemOpen
  \bibfield  {author} {\bibinfo {author} {\bibfnamefont {B.}~\bibnamefont {Yang}}, \bibinfo {author} {\bibfnamefont {H.}~\bibnamefont {Yoon}}, \bibinfo {author} {\bibfnamefont {M.}~\bibnamefont {Ahn}}, \bibinfo {author} {\bibfnamefont {Y.}~\bibnamefont {Lee}},\ and\ \bibinfo {author} {\bibfnamefont {J.}~\bibnamefont {Yoo}},\ }\bibfield  {title} {\bibinfo {title} {Extended axion dark matter search using the capp18t haloscope},\ }\bibfield  {journal} {\bibinfo  {journal} {Physical Review Letters}\ }\textbf {\bibinfo {volume} {131}},\ \href {https://doi.org/10.1103/physrevlett.131.081801} {10.1103/physrevlett.131.081801} (\bibinfo {year} {2023})\BibitemShut {NoStop}%
\bibitem [{\citenamefont {Kim}\ \emph {et~al.}(2024)\citenamefont {Kim}, \citenamefont {Jeong}, \citenamefont {Youn}, \citenamefont {Bae}, \citenamefont {Lee}, \citenamefont {van Loo}, \citenamefont {Nakamura}, \citenamefont {Oh}, \citenamefont {Seong}, \citenamefont {Uchaikin}, \citenamefont {Kim},\ and\ \citenamefont {Semertzidis}}]{kim2024experimentalsearchinvisibledark}%
  \BibitemOpen
  \bibfield  {author} {\bibinfo {author} {\bibfnamefont {Y.}~\bibnamefont {Kim}}, \bibinfo {author} {\bibfnamefont {J.}~\bibnamefont {Jeong}}, \bibinfo {author} {\bibfnamefont {S.}~\bibnamefont {Youn}}, \bibinfo {author} {\bibfnamefont {S.}~\bibnamefont {Bae}}, \bibinfo {author} {\bibfnamefont {K.}~\bibnamefont {Lee}}, \bibinfo {author} {\bibfnamefont {A.~F.}\ \bibnamefont {van Loo}}, \bibinfo {author} {\bibfnamefont {Y.}~\bibnamefont {Nakamura}}, \bibinfo {author} {\bibfnamefont {S.}~\bibnamefont {Oh}}, \bibinfo {author} {\bibfnamefont {T.}~\bibnamefont {Seong}}, \bibinfo {author} {\bibfnamefont {S.}~\bibnamefont {Uchaikin}}, \bibinfo {author} {\bibfnamefont {J.~E.}\ \bibnamefont {Kim}},\ and\ \bibinfo {author} {\bibfnamefont {Y.~K.}\ \bibnamefont {Semertzidis}},\ }\href {https://arxiv.org/abs/2312.11003} {\bibinfo {title} {Experimental search for invisible axions as a test of axion cosmology around 22 $\mu$ev}} (\bibinfo {year} {2024}),\ \Eprint {https://arxiv.org/abs/2312.11003} {arXiv:2312.11003 [hep-ex]}
  \BibitemShut {NoStop}%
\bibitem [{\citenamefont {Ahn}\ \emph {et~al.}(2024)\citenamefont {Ahn}, \citenamefont {Kim}, \citenamefont {Ivanov}, \citenamefont {Kwon}, \citenamefont {Byun}, \citenamefont {Van~Loo}, \citenamefont {Park}, \citenamefont {Jeong}, \citenamefont {Lee}, \citenamefont {Kim} \emph {et~al.}}]{ahn2024extensivesearchaxiondark}%
  \BibitemOpen
  \bibfield  {author} {\bibinfo {author} {\bibfnamefont {S.}~\bibnamefont {Ahn}}, \bibinfo {author} {\bibfnamefont {J.}~\bibnamefont {Kim}}, \bibinfo {author} {\bibfnamefont {B.~I.}\ \bibnamefont {Ivanov}}, \bibinfo {author} {\bibfnamefont {O.}~\bibnamefont {Kwon}}, \bibinfo {author} {\bibfnamefont {H.}~\bibnamefont {Byun}}, \bibinfo {author} {\bibfnamefont {A.~F.}\ \bibnamefont {Van~Loo}}, \bibinfo {author} {\bibfnamefont {S.}~\bibnamefont {Park}}, \bibinfo {author} {\bibfnamefont {J.}~\bibnamefont {Jeong}}, \bibinfo {author} {\bibfnamefont {S.}~\bibnamefont {Lee}}, \bibinfo {author} {\bibfnamefont {J.}~\bibnamefont {Kim}}, \emph {et~al.},\ }\href {https://arxiv.org/abs/2402.12892} {\bibinfo {title} {Extensive search for axion dark matter over 1\,ghz with capp's main axion experiment}} (\bibinfo {year} {2024}),\ \Eprint {https://arxiv.org/abs/2402.12892} {arXiv:2402.12892 [hep-ex]} \BibitemShut {NoStop}%
\bibitem [{\citenamefont {Kim}(1979)}]{PhysRevLett.43.103}%
  \BibitemOpen
  \bibfield  {author} {\bibinfo {author} {\bibfnamefont {J.~E.}\ \bibnamefont {Kim}},\ }\bibfield  {title} {\bibinfo {title} {Weak-interaction singlet and strong $\mathrm{CP}$ invariance},\ }\href {https://doi.org/10.1103/PhysRevLett.43.103} {\bibfield  {journal} {\bibinfo  {journal} {Phys. Rev. Lett.}\ }\textbf {\bibinfo {volume} {43}},\ \bibinfo {pages} {103} (\bibinfo {year} {1979})}\BibitemShut {NoStop}%
\bibitem [{\citenamefont {Shifman}\ \emph {et~al.}(1980)\citenamefont {Shifman}, \citenamefont {Vainshtein},\ and\ \citenamefont {Zakharov}}]{SHIFMAN1980493}%
  \BibitemOpen
  \bibfield  {author} {\bibinfo {author} {\bibfnamefont {M.}~\bibnamefont {Shifman}}, \bibinfo {author} {\bibfnamefont {A.}~\bibnamefont {Vainshtein}},\ and\ \bibinfo {author} {\bibfnamefont {V.}~\bibnamefont {Zakharov}},\ }\bibfield  {title} {\bibinfo {title} {Can confinement ensure natural cp invariance of strong interactions?},\ }\href {https://doi.org/https://doi.org/10.1016/0550-3213(80)90209-6} {\bibfield  {journal} {\bibinfo  {journal} {Nuclear Physics B}\ }\textbf {\bibinfo {volume} {166}},\ \bibinfo {pages} {493} (\bibinfo {year} {1980})}\BibitemShut {NoStop}%
\bibitem [{\citenamefont {Dine}\ \emph {et~al.}(1981)\citenamefont {Dine}, \citenamefont {Fischler},\ and\ \citenamefont {Srednicki}}]{DINE1981199}%
  \BibitemOpen
  \bibfield  {author} {\bibinfo {author} {\bibfnamefont {M.}~\bibnamefont {Dine}}, \bibinfo {author} {\bibfnamefont {W.}~\bibnamefont {Fischler}},\ and\ \bibinfo {author} {\bibfnamefont {M.}~\bibnamefont {Srednicki}},\ }\bibfield  {title} {\bibinfo {title} {A simple solution to the strong cp problem with a harmless axion},\ }\href {https://doi.org/https://doi.org/10.1016/0370-2693(81)90590-6} {\bibfield  {journal} {\bibinfo  {journal} {Physics Letters B}\ }\textbf {\bibinfo {volume} {104}},\ \bibinfo {pages} {199} (\bibinfo {year} {1981})}\BibitemShut {NoStop}%
\bibitem [{\citenamefont {Horns}\ \emph {et~al.}(2013)\citenamefont {Horns}, \citenamefont {Jaeckel}, \citenamefont {Lindner}, \citenamefont {Lobanov}, \citenamefont {Redondo},\ and\ \citenamefont {Ringwald}}]{Horns:2012jf}%
  \BibitemOpen
  \bibfield  {author} {\bibinfo {author} {\bibfnamefont {D.}~\bibnamefont {Horns}}, \bibinfo {author} {\bibfnamefont {J.}~\bibnamefont {Jaeckel}}, \bibinfo {author} {\bibfnamefont {A.}~\bibnamefont {Lindner}}, \bibinfo {author} {\bibfnamefont {A.}~\bibnamefont {Lobanov}}, \bibinfo {author} {\bibfnamefont {J.}~\bibnamefont {Redondo}},\ and\ \bibinfo {author} {\bibfnamefont {A.}~\bibnamefont {Ringwald}},\ }\bibfield  {title} {\bibinfo {title} {Searching for wispy cold dark matter with a dish antenna},\ }\href {https://doi.org/10.1088/1475-7516/2013/04/016} {\bibfield  {journal} {\bibinfo  {journal} {Journal of Cosmology and Astroparticle Physics}\ }\textbf {\bibinfo {volume} {2013}}\bibinfo  {number} { (04)},\ \bibinfo {pages} {016}}\BibitemShut {NoStop}%
\bibitem [{\citenamefont {Liu}\ \emph {et~al.}(2022)\citenamefont {Liu} \emph {et~al.}}]{BREAD:2021tpx}%
  \BibitemOpen
\bibfield  {number} {  }\bibfield  {author} {\bibinfo {author} {\bibfnamefont {J.}~\bibnamefont {Liu}} \emph {et~al.} (\bibinfo {collaboration} {BREAD}),\ }\bibfield  {title} {\bibinfo {title} {{Broadband Solenoidal Haloscope for Terahertz Axion Detection}},\ }\href {https://doi.org/10.1103/PhysRevLett.128.131801} {\bibfield  {journal} {\bibinfo  {journal} {Phys. Rev. Lett.}\ }\textbf {\bibinfo {volume} {128}},\ \bibinfo {pages} {131801} (\bibinfo {year} {2022})},\ \Eprint {https://arxiv.org/abs/2111.12103} {arXiv:2111.12103 [physics.ins-det]} \BibitemShut {NoStop}%
\bibitem [{\citenamefont {Turner}(1990)}]{Turner:1990qx}%
  \BibitemOpen
  \bibfield  {author} {\bibinfo {author} {\bibfnamefont {M.~S.}\ \bibnamefont {Turner}},\ }\bibfield  {title} {\bibinfo {title} {{Periodic signatures for the detection of cosmic axions}},\ }\href {https://doi.org/10.1103/PhysRevD.42.3572} {\bibfield  {journal} {\bibinfo  {journal} {Phys. Rev. D}\ }\textbf {\bibinfo {volume} {42}},\ \bibinfo {pages} {3572} (\bibinfo {year} {1990})}\BibitemShut {NoStop}%
\bibitem [{\citenamefont {Knirck}\ \emph {et~al.}(2018)\citenamefont {Knirck}, \citenamefont {Millar}, \citenamefont {O'Hare}, \citenamefont {Redondo},\ and\ \citenamefont {Steffen}}]{Knirck:2018knd}%
  \BibitemOpen
  \bibfield  {author} {\bibinfo {author} {\bibfnamefont {S.}~\bibnamefont {Knirck}}, \bibinfo {author} {\bibfnamefont {A.~J.}\ \bibnamefont {Millar}}, \bibinfo {author} {\bibfnamefont {C.~A.~J.}\ \bibnamefont {O'Hare}}, \bibinfo {author} {\bibfnamefont {J.}~\bibnamefont {Redondo}},\ and\ \bibinfo {author} {\bibfnamefont {F.~D.}\ \bibnamefont {Steffen}},\ }\bibfield  {title} {\bibinfo {title} {{Directional axion detection}},\ }\href {https://doi.org/10.1088/1475-7516/2018/11/051} {\bibfield  {journal} {\bibinfo  {journal} {JCAP}\ }\textbf {\bibinfo {volume} {11}}\bibfield  {number} {\bibinfo  {number} { (11)},\ \bibinfo {pages} {051}},\ }\Eprint {https://arxiv.org/abs/1806.05927} {arXiv:1806.05927 [astro-ph.CO]} \BibitemShut {NoStop}%
%%CITATION = ARXIV:1806.05927;%%
\bibitem [{\citenamefont {Suzuki}\ \emph {et~al.}(2015)\citenamefont {Suzuki}, \citenamefont {Horie}, \citenamefont {Inoue},\ and\ \citenamefont {Minowa}}]{Suzuki:2015sza}%
  \BibitemOpen
  \bibfield  {author} {\bibinfo {author} {\bibfnamefont {J.}~\bibnamefont {Suzuki}}, \bibinfo {author} {\bibfnamefont {T.}~\bibnamefont {Horie}}, \bibinfo {author} {\bibfnamefont {Y.}~\bibnamefont {Inoue}},\ and\ \bibinfo {author} {\bibfnamefont {M.}~\bibnamefont {Minowa}},\ }\bibfield  {title} {\bibinfo {title} {{Experimental Search for Hidden Photon CDM in the eV mass range with a Dish Antenna}},\ }\href {https://doi.org/10.1088/1475-7516/2015/09/042} {\bibfield  {journal} {\bibinfo  {journal} {JCAP}\ }\textbf {\bibinfo {volume} {09}},\ \bibinfo {pages} {042}},\ \Eprint {https://arxiv.org/abs/1504.00118} {arXiv:1504.00118 [hep-ex]} \BibitemShut {NoStop}%
\bibitem [{\citenamefont {Brun}\ \emph {et~al.}(2019)\citenamefont {Brun}, \citenamefont {Chevalier},\ and\ \citenamefont {Flouzat}}]{Brun:2019kak}%
  \BibitemOpen
  \bibfield  {author} {\bibinfo {author} {\bibfnamefont {P.}~\bibnamefont {Brun}}, \bibinfo {author} {\bibfnamefont {L.}~\bibnamefont {Chevalier}},\ and\ \bibinfo {author} {\bibfnamefont {C.}~\bibnamefont {Flouzat}},\ }\bibfield  {title} {\bibinfo {title} {{Direct Searches for Hidden-Photon Dark Matter with the SHUKET Experiment}},\ }\href {https://doi.org/10.1103/PhysRevLett.122.201801} {\bibfield  {journal} {\bibinfo  {journal} {Phys. Rev. Lett.}\ }\textbf {\bibinfo {volume} {122}},\ \bibinfo {pages} {201801} (\bibinfo {year} {2019})},\ \Eprint {https://arxiv.org/abs/1905.05579} {arXiv:1905.05579 [hep-ex]} \BibitemShut {NoStop}%
\bibitem [{\citenamefont {Tomita}\ \emph {et~al.}(2020)\citenamefont {Tomita}, \citenamefont {Oguri}, \citenamefont {Inoue}, \citenamefont {Minowa}, \citenamefont {Nagasaki}, \citenamefont {Suzuki},\ and\ \citenamefont {Tajima}}]{Tomita:2020usq}%
  \BibitemOpen
  \bibfield  {author} {\bibinfo {author} {\bibfnamefont {N.}~\bibnamefont {Tomita}}, \bibinfo {author} {\bibfnamefont {S.}~\bibnamefont {Oguri}}, \bibinfo {author} {\bibfnamefont {Y.}~\bibnamefont {Inoue}}, \bibinfo {author} {\bibfnamefont {M.}~\bibnamefont {Minowa}}, \bibinfo {author} {\bibfnamefont {T.}~\bibnamefont {Nagasaki}}, \bibinfo {author} {\bibfnamefont {J.}~\bibnamefont {Suzuki}},\ and\ \bibinfo {author} {\bibfnamefont {O.}~\bibnamefont {Tajima}},\ }\bibfield  {title} {\bibinfo {title} {{Search for hidden-photon cold dark matter using a K-band cryogenic receiver}},\ }\href {https://doi.org/10.1088/1475-7516/2020/09/012} {\bibfield  {journal} {\bibinfo  {journal} {JCAP}\ }\textbf {\bibinfo {volume} {09}},\ \bibinfo {pages} {012}},\ \Eprint {https://arxiv.org/abs/2006.02828} {arXiv:2006.02828 [hep-ex]} \BibitemShut {NoStop}%
\bibitem [{\citenamefont {Ramanathan}\ \emph {et~al.}(2023)\citenamefont {Ramanathan}, \citenamefont {Klimovich}, \citenamefont {Basu~Thakur}, \citenamefont {Eom}, \citenamefont {LeDuc}, \citenamefont {Shu}, \citenamefont {Beyer},\ and\ \citenamefont {Day}}]{Ramanathan:2022egk}%
  \BibitemOpen
  \bibfield  {author} {\bibinfo {author} {\bibfnamefont {K.}~\bibnamefont {Ramanathan}}, \bibinfo {author} {\bibfnamefont {N.}~\bibnamefont {Klimovich}}, \bibinfo {author} {\bibfnamefont {R.}~\bibnamefont {Basu~Thakur}}, \bibinfo {author} {\bibfnamefont {B.~H.}\ \bibnamefont {Eom}}, \bibinfo {author} {\bibfnamefont {H.~G.}\ \bibnamefont {LeDuc}}, \bibinfo {author} {\bibfnamefont {S.}~\bibnamefont {Shu}}, \bibinfo {author} {\bibfnamefont {A.~D.}\ \bibnamefont {Beyer}},\ and\ \bibinfo {author} {\bibfnamefont {P.~K.}\ \bibnamefont {Day}},\ }\bibfield  {title} {\bibinfo {title} {{Wideband Direct Detection Constraints on Hidden Photon Dark Matter with the QUALIPHIDE Experiment}},\ }\href {https://doi.org/10.1103/PhysRevLett.130.231001} {\bibfield  {journal} {\bibinfo  {journal} {Phys. Rev. Lett.}\ }\textbf {\bibinfo {volume} {130}},\ \bibinfo {pages} {231001} (\bibinfo {year} {2023})},\ \Eprint {https://arxiv.org/abs/2209.03419} {arXiv:2209.03419 [astro-ph.CO]} \BibitemShut {NoStop}%
\bibitem [{\citenamefont {Kotaka}\ \emph {et~al.}(2023)\citenamefont {Kotaka} \emph {et~al.}}]{DOSUE-RR:2022ise}%
  \BibitemOpen
  \bibfield  {author} {\bibinfo {author} {\bibfnamefont {S.}~\bibnamefont {Kotaka}} \emph {et~al.} (\bibinfo {collaboration} {DOSUE-RR}),\ }\bibfield  {title} {\bibinfo {title} {{Search for Dark Photon Dark Matter in the Mass Range 74\textendash{}110\,\,\ensuremath{\mu}eV with a Cryogenic Millimeter-Wave Receiver}},\ }\href {https://doi.org/10.1103/PhysRevLett.130.071805} {\bibfield  {journal} {\bibinfo  {journal} {Phys. Rev. Lett.}\ }\textbf {\bibinfo {volume} {130}},\ \bibinfo {pages} {071805} (\bibinfo {year} {2023})},\ \Eprint {https://arxiv.org/abs/2205.03679} {arXiv:2205.03679 [hep-ex]} \BibitemShut {NoStop}%
\bibitem [{\citenamefont {Bajjali}\ \emph {et~al.}(2023)\citenamefont {Bajjali} \emph {et~al.}}]{Bajjali:2023uis}%
  \BibitemOpen
  \bibfield  {author} {\bibinfo {author} {\bibfnamefont {F.}~\bibnamefont {Bajjali}} \emph {et~al.},\ }\bibfield  {title} {\bibinfo {title} {{First results from BRASS-p broadband searches for hidden photon dark matter}},\ }\href@noop {} {\  (\bibinfo {year} {2023})},\ \Eprint {https://arxiv.org/abs/2306.05934} {arXiv:2306.05934 [hep-ex]} \BibitemShut {NoStop}%
\bibitem [{\citenamefont {Adachi}\ \emph {et~al.}(2023)\citenamefont {Adachi}, \citenamefont {Fujinaka}, \citenamefont {Honda}, \citenamefont {Muto}, \citenamefont {Nakata}, \citenamefont {Sueno}, \citenamefont {Sumida}, \citenamefont {Suzuki}, \citenamefont {Tajima},\ and\ \citenamefont {Takeuchi}}]{Adachi:2023wuo}%
  \BibitemOpen
  \bibfield  {author} {\bibinfo {author} {\bibfnamefont {S.}~\bibnamefont {Adachi}}, \bibinfo {author} {\bibfnamefont {F.}~\bibnamefont {Fujinaka}}, \bibinfo {author} {\bibfnamefont {S.}~\bibnamefont {Honda}}, \bibinfo {author} {\bibfnamefont {Y.}~\bibnamefont {Muto}}, \bibinfo {author} {\bibfnamefont {H.}~\bibnamefont {Nakata}}, \bibinfo {author} {\bibfnamefont {Y.}~\bibnamefont {Sueno}}, \bibinfo {author} {\bibfnamefont {T.}~\bibnamefont {Sumida}}, \bibinfo {author} {\bibfnamefont {J.}~\bibnamefont {Suzuki}}, \bibinfo {author} {\bibfnamefont {O.}~\bibnamefont {Tajima}},\ and\ \bibinfo {author} {\bibfnamefont {H.}~\bibnamefont {Takeuchi}},\ }\bibfield  {title} {\bibinfo {title} {{Search for Dark Photon Dark Matter in the Mass Range 41--74 $\mu\mathrm{eV}$ using Millimeter-Wave Receiver and Radioshielding Box}},\ }\href@noop {} {\  (\bibinfo {year} {2023})},\ \Eprint {https://arxiv.org/abs/2308.14656} {arXiv:2308.14656 [hep-ex]} \BibitemShut {NoStop}%
\bibitem [{\citenamefont {Knirck}\ \emph {et~al.}(2024)\citenamefont {Knirck}, \citenamefont {Hoshino}, \citenamefont {Awida}, \citenamefont {Cancelo}, \citenamefont {Di~Federico}, \citenamefont {Knepper}, \citenamefont {Lapuente}, \citenamefont {Littmann}, \citenamefont {Miller}, \citenamefont {Mitchell} \emph {et~al.}}]{knirck2024first}%
  \BibitemOpen
  \bibfield  {author} {\bibinfo {author} {\bibfnamefont {S.}~\bibnamefont {Knirck}}, \bibinfo {author} {\bibfnamefont {G.}~\bibnamefont {Hoshino}}, \bibinfo {author} {\bibfnamefont {M.~H.}\ \bibnamefont {Awida}}, \bibinfo {author} {\bibfnamefont {G.~I.}\ \bibnamefont {Cancelo}}, \bibinfo {author} {\bibfnamefont {M.}~\bibnamefont {Di~Federico}}, \bibinfo {author} {\bibfnamefont {B.}~\bibnamefont {Knepper}}, \bibinfo {author} {\bibfnamefont {A.}~\bibnamefont {Lapuente}}, \bibinfo {author} {\bibfnamefont {M.}~\bibnamefont {Littmann}}, \bibinfo {author} {\bibfnamefont {D.~W.}\ \bibnamefont {Miller}}, \bibinfo {author} {\bibfnamefont {D.~V.}\ \bibnamefont {Mitchell}}, \emph {et~al.} (\bibinfo {collaboration} {BREAD Collaboration}),\ }\bibfield  {title} {\bibinfo {title} {First results from a broadband search for dark photon dark matter in the 44 to $52\text{ }\text{ }\mathrm{\ensuremath{\mu}}\mathrm{eV}$ range with a coaxial dish antenna},\ }\href {https://doi.org/10.1103/PhysRevLett.132.131004} {\bibfield
  {journal} {\bibinfo  {journal} {Phys. Rev. Lett.}\ }\textbf {\bibinfo {volume} {132}},\ \bibinfo {pages} {131004} (\bibinfo {year} {2024})}\BibitemShut {NoStop}%
\bibitem [{\citenamefont {Barros}\ \emph {et~al.}(2013)\citenamefont {Barros}, \citenamefont {Silva}, \citenamefont {Fonseca}, \citenamefont {Zang},\ and\ \citenamefont {Bergmann}}]{Barros:2013coax}%
  \BibitemOpen
  \bibfield  {author} {\bibinfo {author} {\bibfnamefont {F.~J.~B.}\ \bibnamefont {Barros}}, \bibinfo {author} {\bibfnamefont {S.~P.}\ \bibnamefont {Silva}}, \bibinfo {author} {\bibfnamefont {W.~S.}\ \bibnamefont {Fonseca}}, \bibinfo {author} {\bibfnamefont {S.~R.}\ \bibnamefont {Zang}},\ and\ \bibinfo {author} {\bibfnamefont {J.~R.}\ \bibnamefont {Bergmann}},\ }\bibfield  {title} {\bibinfo {title} {{Analysis of a coaxial horn antenna using FDTD bidimensional method}},\ }in\ \href {https://doi.org/10.1109/IMOC.2013.6646569} {\emph {\bibinfo {booktitle} {2013 SBMO/IEEE MTT-S International Microwave Optoelectronics Conference (IMOC)}}}\ (\bibinfo {year} {2013})\ pp.\ \bibinfo {pages} {1--4}\BibitemShut {NoStop}%
\bibitem [{\citenamefont {Bykov}\ \emph {et~al.}(2008)\citenamefont {Bykov}, \citenamefont {Bykov}, \citenamefont {Klimov}, \citenamefont {Kurkan},\ and\ \citenamefont {Rostov}}]{Bykov:2008coax}%
  \BibitemOpen
  \bibfield  {author} {\bibinfo {author} {\bibfnamefont {D.~N.}\ \bibnamefont {Bykov}}, \bibinfo {author} {\bibfnamefont {N.~M.}\ \bibnamefont {Bykov}}, \bibinfo {author} {\bibfnamefont {A.~I.}\ \bibnamefont {Klimov}}, \bibinfo {author} {\bibfnamefont {I.~K.}\ \bibnamefont {Kurkan}},\ and\ \bibinfo {author} {\bibfnamefont {V.~V.}\ \bibnamefont {Rostov}},\ }\bibfield  {title} {\bibinfo {title} {A wideband converter of the main mode of the coaxial line into the lowest symmetric mode of a circular waveguide},\ }\href {https://doi.org/10.1134/S0020441208050126} {\bibfield  {journal} {\bibinfo  {journal} {Instrum. Exp. Tech.}\ }\textbf {\bibinfo {volume} {51}},\ \bibinfo {pages} {724} (\bibinfo {year} {2008})}\BibitemShut {NoStop}%
\bibitem [{\citenamefont {{Xilinx}}()}]{xilinx}%
  \BibitemOpen
  \bibfield  {author} {\bibinfo {author} {\bibnamefont {{Xilinx}}},\ }\href@noop {} {\bibinfo {title} {{RFSoC 4x2 Kit}}},\ \bibinfo {howpublished} {\url{https://www.amd.com/en/corporate/university-program/aup-boards/rfsoc4x2.html}},\ \bibinfo {note} {[Online; accessed 11-12-2024]}\BibitemShut {NoStop}%
\bibitem [{COM()}]{COMSOL}%
  \BibitemOpen
  \href {www.COMSOL.com} {\bibinfo {title} {{COMSOL Multiphysics\textregistered~~v. 6.1 COMSOL AB, Stockholm, Sweden}}}\BibitemShut {NoStop}%
\bibitem [{\citenamefont {{Keysight Technologies}}()}]{y_factor}%
  \BibitemOpen
  \bibfield  {author} {\bibinfo {author} {\bibnamefont {{Keysight Technologies}}},\ }\href@noop {} {\bibinfo {title} {{Noise Figure Measurement Accuracy: The Y-Factor Method, Application Note 57-2}}},\ \bibinfo {howpublished} {\url{https://www.keysight.com/us/en/assets/7018-06829/application-notes/5952-3706.pdf}},\ \bibinfo {note} {[Online; accessed 1-August-2023]}\BibitemShut {NoStop}%
\bibitem [{\citenamefont {Stefanazzi}\ \emph {et~al.}(2022)\citenamefont {Stefanazzi} \emph {et~al.}}]{Stefanazzi:2021otz}%
  \BibitemOpen
  \bibfield  {author} {\bibinfo {author} {\bibfnamefont {L.}~\bibnamefont {Stefanazzi}} \emph {et~al.},\ }\bibfield  {title} {\bibinfo {title} {{The QICK (Quantum Instrumentation Control Kit): Readout and control for qubits and detectors}},\ }\href {https://doi.org/10.1063/5.0076249} {\bibfield  {journal} {\bibinfo  {journal} {Rev. Sci. Instrum.}\ }\textbf {\bibinfo {volume} {93}},\ \bibinfo {pages} {044709} (\bibinfo {year} {2022})},\ \Eprint {https://arxiv.org/abs/2110.00557} {arXiv:2110.00557 [quant-ph]} \BibitemShut {NoStop}%
\bibitem [{\citenamefont {{AMD\textregistered}}()}]{pynq}%
  \BibitemOpen
  \bibfield  {author} {\bibinfo {author} {\bibnamefont {{AMD\textregistered}}},\ }\href@noop {} {\bibinfo {title} {{Pynq: Python Productivity}}},\ \bibinfo {howpublished} {\url{http://www.pynq.io/}},\ \bibinfo {note} {[Online; accessed 2-August-2023]}\BibitemShut {NoStop}%
\bibitem [{\citenamefont {Bartram}\ \emph {et~al.}(2021{\natexlab{b}})\citenamefont {Bartram} \emph {et~al.}}]{ADMX:2020hay}%
  \BibitemOpen
  \bibfield  {author} {\bibinfo {author} {\bibfnamefont {C.}~\bibnamefont {Bartram}} \emph {et~al.} (\bibinfo {collaboration} {ADMX}),\ }\bibfield  {title} {\bibinfo {title} {{Axion dark matter experiment: Run 1B analysis details}},\ }\href {https://doi.org/10.1103/PhysRevD.103.032002} {\bibfield  {journal} {\bibinfo  {journal} {Phys. Rev. D}\ }\textbf {\bibinfo {volume} {103}},\ \bibinfo {pages} {032002} (\bibinfo {year} {2021}{\natexlab{b}})},\ \Eprint {https://arxiv.org/abs/2010.06183} {arXiv:2010.06183 [astro-ph.CO]} \BibitemShut {NoStop}%
\bibitem [{\citenamefont {Savitzky}\ and\ \citenamefont {Golay}(1964)}]{doi:10.1021/ac60214a047}%
  \BibitemOpen
  \bibfield  {author} {\bibinfo {author} {\bibfnamefont {A.}~\bibnamefont {Savitzky}}\ and\ \bibinfo {author} {\bibfnamefont {M.~J.~E.}\ \bibnamefont {Golay}},\ }\bibfield  {title} {\bibinfo {title} {Smoothing and differentiation of data by simplified least squares procedures.},\ }\href {https://doi.org/10.1021/ac60214a047} {\bibfield  {journal} {\bibinfo  {journal} {Analytical Chemistry}\ }\textbf {\bibinfo {volume} {36}},\ \bibinfo {pages} {1627} (\bibinfo {year} {1964})},\ \Eprint {https://arxiv.org/abs/https://doi.org/10.1021/ac60214a047} {https://doi.org/10.1021/ac60214a047} \BibitemShut {NoStop}%
\bibitem [{\citenamefont {Asztalos}\ \emph {et~al.}(2001)\citenamefont {Asztalos} \emph {et~al.}}]{ADMX:2001dbg}%
  \BibitemOpen
  \bibfield  {author} {\bibinfo {author} {\bibfnamefont {S.~J.}\ \bibnamefont {Asztalos}} \emph {et~al.} (\bibinfo {collaboration} {ADMX}),\ }\bibfield  {title} {\bibinfo {title} {{Large scale microwave cavity search for dark matter axions}},\ }\href {https://doi.org/10.1103/PhysRevD.64.092003} {\bibfield  {journal} {\bibinfo  {journal} {Phys. Rev. D}\ }\textbf {\bibinfo {volume} {64}},\ \bibinfo {pages} {092003} (\bibinfo {year} {2001})}\BibitemShut {NoStop}%
\bibitem [{\citenamefont {Brubaker}\ \emph {et~al.}(2017{\natexlab{b}})\citenamefont {Brubaker}, \citenamefont {Zhong}, \citenamefont {Lamoreaux}, \citenamefont {Lehnert},\ and\ \citenamefont {van Bibber}}]{Brubaker:2017rna}%
  \BibitemOpen
  \bibfield  {author} {\bibinfo {author} {\bibfnamefont {B.~M.}\ \bibnamefont {Brubaker}}, \bibinfo {author} {\bibfnamefont {L.}~\bibnamefont {Zhong}}, \bibinfo {author} {\bibfnamefont {S.~K.}\ \bibnamefont {Lamoreaux}}, \bibinfo {author} {\bibfnamefont {K.~W.}\ \bibnamefont {Lehnert}},\ and\ \bibinfo {author} {\bibfnamefont {K.~A.}\ \bibnamefont {van Bibber}},\ }\bibfield  {title} {\bibinfo {title} {{HAYSTAC axion search analysis procedure}},\ }\href {https://doi.org/10.1103/PhysRevD.96.123008} {\bibfield  {journal} {\bibinfo  {journal} {Phys. Rev. D}\ }\textbf {\bibinfo {volume} {96}},\ \bibinfo {pages} {123008} (\bibinfo {year} {2017}{\natexlab{b}})},\ \Eprint {https://arxiv.org/abs/1706.08388} {arXiv:1706.08388 [astro-ph.IM]} \BibitemShut {NoStop}%
\bibitem [{\citenamefont {Arsenovic}\ \emph {et~al.}(2022)\citenamefont {Arsenovic}, \citenamefont {Hillairet}, \citenamefont {Anderson}, \citenamefont {Forstén}, \citenamefont {Rieß}, \citenamefont {Eller}, \citenamefont {Sauber}, \citenamefont {Weikle}, \citenamefont {Barnhart},\ and\ \citenamefont {Forstmayr}}]{skrf}%
  \BibitemOpen
  \bibfield  {author} {\bibinfo {author} {\bibfnamefont {A.}~\bibnamefont {Arsenovic}}, \bibinfo {author} {\bibfnamefont {J.}~\bibnamefont {Hillairet}}, \bibinfo {author} {\bibfnamefont {J.}~\bibnamefont {Anderson}}, \bibinfo {author} {\bibfnamefont {H.}~\bibnamefont {Forstén}}, \bibinfo {author} {\bibfnamefont {V.}~\bibnamefont {Rieß}}, \bibinfo {author} {\bibfnamefont {M.}~\bibnamefont {Eller}}, \bibinfo {author} {\bibfnamefont {N.}~\bibnamefont {Sauber}}, \bibinfo {author} {\bibfnamefont {R.}~\bibnamefont {Weikle}}, \bibinfo {author} {\bibfnamefont {W.}~\bibnamefont {Barnhart}},\ and\ \bibinfo {author} {\bibfnamefont {F.}~\bibnamefont {Forstmayr}},\ }\bibfield  {title} {\bibinfo {title} {scikit-rf: An open source python package for microwave network creation, analysis, and calibration [speaker’s corner]},\ }\href {https://doi.org/10.1109/MMM.2021.3117139} {\bibfield  {journal} {\bibinfo  {journal} {IEEE Microwave Magazine}\ }\textbf {\bibinfo {volume} {23}},\ \bibinfo {pages} {98} (\bibinfo {year}
  {2022})}\BibitemShut {NoStop}%
\bibitem [{\citenamefont {O'Hare}(2020)}]{AxionLimits}%
  \BibitemOpen
  \bibfield  {author} {\bibinfo {author} {\bibfnamefont {C.}~\bibnamefont {O'Hare}},\ }\href {https://doi.org/10.5281/zenodo.3932430} {\bibinfo {title} {cajohare/axionlimits: Axionlimits}},\ \bibinfo {howpublished} {\url{https://cajohare.github.io/AxionLimits/}} (\bibinfo {year} {2020})\BibitemShut {NoStop}%
\bibitem [{\citenamefont {Andriamonje}\ \emph {et~al.}(2007)\citenamefont {Andriamonje} \emph {et~al.}}]{CAST:2007jps}%
  \BibitemOpen
  \bibfield  {author} {\bibinfo {author} {\bibfnamefont {S.}~\bibnamefont {Andriamonje}} \emph {et~al.} (\bibinfo {collaboration} {CAST}),\ }\bibfield  {title} {\bibinfo {title} {{An Improved limit on the axion-photon coupling from the CAST experiment}},\ }\href {https://doi.org/10.1088/1475-7516/2007/04/010} {\bibfield  {journal} {\bibinfo  {journal} {JCAP}\ }\textbf {\bibinfo {volume} {04}},\ \bibinfo {pages} {010}},\ \Eprint {https://arxiv.org/abs/hep-ex/0702006} {arXiv:hep-ex/0702006} \BibitemShut {NoStop}%
\bibitem [{\citenamefont {Anastassopoulos}\ \emph {et~al.}(2017)\citenamefont {Anastassopoulos} \emph {et~al.}}]{Anastassopoulos:2017ftl}%
  \BibitemOpen
  \bibfield  {author} {\bibinfo {author} {\bibfnamefont {V.}~\bibnamefont {Anastassopoulos}} \emph {et~al.} (\bibinfo {collaboration} {CAST}),\ }\bibfield  {title} {\bibinfo {title} {{New CAST Limit on the Axion-Photon Interaction}},\ }\href {https://doi.org/10.1038/nphys4109} {\bibfield  {journal} {\bibinfo  {journal} {Nat. Phys.}\ }\textbf {\bibinfo {volume} {13}},\ \bibinfo {pages} {584} (\bibinfo {year} {2017})},\ \Eprint {https://arxiv.org/abs/1705.02290} {arXiv:1705.02290 [hep-ex]} \BibitemShut {NoStop}%
\bibitem [{\citenamefont {Altenm\"uller}\ \emph {et~al.}(2024)\citenamefont {Altenm\"uller} \emph {et~al.}}]{CAST:2024eil}%
  \BibitemOpen
  \bibfield  {author} {\bibinfo {author} {\bibfnamefont {K.}~\bibnamefont {Altenm\"uller}} \emph {et~al.} (\bibinfo {collaboration} {CAST}),\ }\bibfield  {title} {\bibinfo {title} {{A new upper limit on the axion-photon coupling with an extended CAST run with a Xe-based Micromegas detector}},\ }\href@noop {} {\  (\bibinfo {year} {2024})},\ \Eprint {https://arxiv.org/abs/2406.16840} {arXiv:2406.16840 [hep-ex]} \BibitemShut {NoStop}%
\bibitem [{\citenamefont {Hoof}\ and\ \citenamefont {Schulz}(2022)}]{Hoof:2022xbe}%
  \BibitemOpen
  \bibfield  {author} {\bibinfo {author} {\bibfnamefont {S.}~\bibnamefont {Hoof}}\ and\ \bibinfo {author} {\bibfnamefont {L.}~\bibnamefont {Schulz}},\ }\bibfield  {title} {\bibinfo {title} {{Updated constraints on axion-like particles from temporal information in supernova SN1987A gamma-ray data}},\ }\href@noop {} {\  (\bibinfo {year} {2022})},\ \Eprint {https://arxiv.org/abs/2212.09764} {arXiv:2212.09764 [hep-ph]} \BibitemShut {NoStop}%
\bibitem [{\citenamefont {Manzari}\ \emph {et~al.}(2024)\citenamefont {Manzari}, \citenamefont {Park}, \citenamefont {Safdi},\ and\ \citenamefont {Savoray}}]{Manzari:2024jns}%
  \BibitemOpen
  \bibfield  {author} {\bibinfo {author} {\bibfnamefont {C.~A.}\ \bibnamefont {Manzari}}, \bibinfo {author} {\bibfnamefont {Y.}~\bibnamefont {Park}}, \bibinfo {author} {\bibfnamefont {B.~R.}\ \bibnamefont {Safdi}},\ and\ \bibinfo {author} {\bibfnamefont {I.}~\bibnamefont {Savoray}},\ }\bibfield  {title} {\bibinfo {title} {{Supernova axions convert to gamma-rays in magnetic fields of progenitor stars}},\ }\href@noop {} {\  (\bibinfo {year} {2024})},\ \Eprint {https://arxiv.org/abs/2405.19393} {arXiv:2405.19393 [hep-ph]} \BibitemShut {NoStop}%
\bibitem [{\citenamefont {Ayala}\ \emph {et~al.}(2014)\citenamefont {Ayala}, \citenamefont {Dom\'\i{}nguez}, \citenamefont {Giannotti}, \citenamefont {Mirizzi},\ and\ \citenamefont {Straniero}}]{Ayala:2014pea}%
  \BibitemOpen
  \bibfield  {author} {\bibinfo {author} {\bibfnamefont {A.}~\bibnamefont {Ayala}}, \bibinfo {author} {\bibfnamefont {I.}~\bibnamefont {Dom\'\i{}nguez}}, \bibinfo {author} {\bibfnamefont {M.}~\bibnamefont {Giannotti}}, \bibinfo {author} {\bibfnamefont {A.}~\bibnamefont {Mirizzi}},\ and\ \bibinfo {author} {\bibfnamefont {O.}~\bibnamefont {Straniero}},\ }\bibfield  {title} {\bibinfo {title} {{Revisiting the bound on axion-photon coupling from Globular Clusters}},\ }\href {https://doi.org/10.1103/PhysRevLett.113.191302} {\bibfield  {journal} {\bibinfo  {journal} {Phys. Rev. Lett.}\ }\textbf {\bibinfo {volume} {113}},\ \bibinfo {pages} {191302} (\bibinfo {year} {2014})},\ \Eprint {https://arxiv.org/abs/1406.6053} {arXiv:1406.6053 [astro-ph.SR]} \BibitemShut {NoStop}%
\bibitem [{\citenamefont {Dolan}\ \emph {et~al.}(2022)\citenamefont {Dolan}, \citenamefont {Hiskens},\ and\ \citenamefont {Volkas}}]{Dolan:2022kul}%
  \BibitemOpen
  \bibfield  {author} {\bibinfo {author} {\bibfnamefont {M.~J.}\ \bibnamefont {Dolan}}, \bibinfo {author} {\bibfnamefont {F.~J.}\ \bibnamefont {Hiskens}},\ and\ \bibinfo {author} {\bibfnamefont {R.~R.}\ \bibnamefont {Volkas}},\ }\bibfield  {title} {\bibinfo {title} {{Advancing globular cluster constraints on the axion-photon coupling}},\ }\href {https://doi.org/10.1088/1475-7516/2022/10/096} {\bibfield  {journal} {\bibinfo  {journal} {JCAP}\ }\textbf {\bibinfo {volume} {10}},\ \bibinfo {pages} {096}},\ \Eprint {https://arxiv.org/abs/2207.03102} {arXiv:2207.03102 [hep-ph]} \BibitemShut {NoStop}%
\bibitem [{\citenamefont {Ruz}\ \emph {et~al.}(2024)\citenamefont {Ruz}, \citenamefont {Todarello}, \citenamefont {Vogel}, \citenamefont {Giannotti}, \citenamefont {Grefenstette}, \citenamefont {Hudson}, \citenamefont {Hannah}, \citenamefont {Irastorza}, \citenamefont {Kim}, \citenamefont {O'Shea}, \citenamefont {Regis}, \citenamefont {Smith}, \citenamefont {Taoso},\ and\ \citenamefont {Bueno}}]{ruz2024nustaraxionhelioscope}%
  \BibitemOpen
  \bibfield  {author} {\bibinfo {author} {\bibfnamefont {J.}~\bibnamefont {Ruz}}, \bibinfo {author} {\bibfnamefont {E.}~\bibnamefont {Todarello}}, \bibinfo {author} {\bibfnamefont {J.~K.}\ \bibnamefont {Vogel}}, \bibinfo {author} {\bibfnamefont {M.}~\bibnamefont {Giannotti}}, \bibinfo {author} {\bibfnamefont {B.}~\bibnamefont {Grefenstette}}, \bibinfo {author} {\bibfnamefont {H.~S.}\ \bibnamefont {Hudson}}, \bibinfo {author} {\bibfnamefont {I.~G.}\ \bibnamefont {Hannah}}, \bibinfo {author} {\bibfnamefont {I.~G.}\ \bibnamefont {Irastorza}}, \bibinfo {author} {\bibfnamefont {C.~S.}\ \bibnamefont {Kim}}, \bibinfo {author} {\bibfnamefont {T.}~\bibnamefont {O'Shea}}, \bibinfo {author} {\bibfnamefont {M.}~\bibnamefont {Regis}}, \bibinfo {author} {\bibfnamefont {D.~M.}\ \bibnamefont {Smith}}, \bibinfo {author} {\bibfnamefont {M.}~\bibnamefont {Taoso}},\ and\ \bibinfo {author} {\bibfnamefont {J.~T.}\ \bibnamefont {Bueno}},\ }\href {https://arxiv.org/abs/2407.03828} {\bibinfo {title} {Nustar as an axion helioscope}}
  (\bibinfo {year} {2024}),\ \Eprint {https://arxiv.org/abs/2407.03828} {arXiv:2407.03828 [astro-ph.CO]} \BibitemShut {NoStop}%
\bibitem [{\citenamefont {Noordhuis}\ \emph {et~al.}(2023)\citenamefont {Noordhuis}, \citenamefont {Prabhu}, \citenamefont {Witte}, \citenamefont {Chen}, \citenamefont {Cruz},\ and\ \citenamefont {Weniger}}]{PhysRevLett.131.111004}%
  \BibitemOpen
  \bibfield  {author} {\bibinfo {author} {\bibfnamefont {D.}~\bibnamefont {Noordhuis}}, \bibinfo {author} {\bibfnamefont {A.}~\bibnamefont {Prabhu}}, \bibinfo {author} {\bibfnamefont {S.~J.}\ \bibnamefont {Witte}}, \bibinfo {author} {\bibfnamefont {A.~Y.}\ \bibnamefont {Chen}}, \bibinfo {author} {\bibfnamefont {F.}~\bibnamefont {Cruz}},\ and\ \bibinfo {author} {\bibfnamefont {C.}~\bibnamefont {Weniger}},\ }\bibfield  {title} {\bibinfo {title} {Novel constraints on axions produced in pulsar polar-cap cascades},\ }\href {https://doi.org/10.1103/PhysRevLett.131.111004} {\bibfield  {journal} {\bibinfo  {journal} {Phys. Rev. Lett.}\ }\textbf {\bibinfo {volume} {131}},\ \bibinfo {pages} {111004} (\bibinfo {year} {2023})}\BibitemShut {NoStop}%
\bibitem [{\citenamefont {Foster}\ \emph {et~al.}(2020)\citenamefont {Foster}, \citenamefont {Kahn}, \citenamefont {Macias}, \citenamefont {Sun}, \citenamefont {Eatough}, \citenamefont {Kondratiev}, \citenamefont {Peters}, \citenamefont {Weniger},\ and\ \citenamefont {Safdi}}]{PhysRevLett.125.171301}%
  \BibitemOpen
  \bibfield  {author} {\bibinfo {author} {\bibfnamefont {J.~W.}\ \bibnamefont {Foster}}, \bibinfo {author} {\bibfnamefont {Y.}~\bibnamefont {Kahn}}, \bibinfo {author} {\bibfnamefont {O.}~\bibnamefont {Macias}}, \bibinfo {author} {\bibfnamefont {Z.}~\bibnamefont {Sun}}, \bibinfo {author} {\bibfnamefont {R.~P.}\ \bibnamefont {Eatough}}, \bibinfo {author} {\bibfnamefont {V.~I.}\ \bibnamefont {Kondratiev}}, \bibinfo {author} {\bibfnamefont {W.~M.}\ \bibnamefont {Peters}}, \bibinfo {author} {\bibfnamefont {C.}~\bibnamefont {Weniger}},\ and\ \bibinfo {author} {\bibfnamefont {B.~R.}\ \bibnamefont {Safdi}},\ }\bibfield  {title} {\bibinfo {title} {Green bank and effelsberg radio telescope searches for axion dark matter conversion in neutron star magnetospheres},\ }\href {https://doi.org/10.1103/PhysRevLett.125.171301} {\bibfield  {journal} {\bibinfo  {journal} {Phys. Rev. Lett.}\ }\textbf {\bibinfo {volume} {125}},\ \bibinfo {pages} {171301} (\bibinfo {year} {2020})}\BibitemShut {NoStop}%
\bibitem [{\citenamefont {Darling}(2020)}]{Darling_2020}%
  \BibitemOpen
  \bibfield  {author} {\bibinfo {author} {\bibfnamefont {J.}~\bibnamefont {Darling}},\ }\bibfield  {title} {\bibinfo {title} {New limits on axionic dark matter from the magnetar psr j1745-2900},\ }\href {https://doi.org/10.3847/2041-8213/abb23f} {\bibfield  {journal} {\bibinfo  {journal} {The Astrophysical Journal Letters}\ }\textbf {\bibinfo {volume} {900}},\ \bibinfo {pages} {L28} (\bibinfo {year} {2020})}\BibitemShut {NoStop}%
\bibitem [{\citenamefont {Battye}\ \emph {et~al.}(2022)\citenamefont {Battye}, \citenamefont {Darling}, \citenamefont {McDonald},\ and\ \citenamefont {Srinivasan}}]{PhysRevD.105.L021305}%
  \BibitemOpen
  \bibfield  {author} {\bibinfo {author} {\bibfnamefont {R.~A.}\ \bibnamefont {Battye}}, \bibinfo {author} {\bibfnamefont {J.}~\bibnamefont {Darling}}, \bibinfo {author} {\bibfnamefont {J.~I.}\ \bibnamefont {McDonald}},\ and\ \bibinfo {author} {\bibfnamefont {S.}~\bibnamefont {Srinivasan}},\ }\bibfield  {title} {\bibinfo {title} {Towards robust constraints on axion dark matter using psr j1745-2900},\ }\href {https://doi.org/10.1103/PhysRevD.105.L021305} {\bibfield  {journal} {\bibinfo  {journal} {Phys. Rev. D}\ }\textbf {\bibinfo {volume} {105}},\ \bibinfo {pages} {L021305} (\bibinfo {year} {2022})}\BibitemShut {NoStop}%
\bibitem [{\citenamefont {Foster}\ \emph {et~al.}(2022)\citenamefont {Foster}, \citenamefont {Witte}, \citenamefont {Lawson}, \citenamefont {Linden}, \citenamefont {Gajjar}, \citenamefont {Weniger},\ and\ \citenamefont {Safdi}}]{PhysRevLett.129.251102}%
  \BibitemOpen
  \bibfield  {author} {\bibinfo {author} {\bibfnamefont {J.~W.}\ \bibnamefont {Foster}}, \bibinfo {author} {\bibfnamefont {S.~J.}\ \bibnamefont {Witte}}, \bibinfo {author} {\bibfnamefont {M.}~\bibnamefont {Lawson}}, \bibinfo {author} {\bibfnamefont {T.}~\bibnamefont {Linden}}, \bibinfo {author} {\bibfnamefont {V.}~\bibnamefont {Gajjar}}, \bibinfo {author} {\bibfnamefont {C.}~\bibnamefont {Weniger}},\ and\ \bibinfo {author} {\bibfnamefont {B.~R.}\ \bibnamefont {Safdi}},\ }\bibfield  {title} {\bibinfo {title} {Extraterrestrial axion search with the breakthrough listen galactic center survey},\ }\href {https://doi.org/10.1103/PhysRevLett.129.251102} {\bibfield  {journal} {\bibinfo  {journal} {Phys. Rev. Lett.}\ }\textbf {\bibinfo {volume} {129}},\ \bibinfo {pages} {251102} (\bibinfo {year} {2022})}\BibitemShut {NoStop}%
\bibitem [{\citenamefont {Battye}\ \emph {et~al.}(2023)\citenamefont {Battye}, \citenamefont {Keith}, \citenamefont {McDonald}, \citenamefont {Srinivasan}, \citenamefont {Stappers},\ and\ \citenamefont {Weltevrede}}]{Battye_2023}%
  \BibitemOpen
  \bibfield  {author} {\bibinfo {author} {\bibfnamefont {R.}~\bibnamefont {Battye}}, \bibinfo {author} {\bibfnamefont {M.}~\bibnamefont {Keith}}, \bibinfo {author} {\bibfnamefont {J.}~\bibnamefont {McDonald}}, \bibinfo {author} {\bibfnamefont {S.}~\bibnamefont {Srinivasan}}, \bibinfo {author} {\bibfnamefont {B.}~\bibnamefont {Stappers}},\ and\ \bibinfo {author} {\bibfnamefont {P.}~\bibnamefont {Weltevrede}},\ }\bibfield  {title} {\bibinfo {title} {Searching for time-dependent axion dark matter signals in pulsars},\ }\bibfield  {journal} {\bibinfo  {journal} {Physical Review D}\ }\textbf {\bibinfo {volume} {108}},\ \href {https://doi.org/10.1103/physrevd.108.063001} {10.1103/physrevd.108.063001} (\bibinfo {year} {2023})\BibitemShut {NoStop}%
\bibitem [{\citenamefont {Boutan}\ \emph {et~al.}(2018)\citenamefont {Boutan}, \citenamefont {Jones}, \citenamefont {LaRoque}, \citenamefont {Oblath}, \citenamefont {Cervantes}, \citenamefont {Du}, \citenamefont {Force}, \citenamefont {Kimes}, \citenamefont {Ottens}, \citenamefont {Rosenberg}, \citenamefont {Rybka}, \citenamefont {Yang}, \citenamefont {Carosi}, \citenamefont {Woollett}, \citenamefont {Bowring}, \citenamefont {Chou}, \citenamefont {Khatiwada}, \citenamefont {Sonnenschein}, \citenamefont {Wester}, \citenamefont {Bradley}, \citenamefont {Daw}, \citenamefont {Agrawal}, \citenamefont {Dixit}, \citenamefont {Clarke}, \citenamefont {O'Kelley}, \citenamefont {Crisosto}, \citenamefont {Gleason}, \citenamefont {Jois}, \citenamefont {Sikivie}, \citenamefont {Stern}, \citenamefont {Sullivan}, \citenamefont {Tanner}, \citenamefont {Harrington},\ and\ \citenamefont {Lentz}}]{PhysRevLett.121.261302}%
  \BibitemOpen
  \bibfield  {author} {\bibinfo {author} {\bibfnamefont {C.}~\bibnamefont {Boutan}}, \bibinfo {author} {\bibfnamefont {M.}~\bibnamefont {Jones}}, \bibinfo {author} {\bibfnamefont {B.~H.}\ \bibnamefont {LaRoque}}, \bibinfo {author} {\bibfnamefont {N.~S.}\ \bibnamefont {Oblath}}, \bibinfo {author} {\bibfnamefont {R.}~\bibnamefont {Cervantes}}, \bibinfo {author} {\bibfnamefont {N.}~\bibnamefont {Du}}, \bibinfo {author} {\bibfnamefont {N.}~\bibnamefont {Force}}, \bibinfo {author} {\bibfnamefont {S.}~\bibnamefont {Kimes}}, \bibinfo {author} {\bibfnamefont {R.}~\bibnamefont {Ottens}}, \bibinfo {author} {\bibfnamefont {L.~J.}\ \bibnamefont {Rosenberg}}, \bibinfo {author} {\bibfnamefont {G.}~\bibnamefont {Rybka}}, \bibinfo {author} {\bibfnamefont {J.}~\bibnamefont {Yang}}, \bibinfo {author} {\bibfnamefont {G.}~\bibnamefont {Carosi}}, \bibinfo {author} {\bibfnamefont {N.}~\bibnamefont {Woollett}}, \bibinfo {author} {\bibfnamefont {D.}~\bibnamefont {Bowring}}, \bibinfo {author} {\bibfnamefont {A.~S.}\ \bibnamefont
  {Chou}}, \bibinfo {author} {\bibfnamefont {R.}~\bibnamefont {Khatiwada}}, \bibinfo {author} {\bibfnamefont {A.}~\bibnamefont {Sonnenschein}}, \bibinfo {author} {\bibfnamefont {W.}~\bibnamefont {Wester}}, \bibinfo {author} {\bibfnamefont {R.}~\bibnamefont {Bradley}}, \bibinfo {author} {\bibfnamefont {E.~J.}\ \bibnamefont {Daw}}, \bibinfo {author} {\bibfnamefont {A.}~\bibnamefont {Agrawal}}, \bibinfo {author} {\bibfnamefont {A.~V.}\ \bibnamefont {Dixit}}, \bibinfo {author} {\bibfnamefont {J.}~\bibnamefont {Clarke}}, \bibinfo {author} {\bibfnamefont {S.~R.}\ \bibnamefont {O'Kelley}}, \bibinfo {author} {\bibfnamefont {N.}~\bibnamefont {Crisosto}}, \bibinfo {author} {\bibfnamefont {J.~R.}\ \bibnamefont {Gleason}}, \bibinfo {author} {\bibfnamefont {S.}~\bibnamefont {Jois}}, \bibinfo {author} {\bibfnamefont {P.}~\bibnamefont {Sikivie}}, \bibinfo {author} {\bibfnamefont {I.}~\bibnamefont {Stern}}, \bibinfo {author} {\bibfnamefont {N.~S.}\ \bibnamefont {Sullivan}}, \bibinfo {author} {\bibfnamefont {D.~B.}\
  \bibnamefont {Tanner}}, \bibinfo {author} {\bibfnamefont {P.~M.}\ \bibnamefont {Harrington}},\ and\ \bibinfo {author} {\bibfnamefont {E.}~\bibnamefont {Lentz}} (\bibinfo {collaboration} {ADMX Collaboration}),\ }\bibfield  {title} {\bibinfo {title} {Piezoelectrically tuned multimode cavity search for axion dark matter},\ }\href {https://doi.org/10.1103/PhysRevLett.121.261302} {\bibfield  {journal} {\bibinfo  {journal} {Phys. Rev. Lett.}\ }\textbf {\bibinfo {volume} {121}},\ \bibinfo {pages} {261302} (\bibinfo {year} {2018})}\BibitemShut {NoStop}%
\bibitem [{\citenamefont {Bartram}\ \emph {et~al.}(2023)\citenamefont {Bartram}, \citenamefont {Braine}, \citenamefont {Cervantes}, \citenamefont {Crisosto}, \citenamefont {Du}, \citenamefont {Leum}, \citenamefont {Mohapatra}, \citenamefont {Nitta}, \citenamefont {Rosenberg}, \citenamefont {Rybka}, \citenamefont {Yang}, \citenamefont {Clarke}, \citenamefont {Siddiqi}, \citenamefont {Agrawal}, \citenamefont {Dixit}, \citenamefont {Awida}, \citenamefont {Chou}, \citenamefont {Hollister}, \citenamefont {Knirck}, \citenamefont {Sonnenschein}, \citenamefont {Wester}, \citenamefont {Gleason}, \citenamefont {Hipp}, \citenamefont {Jois}, \citenamefont {Sikivie}, \citenamefont {Sullivan}, \citenamefont {Tanner}, \citenamefont {Lentz}, \citenamefont {Khatiwada}, \citenamefont {Carosi}, \citenamefont {Cisneros}, \citenamefont {Robertson}, \citenamefont {Woollett}, \citenamefont {Duffy}, \citenamefont {Boutan}, \citenamefont {Jones}, \citenamefont {LaRoque}, \citenamefont {Oblath}, \citenamefont {Taubman}, \citenamefont
  {Daw}, \citenamefont {Perry}, \citenamefont {Buckley}, \citenamefont {Gaikwad}, \citenamefont {Hoffman}, \citenamefont {Murch}, \citenamefont {Goryachev}, \citenamefont {McAllister}, \citenamefont {Quiskamp}, \citenamefont {Thomson}, \citenamefont {Tobar}, \citenamefont {Bolkhovsky}, \citenamefont {Calusine}, \citenamefont {Oliver},\ and\ \citenamefont {Serniak}}]{Bartram_2023}%
  \BibitemOpen
  \bibfield  {author} {\bibinfo {author} {\bibfnamefont {C.}~\bibnamefont {Bartram}}, \bibinfo {author} {\bibfnamefont {T.}~\bibnamefont {Braine}}, \bibinfo {author} {\bibfnamefont {R.}~\bibnamefont {Cervantes}}, \bibinfo {author} {\bibfnamefont {N.}~\bibnamefont {Crisosto}}, \bibinfo {author} {\bibfnamefont {N.}~\bibnamefont {Du}}, \bibinfo {author} {\bibfnamefont {G.}~\bibnamefont {Leum}}, \bibinfo {author} {\bibfnamefont {P.}~\bibnamefont {Mohapatra}}, \bibinfo {author} {\bibfnamefont {T.}~\bibnamefont {Nitta}}, \bibinfo {author} {\bibfnamefont {L.~J.}\ \bibnamefont {Rosenberg}}, \bibinfo {author} {\bibfnamefont {G.}~\bibnamefont {Rybka}}, \bibinfo {author} {\bibfnamefont {J.}~\bibnamefont {Yang}}, \bibinfo {author} {\bibfnamefont {J.}~\bibnamefont {Clarke}}, \bibinfo {author} {\bibfnamefont {I.}~\bibnamefont {Siddiqi}}, \bibinfo {author} {\bibfnamefont {A.}~\bibnamefont {Agrawal}}, \bibinfo {author} {\bibfnamefont {A.~V.}\ \bibnamefont {Dixit}}, \bibinfo {author} {\bibfnamefont {M.~H.}\ \bibnamefont
  {Awida}}, \bibinfo {author} {\bibfnamefont {A.~S.}\ \bibnamefont {Chou}}, \bibinfo {author} {\bibfnamefont {M.}~\bibnamefont {Hollister}}, \bibinfo {author} {\bibfnamefont {S.}~\bibnamefont {Knirck}}, \bibinfo {author} {\bibfnamefont {A.}~\bibnamefont {Sonnenschein}}, \bibinfo {author} {\bibfnamefont {W.}~\bibnamefont {Wester}}, \bibinfo {author} {\bibfnamefont {J.~R.}\ \bibnamefont {Gleason}}, \bibinfo {author} {\bibfnamefont {A.~T.}\ \bibnamefont {Hipp}}, \bibinfo {author} {\bibfnamefont {S.}~\bibnamefont {Jois}}, \bibinfo {author} {\bibfnamefont {P.}~\bibnamefont {Sikivie}}, \bibinfo {author} {\bibfnamefont {N.~S.}\ \bibnamefont {Sullivan}}, \bibinfo {author} {\bibfnamefont {D.~B.}\ \bibnamefont {Tanner}}, \bibinfo {author} {\bibfnamefont {E.}~\bibnamefont {Lentz}}, \bibinfo {author} {\bibfnamefont {R.}~\bibnamefont {Khatiwada}}, \bibinfo {author} {\bibfnamefont {G.}~\bibnamefont {Carosi}}, \bibinfo {author} {\bibfnamefont {C.}~\bibnamefont {Cisneros}}, \bibinfo {author} {\bibfnamefont {N.}~\bibnamefont
  {Robertson}}, \bibinfo {author} {\bibfnamefont {N.}~\bibnamefont {Woollett}}, \bibinfo {author} {\bibfnamefont {L.~D.}\ \bibnamefont {Duffy}}, \bibinfo {author} {\bibfnamefont {C.}~\bibnamefont {Boutan}}, \bibinfo {author} {\bibfnamefont {M.}~\bibnamefont {Jones}}, \bibinfo {author} {\bibfnamefont {B.~H.}\ \bibnamefont {LaRoque}}, \bibinfo {author} {\bibfnamefont {N.~S.}\ \bibnamefont {Oblath}}, \bibinfo {author} {\bibfnamefont {M.~S.}\ \bibnamefont {Taubman}}, \bibinfo {author} {\bibfnamefont {E.~J.}\ \bibnamefont {Daw}}, \bibinfo {author} {\bibfnamefont {M.~G.}\ \bibnamefont {Perry}}, \bibinfo {author} {\bibfnamefont {J.~H.}\ \bibnamefont {Buckley}}, \bibinfo {author} {\bibfnamefont {C.}~\bibnamefont {Gaikwad}}, \bibinfo {author} {\bibfnamefont {J.}~\bibnamefont {Hoffman}}, \bibinfo {author} {\bibfnamefont {K.}~\bibnamefont {Murch}}, \bibinfo {author} {\bibfnamefont {M.}~\bibnamefont {Goryachev}}, \bibinfo {author} {\bibfnamefont {B.~T.}\ \bibnamefont {McAllister}}, \bibinfo {author} {\bibfnamefont
  {A.}~\bibnamefont {Quiskamp}}, \bibinfo {author} {\bibfnamefont {C.}~\bibnamefont {Thomson}}, \bibinfo {author} {\bibfnamefont {M.~E.}\ \bibnamefont {Tobar}}, \bibinfo {author} {\bibfnamefont {V.}~\bibnamefont {Bolkhovsky}}, \bibinfo {author} {\bibfnamefont {G.}~\bibnamefont {Calusine}}, \bibinfo {author} {\bibfnamefont {W.}~\bibnamefont {Oliver}},\ and\ \bibinfo {author} {\bibfnamefont {K.}~\bibnamefont {Serniak}},\ }\bibfield  {title} {\bibinfo {title} {Dark matter axion search using a josephson traveling wave parametric amplifier},\ }\bibfield  {journal} {\bibinfo  {journal} {Review of Scientific Instruments}\ }\textbf {\bibinfo {volume} {94}},\ \href {https://doi.org/10.1063/5.0122907} {10.1063/5.0122907} (\bibinfo {year} {2023})\BibitemShut {NoStop}%
\bibitem [{\citenamefont {Adair}\ \emph {et~al.}(2022)\citenamefont {Adair}, \citenamefont {Altenm{\"u}ller}, \citenamefont {Anastassopoulos}, \citenamefont {Arguedas~Cuendis}, \citenamefont {Baier}, \citenamefont {Barth}, \citenamefont {Belov}, \citenamefont {Bozicevic}, \citenamefont {Br{\"a}uninger}, \citenamefont {Cantatore} \emph {et~al.}}]{Adair_2022}%
  \BibitemOpen
  \bibfield  {author} {\bibinfo {author} {\bibfnamefont {C.}~\bibnamefont {Adair}}, \bibinfo {author} {\bibfnamefont {K.}~\bibnamefont {Altenm{\"u}ller}}, \bibinfo {author} {\bibfnamefont {V.}~\bibnamefont {Anastassopoulos}}, \bibinfo {author} {\bibfnamefont {S.}~\bibnamefont {Arguedas~Cuendis}}, \bibinfo {author} {\bibfnamefont {J.}~\bibnamefont {Baier}}, \bibinfo {author} {\bibfnamefont {K.}~\bibnamefont {Barth}}, \bibinfo {author} {\bibfnamefont {A.}~\bibnamefont {Belov}}, \bibinfo {author} {\bibfnamefont {D.}~\bibnamefont {Bozicevic}}, \bibinfo {author} {\bibfnamefont {H.}~\bibnamefont {Br{\"a}uninger}}, \bibinfo {author} {\bibfnamefont {G.}~\bibnamefont {Cantatore}}, \emph {et~al.},\ }\bibfield  {title} {\bibinfo {title} {Search for dark matter axions with cast-capp},\ }\bibfield  {journal} {\bibinfo  {journal} {Nature Communications}\ }\textbf {\bibinfo {volume} {13}},\ \href {https://doi.org/10.1038/s41467-022-33913-6} {10.1038/s41467-022-33913-6} (\bibinfo {year} {2022})\BibitemShut {NoStop}%
\bibitem [{\citenamefont {Grenet}\ \emph {et~al.}(2021)\citenamefont {Grenet}, \citenamefont {Ballou}, \citenamefont {Basto}, \citenamefont {Martineau}, \citenamefont {Perrier}, \citenamefont {Pugnat}, \citenamefont {Quevillon}, \citenamefont {Roch},\ and\ \citenamefont {Smith}}]{grenet2021grenobleaxionhaloscopeplatform}%
  \BibitemOpen
  \bibfield  {author} {\bibinfo {author} {\bibfnamefont {T.}~\bibnamefont {Grenet}}, \bibinfo {author} {\bibfnamefont {R.}~\bibnamefont {Ballou}}, \bibinfo {author} {\bibfnamefont {Q.}~\bibnamefont {Basto}}, \bibinfo {author} {\bibfnamefont {K.}~\bibnamefont {Martineau}}, \bibinfo {author} {\bibfnamefont {P.}~\bibnamefont {Perrier}}, \bibinfo {author} {\bibfnamefont {P.}~\bibnamefont {Pugnat}}, \bibinfo {author} {\bibfnamefont {J.}~\bibnamefont {Quevillon}}, \bibinfo {author} {\bibfnamefont {N.}~\bibnamefont {Roch}},\ and\ \bibinfo {author} {\bibfnamefont {C.}~\bibnamefont {Smith}},\ }\href {https://arxiv.org/abs/2110.14406} {\bibinfo {title} {The grenoble axion haloscope platform (grahal): development plan and first results}} (\bibinfo {year} {2021}),\ \Eprint {https://arxiv.org/abs/2110.14406} {arXiv:2110.14406 [hep-ex]} \BibitemShut {NoStop}%
\bibitem [{\citenamefont {Alesini}\ \emph {et~al.}(2019)\citenamefont {Alesini} \emph {et~al.}}]{Alesini:2019ajt}%
  \BibitemOpen
  \bibfield  {author} {\bibinfo {author} {\bibfnamefont {D.}~\bibnamefont {Alesini}} \emph {et~al.},\ }\bibfield  {title} {\bibinfo {title} {{Galactic axions search with a superconducting resonant cavity}},\ }\href {https://doi.org/10.1103/PhysRevD.99.101101} {\bibfield  {journal} {\bibinfo  {journal} {Phys. Rev. D}\ }\textbf {\bibinfo {volume} {99}},\ \bibinfo {pages} {101101} (\bibinfo {year} {2019})},\ \Eprint {https://arxiv.org/abs/1903.06547} {arXiv:1903.06547 [physics.ins-det]} \BibitemShut {NoStop}%
\bibitem [{\citenamefont {Alesini}\ \emph {et~al.}(2021)\citenamefont {Alesini} \emph {et~al.}}]{Alesini:2020vny}%
  \BibitemOpen
  \bibfield  {author} {\bibinfo {author} {\bibfnamefont {D.}~\bibnamefont {Alesini}} \emph {et~al.},\ }\bibfield  {title} {\bibinfo {title} {{Search for invisible axion dark matter of mass m$_a=43~\mu$eV with the QUAX--$a\gamma$ experiment}},\ }\href {https://doi.org/10.1103/PhysRevD.103.102004} {\bibfield  {journal} {\bibinfo  {journal} {Phys. Rev. D}\ }\textbf {\bibinfo {volume} {103}},\ \bibinfo {pages} {102004} (\bibinfo {year} {2021})},\ \Eprint {https://arxiv.org/abs/2012.09498} {arXiv:2012.09498 [hep-ex]} \BibitemShut {NoStop}%
\bibitem [{\citenamefont {Alesini}\ \emph {et~al.}(2022)\citenamefont {Alesini} \emph {et~al.}}]{Alesini:2022lnp}%
  \BibitemOpen
  \bibfield  {author} {\bibinfo {author} {\bibfnamefont {D.}~\bibnamefont {Alesini}} \emph {et~al.},\ }\bibfield  {title} {\bibinfo {title} {{Search for Galactic axions with a high-Q dielectric cavity}},\ }\href {https://doi.org/10.1103/PhysRevD.106.052007} {\bibfield  {journal} {\bibinfo  {journal} {Phys. Rev. D}\ }\textbf {\bibinfo {volume} {106}},\ \bibinfo {pages} {052007} (\bibinfo {year} {2022})},\ \Eprint {https://arxiv.org/abs/2208.12670} {arXiv:2208.12670 [hep-ex]} \BibitemShut {NoStop}%
\bibitem [{\citenamefont {Di~Vora}\ \emph {et~al.}(2023)\citenamefont {Di~Vora} \emph {et~al.}}]{QUAX:2023gop}%
  \BibitemOpen
  \bibfield  {author} {\bibinfo {author} {\bibfnamefont {R.}~\bibnamefont {Di~Vora}} \emph {et~al.} (\bibinfo {collaboration} {QUAX}),\ }\bibfield  {title} {\bibinfo {title} {{Search for galactic axions with a traveling wave parametric amplifier}},\ }\href {https://doi.org/10.1103/PhysRevD.108.062005} {\bibfield  {journal} {\bibinfo  {journal} {Phys. Rev. D}\ }\textbf {\bibinfo {volume} {108}},\ \bibinfo {pages} {062005} (\bibinfo {year} {2023})},\ \Eprint {https://arxiv.org/abs/2304.07505} {arXiv:2304.07505 [hep-ex]} \BibitemShut {NoStop}%
\bibitem [{\citenamefont {Rettaroli}\ \emph {et~al.}(2024)\citenamefont {Rettaroli} \emph {et~al.}}]{QUAX:2024fut}%
  \BibitemOpen
  \bibfield  {author} {\bibinfo {author} {\bibfnamefont {A.}~\bibnamefont {Rettaroli}} \emph {et~al.} (\bibinfo {collaboration} {QUAX}),\ }\bibfield  {title} {\bibinfo {title} {{Search for axion dark matter with the QUAX\textendash{}LNF tunable haloscope}},\ }\href {https://doi.org/10.1103/PhysRevD.110.022008} {\bibfield  {journal} {\bibinfo  {journal} {Phys. Rev. D}\ }\textbf {\bibinfo {volume} {110}},\ \bibinfo {pages} {022008} (\bibinfo {year} {2024})},\ \Eprint {https://arxiv.org/abs/2402.19063} {arXiv:2402.19063 [physics.ins-det]} \BibitemShut {NoStop}%
\bibitem [{\citenamefont {{\'A}lvarez~Melc{\'o}n}\ \emph {et~al.}(2021)\citenamefont {{\'A}lvarez~Melc{\'o}n}, \citenamefont {Arguedas~Cuendis}, \citenamefont {Baier}, \citenamefont {Barth}, \citenamefont {Br{\"a}uninger}, \citenamefont {Calatroni}, \citenamefont {Cantatore}, \citenamefont {Caspers}, \citenamefont {Castel}, \citenamefont {Cetin} \emph {et~al.}}]{alvarez2021first}%
  \BibitemOpen
  \bibfield  {author} {\bibinfo {author} {\bibfnamefont {A.}~\bibnamefont {{\'A}lvarez~Melc{\'o}n}}, \bibinfo {author} {\bibfnamefont {S.}~\bibnamefont {Arguedas~Cuendis}}, \bibinfo {author} {\bibfnamefont {J.}~\bibnamefont {Baier}}, \bibinfo {author} {\bibfnamefont {K.}~\bibnamefont {Barth}}, \bibinfo {author} {\bibfnamefont {H.}~\bibnamefont {Br{\"a}uninger}}, \bibinfo {author} {\bibfnamefont {S.}~\bibnamefont {Calatroni}}, \bibinfo {author} {\bibfnamefont {G.}~\bibnamefont {Cantatore}}, \bibinfo {author} {\bibfnamefont {F.}~\bibnamefont {Caspers}}, \bibinfo {author} {\bibfnamefont {J.}~\bibnamefont {Castel}}, \bibinfo {author} {\bibfnamefont {S.}~\bibnamefont {Cetin}}, \emph {et~al.},\ }\bibfield  {title} {\bibinfo {title} {First results of the cast-rades haloscope search for axions at 34.67 $\mu$ev},\ }\href@noop {} {\bibfield  {journal} {\bibinfo  {journal} {Journal of High Energy Physics}\ }\textbf {\bibinfo {volume} {2021}},\ \bibinfo {pages} {1} (\bibinfo {year} {2021})}\BibitemShut {NoStop}%
\bibitem [{\citenamefont {Ahyoune}\ \emph {et~al.}(2024)\citenamefont {Ahyoune}, \citenamefont {Álvarez Melcón}, \citenamefont {Cuendis}, \citenamefont {Calatroni}, \citenamefont {Cogollos}, \citenamefont {Díaz-Morcillo}, \citenamefont {Döbrich}, \citenamefont {Gallego}, \citenamefont {García-Barceló}, \citenamefont {Gimeno}, \citenamefont {Golm}, \citenamefont {Granados}, \citenamefont {Gutierrez}, \citenamefont {Herwig}, \citenamefont {Irastorza}, \citenamefont {Lamas}, \citenamefont {Lozano-Guerrero}, \citenamefont {Millar}, \citenamefont {Malbrunot}, \citenamefont {Miralda-Escudé}, \citenamefont {Navarro}, \citenamefont {Navarro-Madrid}, \citenamefont {Puig}, \citenamefont {Siodlaczek}, \citenamefont {Telles},\ and\ \citenamefont {Wuensch}}]{ahyoune2024radesaxionsearchresults}%
  \BibitemOpen
  \bibfield  {author} {\bibinfo {author} {\bibfnamefont {S.}~\bibnamefont {Ahyoune}}, \bibinfo {author} {\bibfnamefont {A.}~\bibnamefont {Álvarez Melcón}}, \bibinfo {author} {\bibfnamefont {S.~A.}\ \bibnamefont {Cuendis}}, \bibinfo {author} {\bibfnamefont {S.}~\bibnamefont {Calatroni}}, \bibinfo {author} {\bibfnamefont {C.}~\bibnamefont {Cogollos}}, \bibinfo {author} {\bibfnamefont {A.}~\bibnamefont {Díaz-Morcillo}}, \bibinfo {author} {\bibfnamefont {B.}~\bibnamefont {Döbrich}}, \bibinfo {author} {\bibfnamefont {J.~D.}\ \bibnamefont {Gallego}}, \bibinfo {author} {\bibfnamefont {J.~M.}\ \bibnamefont {García-Barceló}}, \bibinfo {author} {\bibfnamefont {B.}~\bibnamefont {Gimeno}}, \bibinfo {author} {\bibfnamefont {J.}~\bibnamefont {Golm}}, \bibinfo {author} {\bibfnamefont {X.}~\bibnamefont {Granados}}, \bibinfo {author} {\bibfnamefont {J.}~\bibnamefont {Gutierrez}}, \bibinfo {author} {\bibfnamefont {L.}~\bibnamefont {Herwig}}, \bibinfo {author} {\bibfnamefont {I.~G.}\ \bibnamefont {Irastorza}}, \bibinfo
  {author} {\bibfnamefont {N.}~\bibnamefont {Lamas}}, \bibinfo {author} {\bibfnamefont {A.}~\bibnamefont {Lozano-Guerrero}}, \bibinfo {author} {\bibfnamefont {W.~L.}\ \bibnamefont {Millar}}, \bibinfo {author} {\bibfnamefont {C.}~\bibnamefont {Malbrunot}}, \bibinfo {author} {\bibfnamefont {J.}~\bibnamefont {Miralda-Escudé}}, \bibinfo {author} {\bibfnamefont {P.}~\bibnamefont {Navarro}}, \bibinfo {author} {\bibfnamefont {J.~R.}\ \bibnamefont {Navarro-Madrid}}, \bibinfo {author} {\bibfnamefont {T.}~\bibnamefont {Puig}}, \bibinfo {author} {\bibfnamefont {M.}~\bibnamefont {Siodlaczek}}, \bibinfo {author} {\bibfnamefont {G.~T.}\ \bibnamefont {Telles}},\ and\ \bibinfo {author} {\bibfnamefont {W.}~\bibnamefont {Wuensch}},\ }\href {https://arxiv.org/abs/2403.07790} {\bibinfo {title} {Rades axion search results with a high-temperature superconducting cavity in an 11.7 t magnet}} (\bibinfo {year} {2024}),\ \Eprint {https://arxiv.org/abs/2403.07790} {arXiv:2403.07790 [hep-ex]} \BibitemShut {NoStop}%
\bibitem [{\citenamefont {McAllister}\ \emph {et~al.}(2017)\citenamefont {McAllister}, \citenamefont {Flower}, \citenamefont {Ivanov}, \citenamefont {Goryachev}, \citenamefont {Bourhill},\ and\ \citenamefont {Tobar}}]{McAllister:2017lkb}%
  \BibitemOpen
  \bibfield  {author} {\bibinfo {author} {\bibfnamefont {B.~T.}\ \bibnamefont {McAllister}}, \bibinfo {author} {\bibfnamefont {G.}~\bibnamefont {Flower}}, \bibinfo {author} {\bibfnamefont {E.~N.}\ \bibnamefont {Ivanov}}, \bibinfo {author} {\bibfnamefont {M.}~\bibnamefont {Goryachev}}, \bibinfo {author} {\bibfnamefont {J.}~\bibnamefont {Bourhill}},\ and\ \bibinfo {author} {\bibfnamefont {M.~E.}\ \bibnamefont {Tobar}},\ }\bibfield  {title} {\bibinfo {title} {{The ORGAN Experiment: An axion haloscope above 15 GHz}},\ }\href {https://doi.org/10.1016/j.dark.2017.09.010} {\bibfield  {journal} {\bibinfo  {journal} {Phys. Dark Univ.}\ }\textbf {\bibinfo {volume} {18}},\ \bibinfo {pages} {67} (\bibinfo {year} {2017})},\ \Eprint {https://arxiv.org/abs/1706.00209} {arXiv:1706.00209 [physics.ins-det]} \BibitemShut {NoStop}%
\bibitem [{\citenamefont {Quiskamp}\ \emph {et~al.}(2022)\citenamefont {Quiskamp}, \citenamefont {McAllister}, \citenamefont {Altin}, \citenamefont {Ivanov}, \citenamefont {Goryachev},\ and\ \citenamefont {Tobar}}]{Quiskamp:2022pks}%
  \BibitemOpen
  \bibfield  {author} {\bibinfo {author} {\bibfnamefont {A.~P.}\ \bibnamefont {Quiskamp}}, \bibinfo {author} {\bibfnamefont {B.~T.}\ \bibnamefont {McAllister}}, \bibinfo {author} {\bibfnamefont {P.}~\bibnamefont {Altin}}, \bibinfo {author} {\bibfnamefont {E.~N.}\ \bibnamefont {Ivanov}}, \bibinfo {author} {\bibfnamefont {M.}~\bibnamefont {Goryachev}},\ and\ \bibinfo {author} {\bibfnamefont {M.~E.}\ \bibnamefont {Tobar}},\ }\bibfield  {title} {\bibinfo {title} {{Direct search for dark matter axions excluding ALP cogenesis in the 63- to 67-\ensuremath{\mu}eV range with the ORGAN experiment}},\ }\href {https://doi.org/10.1126/sciadv.abq3765} {\bibfield  {journal} {\bibinfo  {journal} {Sci. Adv.}\ }\textbf {\bibinfo {volume} {8}},\ \bibinfo {pages} {abq3765} (\bibinfo {year} {2022})},\ \Eprint {https://arxiv.org/abs/2203.12152} {arXiv:2203.12152 [hep-ex]} \BibitemShut {NoStop}%
\bibitem [{\citenamefont {Quiskamp}\ \emph {et~al.}(2023)\citenamefont {Quiskamp}, \citenamefont {McAllister}, \citenamefont {Altin}, \citenamefont {Ivanov}, \citenamefont {Goryachev},\ and\ \citenamefont {Tobar}}]{quiskamp2023exclusionalpcogenesisdark}%
  \BibitemOpen
  \bibfield  {author} {\bibinfo {author} {\bibfnamefont {A.}~\bibnamefont {Quiskamp}}, \bibinfo {author} {\bibfnamefont {B.~T.}\ \bibnamefont {McAllister}}, \bibinfo {author} {\bibfnamefont {P.}~\bibnamefont {Altin}}, \bibinfo {author} {\bibfnamefont {E.~N.}\ \bibnamefont {Ivanov}}, \bibinfo {author} {\bibfnamefont {M.}~\bibnamefont {Goryachev}},\ and\ \bibinfo {author} {\bibfnamefont {M.~E.}\ \bibnamefont {Tobar}},\ }\href {https://arxiv.org/abs/2310.00904} {\bibinfo {title} {Exclusion of alp cogenesis dark matter in a mass window above 100 $\mu$ev}} (\bibinfo {year} {2023}),\ \Eprint {https://arxiv.org/abs/2310.00904} {arXiv:2310.00904 [hep-ex]} \BibitemShut {NoStop}%
\bibitem [{\citenamefont {Quiskamp}\ \emph {et~al.}(2024)\citenamefont {Quiskamp}, \citenamefont {Flower}, \citenamefont {Samuels}, \citenamefont {McAllister}, \citenamefont {Altin}, \citenamefont {Ivanov}, \citenamefont {Goryachev},\ and\ \citenamefont {Tobar}}]{quiskamp2024nearquantumlimitedaxiondark}%
  \BibitemOpen
  \bibfield  {author} {\bibinfo {author} {\bibfnamefont {A.~P.}\ \bibnamefont {Quiskamp}}, \bibinfo {author} {\bibfnamefont {G.}~\bibnamefont {Flower}}, \bibinfo {author} {\bibfnamefont {S.}~\bibnamefont {Samuels}}, \bibinfo {author} {\bibfnamefont {B.~T.}\ \bibnamefont {McAllister}}, \bibinfo {author} {\bibfnamefont {P.}~\bibnamefont {Altin}}, \bibinfo {author} {\bibfnamefont {E.~N.}\ \bibnamefont {Ivanov}}, \bibinfo {author} {\bibfnamefont {M.}~\bibnamefont {Goryachev}},\ and\ \bibinfo {author} {\bibfnamefont {M.~E.}\ \bibnamefont {Tobar}},\ }\href {https://arxiv.org/abs/2407.18586} {\bibinfo {title} {Near-quantum limited axion dark matter search with the organ experiment around 26 $\mu$ev}} (\bibinfo {year} {2024}),\ \Eprint {https://arxiv.org/abs/2407.18586} {arXiv:2407.18586 [hep-ex]} \BibitemShut {NoStop}%
\bibitem [{\citenamefont {dos Santos~Garcia}\ \emph {et~al.}(2024)\citenamefont {dos Santos~Garcia}, \citenamefont {Bergermann}, \citenamefont {Caldwell}, \citenamefont {Dabhi}, \citenamefont {Diaconu}, \citenamefont {Diehl}, \citenamefont {Dvali}, \citenamefont {Egge}, \citenamefont {Garutti}, \citenamefont {Heyminck}, \citenamefont {Hubaut}, \citenamefont {Ivanov}, \citenamefont {Jochum}, \citenamefont {Knirck}, \citenamefont {Kramer}, \citenamefont {Kreikemeyer-Lorenzo}, \citenamefont {Krieger}, \citenamefont {Lee}, \citenamefont {Leppla-Weber}, \citenamefont {Li}, \citenamefont {Lindner}, \citenamefont {Majorovits}, \citenamefont {Maldonado}, \citenamefont {Martini}, \citenamefont {Miyazaki}, \citenamefont {Öz}, \citenamefont {Pralavorio}, \citenamefont {Raffelt}, \citenamefont {Redondo}, \citenamefont {Ringwald}, \citenamefont {Schaffran}, \citenamefont {Schmidt}, \citenamefont {Steffen}, \citenamefont {Strandhagen}, \citenamefont {Usherov}, \citenamefont {Wang},\ and\ \citenamefont
  {Wieching}}]{garcia2024searchaxiondarkmatter}%
  \BibitemOpen
  \bibfield  {author} {\bibinfo {author} {\bibfnamefont {B.~A.}\ \bibnamefont {dos Santos~Garcia}}, \bibinfo {author} {\bibfnamefont {D.}~\bibnamefont {Bergermann}}, \bibinfo {author} {\bibfnamefont {A.}~\bibnamefont {Caldwell}}, \bibinfo {author} {\bibfnamefont {V.}~\bibnamefont {Dabhi}}, \bibinfo {author} {\bibfnamefont {C.}~\bibnamefont {Diaconu}}, \bibinfo {author} {\bibfnamefont {J.}~\bibnamefont {Diehl}}, \bibinfo {author} {\bibfnamefont {G.}~\bibnamefont {Dvali}}, \bibinfo {author} {\bibfnamefont {J.}~\bibnamefont {Egge}}, \bibinfo {author} {\bibfnamefont {E.}~\bibnamefont {Garutti}}, \bibinfo {author} {\bibfnamefont {S.}~\bibnamefont {Heyminck}}, \bibinfo {author} {\bibfnamefont {F.}~\bibnamefont {Hubaut}}, \bibinfo {author} {\bibfnamefont {A.}~\bibnamefont {Ivanov}}, \bibinfo {author} {\bibfnamefont {J.}~\bibnamefont {Jochum}}, \bibinfo {author} {\bibfnamefont {S.}~\bibnamefont {Knirck}}, \bibinfo {author} {\bibfnamefont {M.}~\bibnamefont {Kramer}}, \bibinfo {author} {\bibfnamefont {D.}~\bibnamefont
  {Kreikemeyer-Lorenzo}}, \bibinfo {author} {\bibfnamefont {C.}~\bibnamefont {Krieger}}, \bibinfo {author} {\bibfnamefont {C.}~\bibnamefont {Lee}}, \bibinfo {author} {\bibfnamefont {D.}~\bibnamefont {Leppla-Weber}}, \bibinfo {author} {\bibfnamefont {X.}~\bibnamefont {Li}}, \bibinfo {author} {\bibfnamefont {A.}~\bibnamefont {Lindner}}, \bibinfo {author} {\bibfnamefont {B.}~\bibnamefont {Majorovits}}, \bibinfo {author} {\bibfnamefont {J.~P.~A.}\ \bibnamefont {Maldonado}}, \bibinfo {author} {\bibfnamefont {A.}~\bibnamefont {Martini}}, \bibinfo {author} {\bibfnamefont {A.}~\bibnamefont {Miyazaki}}, \bibinfo {author} {\bibfnamefont {E.}~\bibnamefont {Öz}}, \bibinfo {author} {\bibfnamefont {P.}~\bibnamefont {Pralavorio}}, \bibinfo {author} {\bibfnamefont {G.}~\bibnamefont {Raffelt}}, \bibinfo {author} {\bibfnamefont {J.}~\bibnamefont {Redondo}}, \bibinfo {author} {\bibfnamefont {A.}~\bibnamefont {Ringwald}}, \bibinfo {author} {\bibfnamefont {J.}~\bibnamefont {Schaffran}}, \bibinfo {author} {\bibfnamefont
  {A.}~\bibnamefont {Schmidt}}, \bibinfo {author} {\bibfnamefont {F.}~\bibnamefont {Steffen}}, \bibinfo {author} {\bibfnamefont {C.}~\bibnamefont {Strandhagen}}, \bibinfo {author} {\bibfnamefont {I.}~\bibnamefont {Usherov}}, \bibinfo {author} {\bibfnamefont {H.}~\bibnamefont {Wang}},\ and\ \bibinfo {author} {\bibfnamefont {G.}~\bibnamefont {Wieching}},\ }\href {https://arxiv.org/abs/2409.11777} {\bibinfo {title} {First search for axion dark matter with a madmax prototype}} (\bibinfo {year} {2024}),\ \Eprint {https://arxiv.org/abs/2409.11777} {arXiv:2409.11777 [hep-ex]} \BibitemShut {NoStop}%
\bibitem [{\citenamefont {Jaeckel}\ and\ \citenamefont {Redondo}(2013)}]{Jaeckel:2013sqa}%
  \BibitemOpen
  \bibfield  {author} {\bibinfo {author} {\bibfnamefont {J.}~\bibnamefont {Jaeckel}}\ and\ \bibinfo {author} {\bibfnamefont {J.}~\bibnamefont {Redondo}},\ }\bibfield  {title} {\bibinfo {title} {{An antenna for directional detection of WISPy dark matter}},\ }\href {https://doi.org/10.1088/1475-7516/2013/11/016} {\bibfield  {journal} {\bibinfo  {journal} {JCAP}\ }\textbf {\bibinfo {volume} {11}},\ \bibinfo {pages} {016}},\ \Eprint {https://arxiv.org/abs/1307.7181} {arXiv:1307.7181 [hep-ph]} \BibitemShut {NoStop}%
%%CITATION = ARXIV:1307.7181;%%
\bibitem [{\citenamefont {Jaeckel}\ and\ \citenamefont {Knirck}(2016)}]{Jaeckel:2015kea}%
  \BibitemOpen
  \bibfield  {author} {\bibinfo {author} {\bibfnamefont {J.}~\bibnamefont {Jaeckel}}\ and\ \bibinfo {author} {\bibfnamefont {S.}~\bibnamefont {Knirck}},\ }\bibfield  {title} {\bibinfo {title} {{Directional Resolution of Dish Antenna Experiments to Search for WISPy Dark Matter}},\ }\href {https://doi.org/10.1088/1475-7516/2016/01/005} {\bibfield  {journal} {\bibinfo  {journal} {JCAP}\ }\textbf {\bibinfo {volume} {01}},\ \bibinfo {pages} {005}},\ \Eprint {https://arxiv.org/abs/1509.00371} {arXiv:1509.00371 [hep-ph]} \BibitemShut {NoStop}%
%%CITATION = ARXIV:1509.00371;%%
\bibitem [{\citenamefont {Jaeckel}\ and\ \citenamefont {Knirck}(2017)}]{Jaeckel:2017sjb}%
  \BibitemOpen
  \bibfield  {author} {\bibinfo {author} {\bibfnamefont {J.}~\bibnamefont {Jaeckel}}\ and\ \bibinfo {author} {\bibfnamefont {S.}~\bibnamefont {Knirck}},\ }\bibfield  {title} {\bibinfo {title} {{Dish Antenna Searches for WISPy Dark Matter: Directional Resolution Small Mass Limitations}},\ }in\ \href {https://doi.org/10.3204/DESY-PROC-2009-03/Knirck_Stefan} {\emph {\bibinfo {booktitle} {{Proceedings, 12th Patras Workshop on Axions, WIMPs and WISPs (PATRAS 2016): Jeju Island, South Korea, June 20-24, 2016}}}}\ (\bibinfo {year} {2017})\ pp.\ \bibinfo {pages} {78--81},\ \Eprint {https://arxiv.org/abs/1702.04381} {arXiv:1702.04381 [hep-ph]} \BibitemShut {NoStop}%
%%CITATION = ARXIV:1702.04381;%%
\bibitem [{\citenamefont {Diehl}\ \emph {et~al.}(2023)\citenamefont {Diehl}, \citenamefont {Knollm\"uller},\ and\ \citenamefont {Schulz}}]{Diehl:2023fuk}%
  \BibitemOpen
  \bibfield  {author} {\bibinfo {author} {\bibfnamefont {J.}~\bibnamefont {Diehl}}, \bibinfo {author} {\bibfnamefont {J.}~\bibnamefont {Knollm\"uller}},\ and\ \bibinfo {author} {\bibfnamefont {O.}~\bibnamefont {Schulz}},\ }\bibfield  {title} {\bibinfo {title} {{Bias-Free Estimation of Signals on Top of Unknown Backgrounds}},\ }\href@noop {} {\  (\bibinfo {year} {2023})},\ \Eprint {https://arxiv.org/abs/2306.17667} {arXiv:2306.17667 [astro-ph.IM]} \BibitemShut {NoStop}%
\bibitem [{\citenamefont {{Knirck}}\ and\ \citenamefont {{ADMX Collaboration Team}}(2023)}]{ADMX-EFR}%
  \BibitemOpen
  \bibfield  {author} {\bibinfo {author} {\bibfnamefont {S.}~\bibnamefont {{Knirck}}}\ and\ \bibinfo {author} {\bibnamefont {{ADMX Collaboration Team}}},\ }\bibfield  {title} {\bibinfo {title} {{ADMX Extended Frequency Range (EFR): Searching for 2-4GHz axions with 18 cavities}},\ }in\ \href@noop {} {\emph {\bibinfo {booktitle} {APS April Meeting Abstracts}}},\ \bibinfo {series} {APS Meeting Abstracts}, Vol.\ \bibinfo {volume} {2023}\ (\bibinfo {year} {2023})\ p.\ \bibinfo {pages} {CCC01.002}\BibitemShut {NoStop}%
\end{thebibliography}%
\end{document}